\documentclass{iopjournal}
\makeatletter
\AtBeginDocument{%
  \pagestyle{plain}%
  \def\@oddhead{}%
  \def\@evenhead{}%
  \def\@oddfoot{\hfil\thepage\hfil}%
  \def\@evenfoot{\hfil\thepage\hfil}%
}
\renewcommand{\orcid}[1]{}%
\makeatother
\usepackage{enumitem}
\usepackage{cite}
\usepackage{amsmath,amssymb,amsfonts}
\usepackage{algorithmic}
\usepackage{graphicx}
\usepackage{textcomp}
\usepackage{xcolor}
\usepackage{physics}
\usepackage{subcaption}
\usepackage{wrapfig,multirow}
\usepackage{xcolor,colortbl}
\usepackage{makecell}
\usepackage{enumitem}
\usepackage{rotating}
\usepackage{array}
\usepackage{makecell}
\usepackage{textcomp}
\usepackage{mathtools}

\usepackage{appendix}
\usepackage{chngcntr}
\usepackage{hyperref}
\usepackage{etoolbox}

\counterwithin*{section}{part}
\counterwithin*{@ppsaveapp}{part}

\def\BibTeX{{\rm B\kern-.05em{\sc i\kern-.025em b}\kern-.08em
    T\kern-.1667em\lower.7ex\hbox{E}\kern-.125emX}}

\newcommand{\Ca}{$^{40}\textrm{Ca}^+$ }

\begin{document}


\title{Theoretical Analysis and Simulations of Memory-based and All-photonic Quantum Repeaters and Networks}

\author{Chuen Hei Chan$^{1,2,\dagger}$\orcid{0009-0001-9890-5582}, Charu Jain$^{2,\dagger}$\orcid{0009-0004-2609-6511}, Ezra Kissel$^2$\orcid{0000-0003-3972-9651}, Wenji Wu$^{2,*}$\orcid{0000-0002-6194-6276}, Edwin Barnes$^1$\orcid{0000-0003-1666-9385}, Sophia E. Economou$^{1,*}$\orcid{0000-0002-1939-5589}, and Inder Monga$^2$\orcid{0000-0003-4524-0457}}

\affil{$^1$Department of Physics, Virginia Tech, Blacksburg, Virginia, USA}

\affil{$^2$Scientific Networking Division, Lawrence Berkeley National Laboratory, Berkeley, USA}

\affil{$^\dagger$These authors contributed equally to this work.}

\affil{$^*$Author to whom any correspondence should be addressed.}

\email{wenji@lbl.gov, economou@vt.edu}

\begin{center}
\small (Dated: \today)
\end{center}
\begin{abstract}
Developing and deploying advanced Quantum Repeater (QR) technologies will be necessary to scale quantum networks to longer distances. Depending on the error mitigation mechanisms adopted to suppress loss and errors, QRs are typically classified into memory-based or all-photonic QRs; and each type of QR may be best suited for a specific type of underlying quantum technology, a particular scale of quantum networks, or a specific regime of operational parameters. We perform theoretical analysis and simulations of quantum repeaters and networks to
investigate the relative performance and resource requirements of different quantum network paradigms. Our results will help guide the optimization of quantum hardware and components and shed light on the role of a robust control plane. We present our  
research findings on theoretical analysis and simulations of memory-based first-generation trapped-ion quantum repeaters and networks, and all-photonic entanglement-based quantum repeaters and networks. We study the relative performance in terms of entanglement generation rate and fidelity, as well as the resource requirements of these two different quantum network paradigms.  
\end{abstract}

\section{Introduction}
Quantum networks are envisioned to achieve novel capabilities that are provably impossible using their classical counterparts and could be transformative to science, the economy, and national security~\cite{azuma2023quantum}. These novel capabilities range from key distribution and other cryptographic tasks \cite{bennett2014quantum}, sensing and metrology \cite{komar2014quantum}, to distributed systems \cite{ben2005fast,denchev2008distributed}. Like classical networks, quantum networks must be large enough to facilitate the desired level of communication, information sharing, or collaboration. Qubits cannot be copied due to the no-cloning theorem, which forbids the creation of identical copies of an arbitrary unknown quantum state. Therefore, qubits cannot be physically transmitted over long distances without being hindered by the effects of signal loss and decoherence inherent in transport mediums such as optical fiber. To scale quantum networks to longer distances, advanced QR technologies are required. The major function of QRs is to divide the end-to-end distance of quantum links into shorter intermediate segments connected by QRs, in which errors from fiber attenuation (i.e., loss) and other sources (e.g., gate operation errors) can be corrected. Depending on the error mitigation mechanisms adopted to suppress loss and operation errors, QRs are typically classified into three generations – 1G, 2G, and 3G \cite{azuma2023quantum, dur1999quantum, jiang2009quantum, muralidharan2016optimal, munro2015inside, munro2012quantum, muralidharan2014ultrafast}. 1G and 2G rely on quantum memories (i.e., matter qubits) to store information for long times. In 2G, quantum error correction is employed, which reduces the requirements for two-way communication between the end-nodes of the network (the requirement that lowers the rates in 1G repeaters). 3G is based on photonic encoding and is typically one-way (i.e., no entanglement is typically established between the nodes). An alternative all-photonic approach~\cite{azuma2015all} relies on entanglement swapping between adjacent nodes. This type of QR is termed \textit{All-Photonic Entanglement-based Quantum Repeater (APE QR)}. Each generation of QRs may be best suited for a specific type of underlying quantum technology, for a particular scale of quantum networks, or for a specific regime of operational parameters such as local gate speed and gate fidelity. This leads to a critical question \textit{“what are the relative performance and resource requirements of different types of QRs and networks under various conditions?"}.   

Ideally, research on quantum networks should be carried out in real network environments, which provide practical experiences and lessons. However, this approach has a few constraints given the scale and state of current testbeds. First, building quantum network testbeds is expensive and time-consuming. To date, only a few quantum network testbeds have been, or are being, built around the world, with each having its own proprietary techniques \cite{monga2023quant, knaut2024entanglement, bersin2024development, pompili2021realization, chen2021integrated}. These testbeds are typically not open to external researchers. Second, due to constraints of existing quantum technologies, many advanced quantum network protocols and schemes, such as 2G, 3G, and APE QR protocols, remain in the conceptual stage and cannot be realized in the current quantum network testbeds. Third, existing quantum network testbeds have only a few nodes and very limited capabilities. Thus, it is impossible to conduct quantum network scalability research in such environments. Instead, theoretical analysis and simulations of quantum networks offer a powerful and cost-effective alternative to study and research quantum networks without requiring physical networks. 


In this paper, we present our research findings on theoretical analysis and simulations of memory-based first generation trapped-ion QRs and networks, and APE (based on deterministic photon emission from emitters) QRs and networks, as well as the relative performance and resource requirements of these two different paradigms. The performance metrics include entanglement generation rate and fidelity between network end nodes. For a fair comparison of resource requirements of different quantum network paradigms, entanglement generation rate is normalized by the total number of matter qubits involved in the end nodes (e.g., trapped ions and/or nuclear spins), which will be defined in later section. 

The contributions of our work are the following:

\begin{itemize}[nosep]
    \item We build theoretical models for performance analysis of different quantum network paradigms. Our theoretical models, which have been verified and validated by using simulation results and/or experimental data, can be used to guide the design and construction of quantum networks.
    \item We study and develop simulation models for different quantum network paradigms, incorporating models for mechanisms of control, entanglement generation and distribution, as well as extensible models for translating low-level physical features to the digital domain of quantum network operations. We carefully verify and validate our simulation models by using theoretical analysis and/or experimental data to ensure correctness and accuracy.  
    \item We employ a rigorous and holistic approach to the simulation and comparison of different quantum network paradigms. A unique advantage of our approach is that the consideration of the analog processes and dynamics of the QRs allows us to identify relative merits of each repeater paradigm for specific physical implementations (trapped ions, quantum dots, etc). 
    \item Our study and analysis identify important quantum network performance metrics including quantum network capacity, resource consumption requirements, and scalability. Such results can guide the research and development of QR technologies and the design and construction of large-scale quantum networks.
    \item Our software is open source and freely available \cite{QNPack}. With such software, new quantum network design, protocols, and architectures can be studied.
\end{itemize} 

The rest of the paper is organized as follows. After an overview of related work in Section \ref{related},  Section~\ref{background} presents memory-based trapped-ion QRs and networks, and all-photonic APE QRs and networks. Theoretical analysis of these two quantum network paradigms is given in Section~\ref{sec:theoretic_analysis}. Section \ref{design} discusses the design and implementation of these two paradigms using NetSquid. Section \ref{analysis} presents simulation analysis. Section \ref{conclusion} concludes the paper.   

\section{Related work}
\label{related}

\subsection{Quantum network modeling and simulation}
Quantum network modeling and simulation have received increasing attention among researchers in recent years. A few quantum network modeling and simulation tools, such as QuISP \cite{satoh2022quisp}, NetSquid \cite{coopmans2021netsquid}, SeQueNCe \cite{wu2021sequence}, QuNetSim \cite{diadamo2021qunetsim}, and SQUANCH \cite{bartlett2018distributed}, have been developed. Researchers have used these tools and numerical simulations to study quantum network architectures and protocol stacks. These studies include quantum network design \cite{huie2021multiplexed}, protocols \cite{kozlowski2020designing, dahlberg2019link}, routing \cite{chakraborty2019distributed, shi2020concurrent}, capacity \cite{pirandola2019end}, benchmarking \cite{helsen2023benchmarking, brito2020statistical}, space-based quantum communication \cite{wang2022exploiting}, error management \cite{nagayama2021towards}, and quantum network coding \cite{pathumsoot2020modeling}.

\subsection{NetSquid}
NetSquid is a software platform for simulating quantum networks and modular quantum computing systems developed at QuTech \cite{coopmans2021netsquid}. It leverages a discrete-event simulation engine and a rich quantum computing library to accurately model time-dependent behaviors and physical imperfections of quantum hardware like repeaters, channels, and memories. NetSquid's modular design, where components can be modeled like building blocks, enables detailed simulations of the physical layer, classical control plane, and application level protocols, allowing researchers to explore network performance, optimize architectures, validate protocols, and evaluate the feasibility of quantum applications on realistic hardware.

\subsection{Research on quantum repeaters and networks}
Networks based on memory-based 1G repeaters have been extensively studied in the literature \cite{azuma2023quantum, bernardes2011rate, wang2023efficient, Avis_2023, PRXQuantum.5.010351, P_rez_Castro_2024}. However, much of this prior work relies on overly simplifying assumptions: the QR dynamics are not considered microscopically, and the detailed interactions between different network components are also not taken into account, potentially leading to an incomplete understanding of network rates and scaling. In contrast, this paper proposes a detailed component model for the QR platforms under evaluation.

 Networks based on APE repeaters, with matter-based emitters deterministically generating photonic resource states, have also been analyzed and compared to memory-based repeaters~\cite{hilaire2021resource}. This paradigm is distinct from fully all-optical repeater proposals, which rely exclusively on single-photon sources and linear optics~\cite{pant2017rate}. While Ref.~\cite{hilaire2021resource} took the first step toward quantifying the performance of APE QRs relying on emitter-generated resource states, it did not incorporate a microscopic description of the QR dynamics. Building on this line of work, our study advances the modeling and comparison in several key respects. First, we extend the performance comparison by accounting for the explicit memory-qubit demands at the end nodes for the APE network. Second, rather than adopting a network-layer model derived solely from an abstract repeater description, we explicitly include the physical communication layer and quantum link layer, thereby providing rate estimates that more directly reflect implementation-level constraints. Finally, we benchmark APE repeaters against a 1G repeater model grounded in practical considerations for a specific matter-qubit platform, enabling a more realistic and technologically informed comparison.

\section{Memory-based and all-photonic quantum repeaters and networks}
\label{background}

In this section, we present memory-based 1G trapped ion QRs and networks, and APE QRs and networks. A few key mechanisms and protocols are presented to illustrate how a 1G trapped ion or an APE quantum network works. 

Before proceeding, we first introduce the major components that constitute a quantum network: (a) \textit{Quantum end nodes (Q-nodes)} are much like their classic counterparts (i.e., nodes), representing the communication parties in a quantum network. (b) \textit{Quantum repeaters} can extend a quantum network to a larger distance by mitigating loss and correcting errors. (c) \textit{Bell-State-Measurement nodes (BSM-nodes)} can perform Bell state measurement (BSM) and local Pauli operations for incoming photons. And (d) \textit{Quantum (classical) channels.} Q-nodes, QRs, and BSM-nodes are connected to each other through optical fibers. Dedicated wavelengths of these fibers are used as quantum (classical) channels to transmit quantum (classical) signals between them.   

    \subsection{1G trapped ion QRs and networks}
    \label{sec:trap_ion_network}

    Memory-based QRs are based on \textit{heralded-entanglement generation} between quantum memories of neighboring QRs to generate entanglements of short distance, and \textit{entanglement swapping} between quantum memories within QRs to concatenate entanglements of short distance into end-to-end entanglements of long distance. Candidates for quantum memories in QRs include trapped ions~\cite{krutyanskiy2023entanglement, stephenson2020high, inlek2017multispecies}, NV centers in diamond~\cite{pompili2021realization,hermans2022qubit}, silicon-vacancy center in diamond~\cite{borregaard2020one, nguyen2019integrated} and defects in silicon carbide~\cite{bourassa2020entanglement,lukin2020integrated,ecker2024quantum,zhou2025silicon}, among which trapped ions are considered excellent candidates for quantum processors \cite{bruzewicz2019trapped, pino2021demonstration} due to their long coherence times, optical interfaces, and mature control implementations. A recent milestone experiment in quantum networks demonstrated the generation of entanglement between two nodes separated by $50$ km mediated through an intermediate repeater node based on trapped \Ca ions~\cite{krutyanskiy2023telecom}. Considering the importance of trapped ions in both quantum computing and quantum networking, we focus our efforts on this platform. Specifically, our analysis and simulation work is based on trapped \Ca ions~\cite{krutyanskiy2023telecom, schupp2021interface, krutyanskiy2023entanglement}, which represent the state of the art in this area. We incorporate the low-level physical processes into the detailed theoretical analysis and simulations of 1G trapped-ion QRs.
    
        \subsubsection{Heralded entanglement generation (HEG)}
        \label{sec:heg}

\begin{figure}[h]
        \centering
    	\includegraphics[width = 0.8\linewidth]{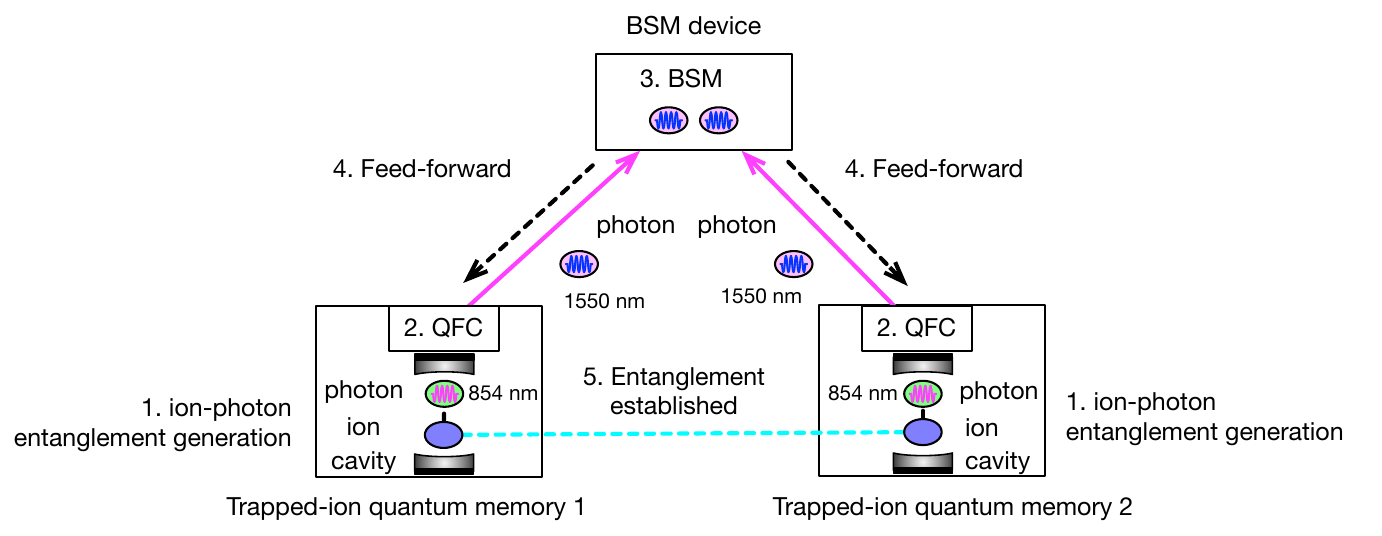}
        \caption{HEG between remote ions~\cite{monga2023quant}.}
        \label{fig:ion_ion_entanglement}
        \vspace{-4mm}
        \end{figure}

        Because entanglement generation is a probabilistic process, heralding allows for confirmation that an entangled state has been successfully created so that it can be used for further operations. A heralded ion-ion entanglement establishing process between two \Ca trapped-ion systems is illustrated in Fig.~\ref{fig:ion_ion_entanglement}. This process relies on the synchronous preparation of entanglement between the electronic degree of freedom of a trapped ion and the polarization of a photon within each trapped-ion system. Each cycle starts with ion state initialization, followed by a Raman excitation that triggers an emission of a 854 nm-photon into the cavity mode. After quantum frequency conversion, the resulting photon in the telecom-band, entangled with the ion, is transmitted towards a BSM device at an intermediate node in an attempt to create remote ion-ion entanglement via a projective measurement in the Bell basis of the polarization states. If a two-fold coincidence has been detected at the BSM device, it signals projection into an ion-ion entangled Bell state and each node is instructed to stop to preserve entanglement. This ion-ion entanglement is stored as a resource for future use. If no Bell state has been detected, each system continues the Raman excitation attempts. The process repeats up to a maximum number of attempts, after which the process exits with failure.

        \subsubsection{Entanglement swapping and end-to-end entanglement generation} 
        \label{sec:ent_swapping}
\begin{figure}[h]
        \centering
    	\includegraphics[width = 0.5\linewidth]{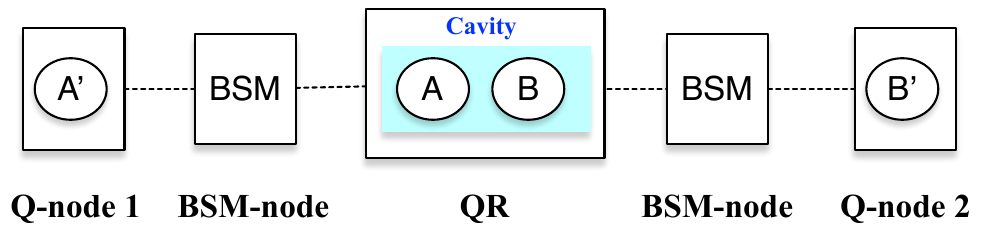}
    	\caption{An 1G quantum repeater and network based on trapped \Ca ions.}
    	\label{fig:trapped_ion_network}
        \end{figure}

        Entanglement swapping concatenates short-distance entanglements into end-to-end long-distance entanglement. Based on the recent experimental realization of a quantum repeater based on trapped \Ca ions~\cite{krutyanskiy2023telecom}, we review how a 1G quantum repeater performs such a function. As illustrated in Fig.~\ref{fig:trapped_ion_network}, an elementary quantum network consists of two Q-nodes, two BSM-nodes, and a QR. Q-nodes 1 and 2 each have one \Ca ion, which we refer to as A' and B', respectively. The QR is based on two \Ca ions, A and B, which are trapped in a linear Paul trap. To achieve greater emission efficiency, the trap is coupled to a high-finesse optical cavity. Because both ion A and B are trapped in a linear Paul trap that is coupled to a cavity, HEG operations that involve ion A and B must be performed in sequence to avoid simultaneous 854 nm-photon emissions into the cavity.
        
\begin{figure}[h]
        \centering
    	\includegraphics[width =0.78\linewidth]{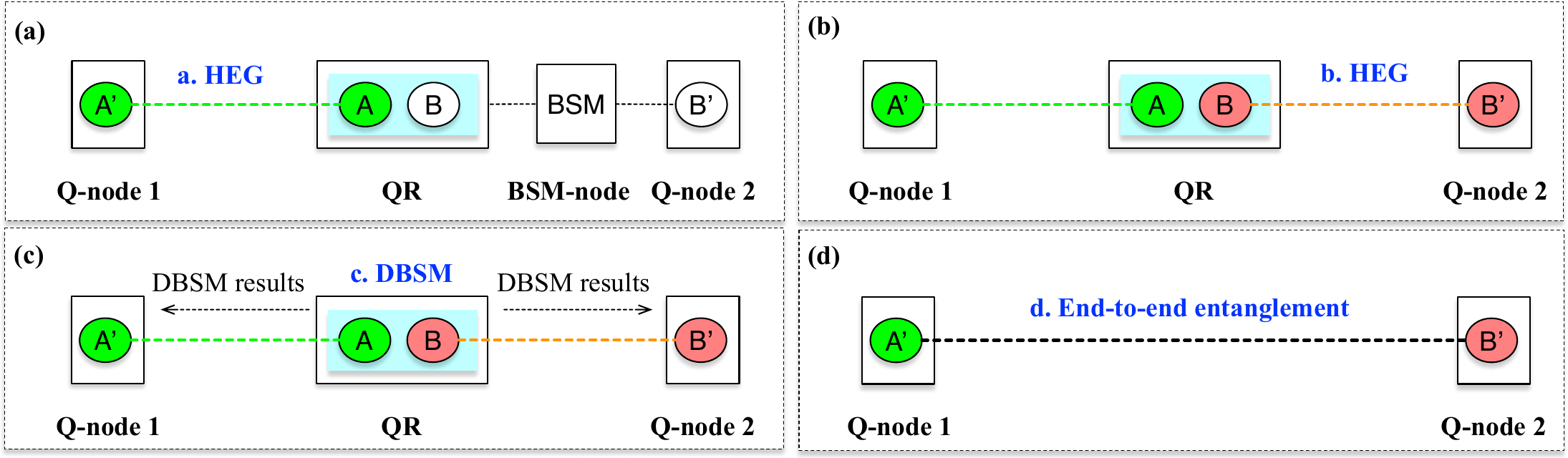}
        \caption{Entanglement swapping and end-to-end entanglement generation.}
    	\label{fig:entanglement_swapping}
        \end{figure}

        As illustrated in Fig.~\ref{fig:entanglement_swapping}, the end-to-end entanglement generation process between Q-node 1 and 2 works as follows. (a) HEG is performed between ion A' of Q-node 1 and ion A of the QR. A heralding signal confirms entanglement A' $\leftrightarrow$ A has been successfully generated. (b) HEG is performed between ion B of the QR and ion B' of Q-node 2. Similarly, a heralding signal confirms entanglement B $\leftrightarrow$ B' has been successfully created. (c) A Deterministic Bell State Measurement (DBSM) is performed between ion A and B at the QR, with measurement outcomes sent to Q-nodes 1 and 2, respectively. The DBSM is performed in two steps. First, a laser-driven two-qubit \text{Mølmer-Sørensen} (MS) logic gate~\cite{PhysRevLett.82.1971} is applied on A and B. Second, the logical state of ions A and B is measured via fluorescence detection. (d) Based on Bell state measurement outcomes, corresponding Pauli operations are applied on ion A' of Q-node 1 and ion B' of Q-node 2, respectively. End-to-end entanglement A' $\leftrightarrow$ B' is successfully established. 
        
       \subsubsection{Parallel HEGs in a repeater chain}
       \label{sec:parallel_hegs}
\begin{figure}[h]
        \centering
        \vspace{-0mm}
        \includegraphics[width =0.68\linewidth]{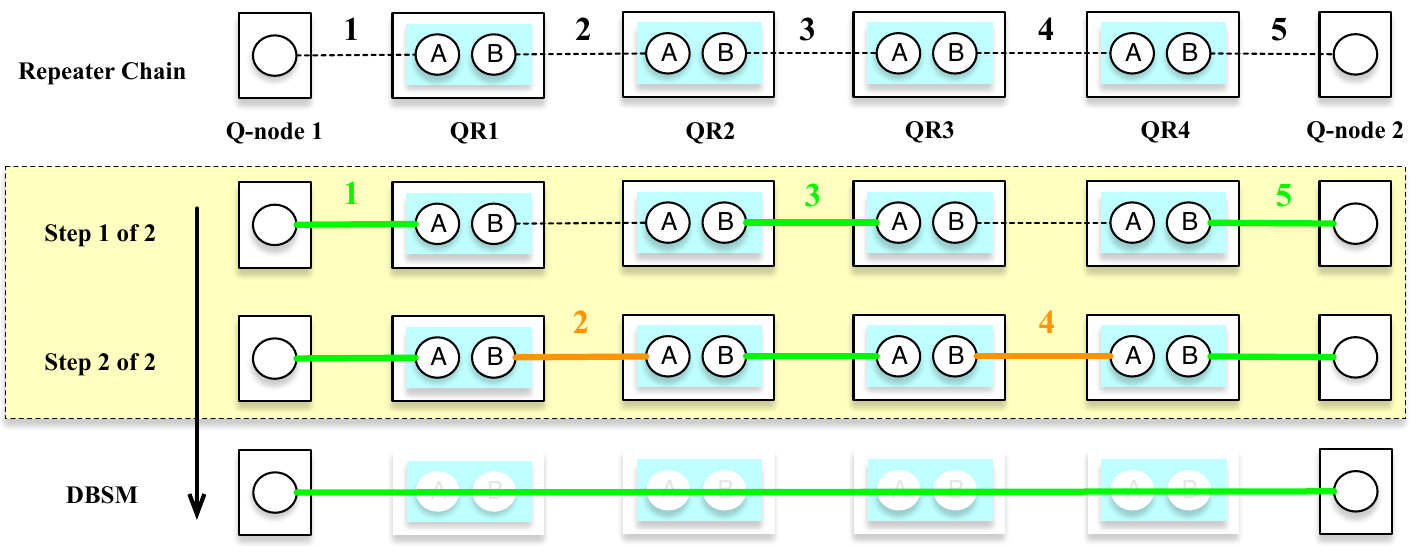}
        \caption{A two-step scheme to execute HEGs in a repeater chain. Parallel HEGs are executed on every other segment. The overall process can be completed in two steps, (1, 3, 5) in the 1st step and (2, 4) in the 2nd step. When any a HEG fails, the overall end-to-end entanglement generation process will restart.}
        \vspace{-2mm}
	\label{fig:HEG_in_parallel}
        \end{figure}

        As discussed in Section~\ref{sec:ent_swapping}, when both ions A and B are trapped in a linear trap that is coupled to a cavity, HEG operations that involve ions A and B must be performed sequentially to avoid simultaneous photon emission into the cavity. Therefore, to satisfy such a constraint, parallel HEG operations are not allowed in the neighboring segments along a QR chain. 
        While linear, hop-by-hop HEG may be an obvious first choice, a more efficient protocol is to conduct parallel HEG operations in every other segment. In this manner, all the HEG operations can be completed in two steps, regardless of the number of QRs involved in the chain. Fig.~\ref{fig:HEG_in_parallel} illustrates such an example. Realizing such a two-step scheme depends on centralized resource scheduling, the knowledge of global network information, and accurate network control.

    \subsection{APE QRs and networks}

     APE QRs are based on the idea that repeaters featuring quantum-error-correction (QEC) codes can be implemented all-photonically~\cite{azuma2015all}. In this approach, the functions of end-to-end entanglement establishment and QEC can be implemented by engineering photonic graph states, and the necessary signaling is less than required for memory-based QRs, improving the rates of information transmission and entanglement generation. 

        \begin{figure}[h]
        \centering
        \vspace{-2mm}
        \includegraphics[width = 0.6\linewidth]{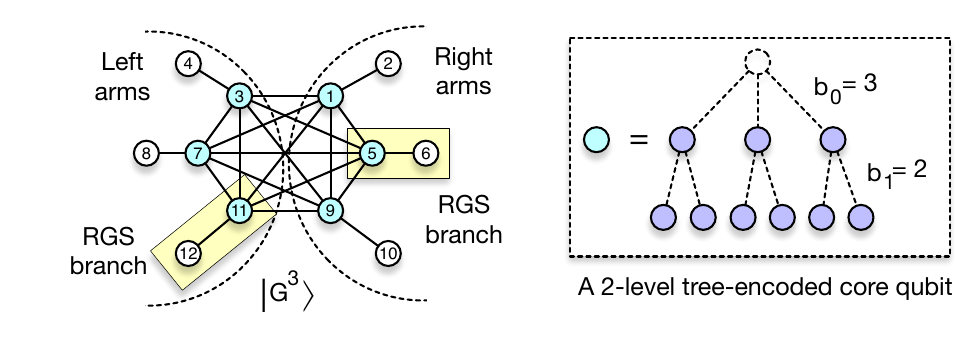}
        \caption{An example RGS $|G^3\rangle$ with RGS parameters $(m,b_0,b_1)=(3,3,2)$. Each RGS branch consists of a core qubit(light blue) and a leaf qubit(white). Each core qubit is encoded by a 2-level tree.} 
        \vspace{-3mm}
        \label{fig:RGS}
        \end{figure}
     
     Previous studies have shown that the generation of photonic graph states using linear optics and measurement-based feed-forward approaches would require formidable resource overheads~\cite{pant2017rate}, i.e., a huge number of single-photon sources are required in each repeater node. To solve this problem, alternative approaches based on quantum emitters have been proposed. It has been shown that the photonic graph states required for APE QRs can be generated by coupling a few quantum emitters to ancillary matter qubits~\cite{buterakos2017deterministic}. Initial quantitative performance analyses have demonstrated the potential of this ancillary-qubit-assisted graph state generation scheme in the realization of APE QRs~\cite{hilaire2021resource}. In this paper, we continue to explore this scheme through detailed theoretical analysis and simulations of APE QRs.  

    \begin{figure}[]
    \centering
	\includegraphics[width = 0.66\linewidth]{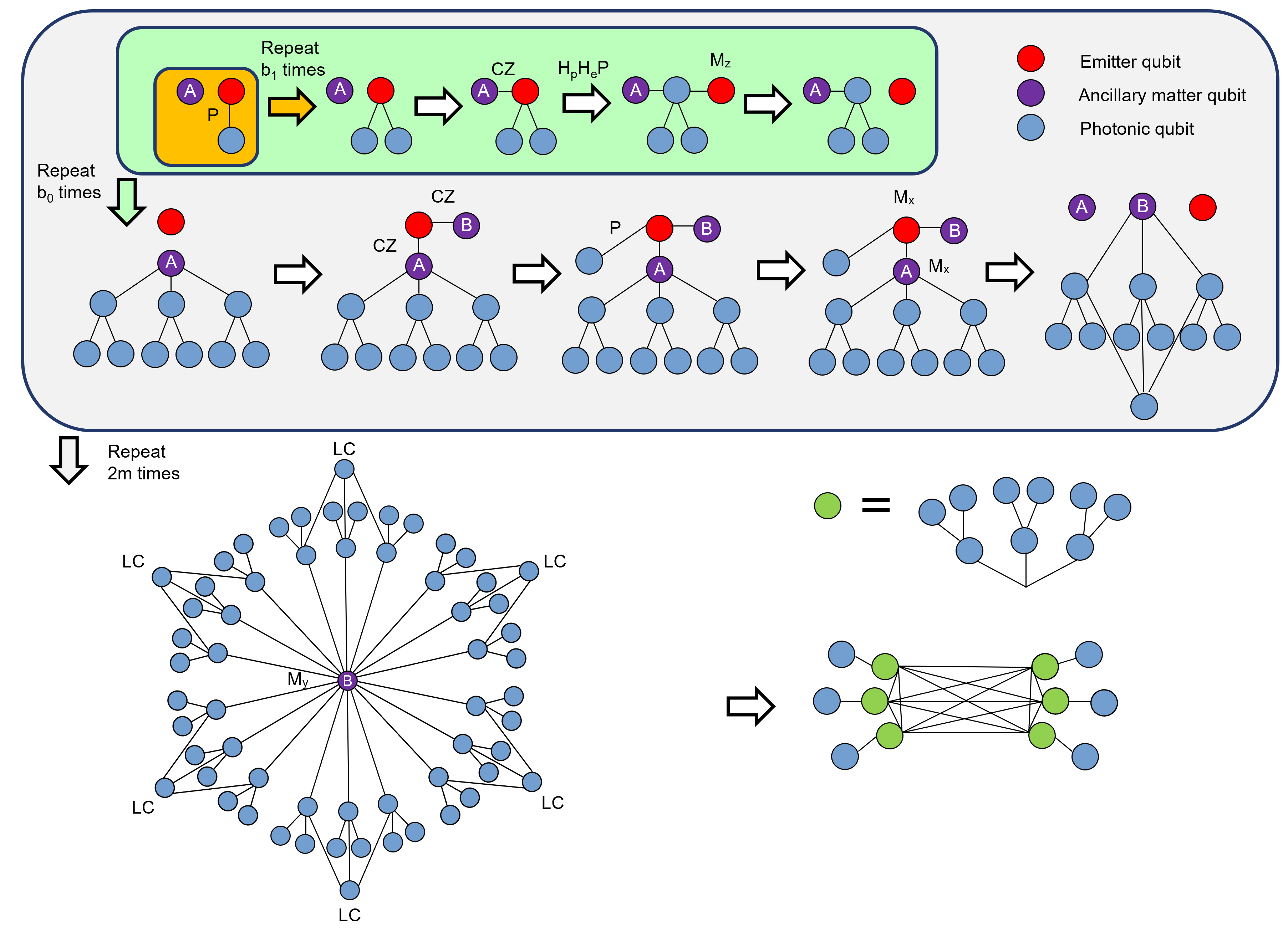}
	\caption{RGS generation from quantum emitters. The RGS parameters are $(m,b_0,b_1)=(3,3,2)$.}
	\label{fig:RGS_generation}
    \vspace{-3mm}
    \end{figure}
        
        \subsubsection{Repeater graph state}
        A graph state $\ket{G}$ is a type of multi-qubit state that can be represented by a set of vertices $V$ and edges $E$. It can be defined as follows:
        \begin{align}
        \ket{G}:=\prod_{(v,w)\in E}CZ_{v,w}\ket{+}^{\otimes |V|}.
        \end{align}
        A repeater graph state (RGS) $|G^m\rangle$ was proposed in \cite{azuma2015all} to be the resource state in an APE QR. It consists of 2$m$ branches, each containing a core qubit and a leaf qubit. The core qubits of all branches are fully connected to one another, and one leaf qubit is connected to each core qubit. In addition, tree-encoding~\cite{varnava2006loss} is used to encode each core qubit to protect against photon loss. For simplicity of notation, an RGS that has $2 \times m$ branches and $2$-level tree-encoded core qubits with branching parameters $b_0$ and $b_1$ is denoted by $(m, b_0, b_1)$. Fig.~\ref{fig:RGS} illustrates an example RGS $|G^3\rangle$, which has $2\times3$ branches and $2$-level tree-encoded core qubits with branching parameters $b_0=3$ and $b_1=2$. Throughout the paper we assume depth-2 trees, which can be generated using three matter qubits (1 or more emitters and the rest ancilla qubits) \cite{buterakos2017deterministic,hilaire2021resource}.
        
        \subsubsection{Tree-based encoding} In this encoding, the $X$ and $Z$ operators of each core photonic qubit in the RGS are replaced by the logical operators $X_L$ and $Z_L$:
    
        \begin{align}
        X_L=&X_j\bigotimes_{k\in C_j}Z_k \;,\label{eq:logic_x} \\
        Z_L=&\bigotimes_{k\in C_0}Z_k \;. \label{eq:logic_z}
        \end{align}
        where $j$ is any qubit in level $1$ of the tree, and $C_j$ is the set of child qubits of $j$. In the case of a $2$-level tree, there are $b_0$ possible choices of $X_L$, and hence the detection of all photons of one $X_L$ would suffice for a logical $X$ measurement. Since the measurement of one $X_L$ would fix all the other $X_L$'s, a majority vote can be performed to provide further protection against error. On the other hand, a logical $Z$ measurement requires all $Z$ measurements of level-$1$ photons to succeed. Each level-$1$ photon can either be measured directly or, in the case of photon loss, indirectly by $X$ measurement of any one of the $b_1$ child qubits \cite{azuma2015all,varnava2006loss}. Similarly, a majority vote can be performed among all indirect $Z$ measurements to further protect against error.      
        \subsubsection{Deterministic graph state generation} The operations required for the scheme include: the $P$ gate, where an entangled photon is emitted from the emitter qubit; the Hadamard gates $H_e$ and $H_p$ on the emitters and the photonic qubits, respectively; the CZ gates between the emitter qubit and ancillary matter qubits; and the local complementation (LC), which is defined as: \begin{align}
        LC_j=\sqrt{-iX_j}\bigotimes_{k\in N_j}\sqrt{iZ_k}  \;.
        \end{align}
        and $N_j$ is the set of neighboring vertices connected to qubit $j$. Fig.~\ref{fig:RGS_generation} shows the protocol for the deterministic generation of an encoded RGS by a single quantum emitter and two ancillary matter qubits.  

        \begin{figure}[t]
        \centering
	    \includegraphics[width = 0.8\linewidth]{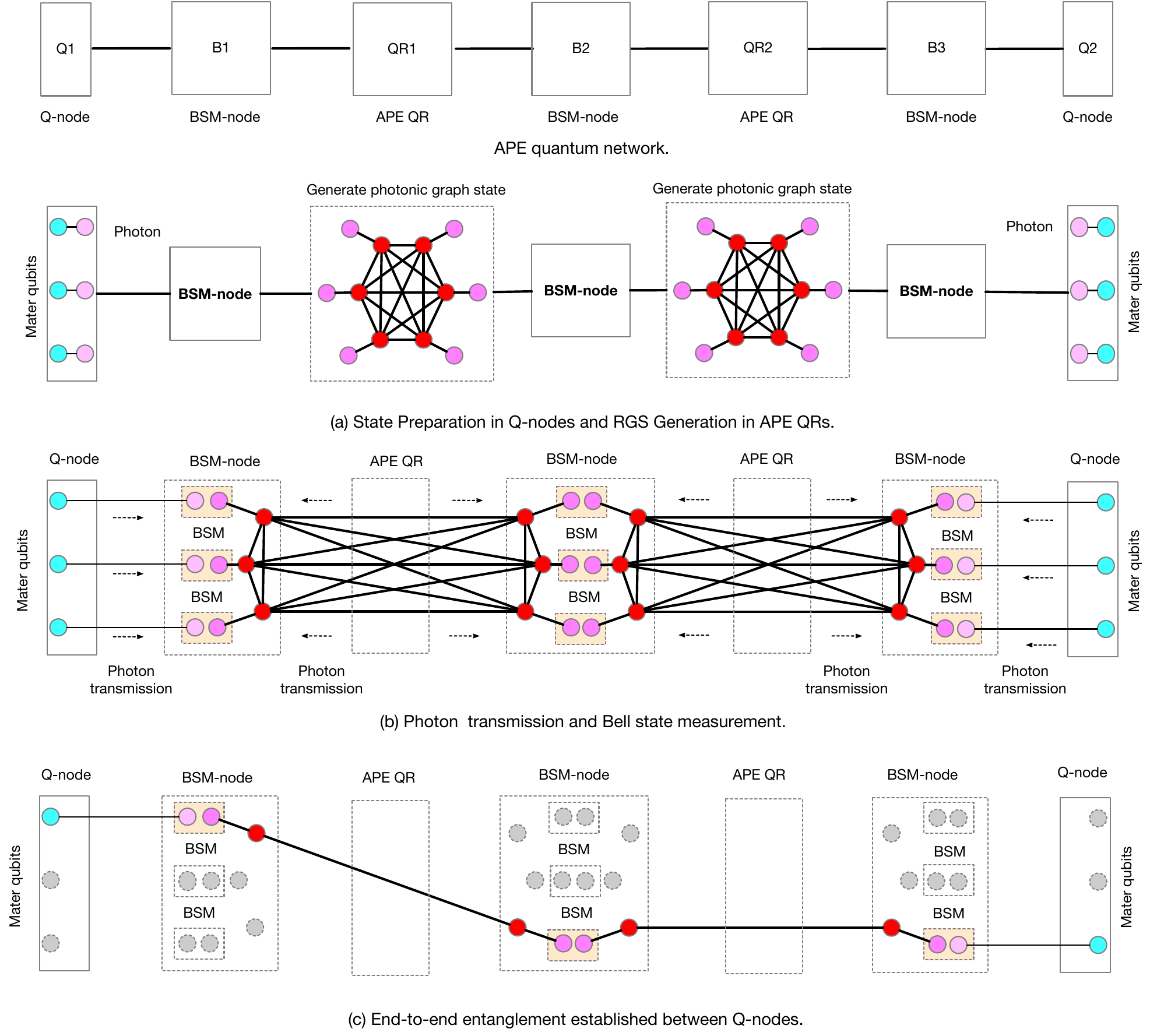}
        \caption{Entanglement generation in an APE quantum network.}
    	\label{fig:ape_entanglement}
        \vspace{-4mm}
        \end{figure}

        \subsubsection{End-to-end entanglement generation} 
        \label{sec:ape_entanglement_gen}
        An APE quantum network aims to establish end-to-end entanglement between Q-nodes. As illustrated in Fig.~\ref{fig:ape_entanglement}, the APE protocol works as follows: first, Q-nodes Q1 and Q2 each prepare $m$ single photons that are maximally entangled with their local matter qubits and sent to the adjacent BSM-nodes B1 and B3, respectively. At the same time, APE QR1 and QR2 each prepare an RGS $|G^m\rangle$, and the left (right) arms are sent to the left-hand (right-hand) adjacent nodes B1(B2) and B2(B3), respectively. Once receiving the single photons, B1, B2, and B3 each apply BSMs on the $m$ pairs of the leaf qubits of the left and right arms. If one of the BSMs succeeds, the encoded core qubits connected to those leaf qubits will be measured in the $X_L$ basis, and all other core qubits will be measured in the $Z_L$ basis; if a BSM fails, the encoded core qubits connected to those leaf qubits will be measured in the $Z_L$ basis. For the APE protocol to successfully generate an entangled pair, it requires at least one BSM in each BSM-node as well as the logical measurements of all encoded core qubits to succeed. Finally, the BSM-nodes announce the measurement outcomes to Q1 and Q2. The protocol succeeds if no failure occurs.

\section{Theoretical study and analysis}
\label{sec:theoretic_analysis}

In this section, we present our theoretical study and analysis of memory-based 1G trapped ion quantum repeaters and networks, and all-photonic APE quantum repeaters and networks. We derive theoretical models for the entanglement generation rate and fidelity between the end nodes of a QR chain. The chain has the form of $Q_1 \leftrightarrow BSM_1 \leftrightarrow QR_1 \dots\ QR_n \leftrightarrow BSM_{n+1} \leftrightarrow Q_2$, which consists of $n$ QRs and $n+1$ BSM-nodes between Q-nodes $Q_1$ and $Q_2$, with the nodes evenly spaced along the chain. For the memory-based scheme, each Q-node has one \Ca ion; each QR has two \Ca ions that are trapped in a linear Paul trap. For the all-photonic scheme, each Q-node has multiple matter qubits to allow continuous end-to-end entanglement generation; each QR generates a ($m,b_0,b_1$) tree-encoded repeater graph state. Table~\ref{tab:parameters} lists the notations used in our analysis.

\begin{table}[t]
\centering
\renewcommand{\arraystretch}{1.3}
\begin{tabular}{|p{0.7cm}|p{1.4cm}|p{11.6cm}|}
\hline
 & \textbf{Symbol} & \textbf{Description} \\ 
\hline
\multirow{0}{*}{\rotatebox[origin=x]{90}{\parbox{3cm}{\centering General parameters}}} 
& $n$ & Number of repeaters in a repeater chain. \\
& $\mu$ & Total photon loss probability between neighboring nodes. \\
& $\mu_{\text{coll}}$ & Collection loss probability of photon generation. \\
& $\mu_{\text{ch}}$ & Loss probability due to fiber attenuation and coupling between neighboring nodes. \\
& $\mu_{\text{QFC}}$ & Loss probability due to quantum frequency conversion. \\
& $\mu_\mathrm{d}$ & Loss probability due to photon detection. \\
& $\bar{F}$ & End-to-end entanglement fidelity of a repeater chain. \\
& $\mathrm{EGR}$ & End-to-end entanglement generation rate of a repeater chain. \\ 
\hline
\multirow{11}{*}{\rotatebox[origin=x]{90}{\parbox{3.2cm}{\centering Trapped ion parameters}}} 
& $P_{\rm HEG1(2)}$ & Success probability of the 1st (2nd) step of the two-step parallel HEG scheme. \\
& $P(h_i)$ & Probability of successfully creating entanglement on attempt $h_i$ at link $i$. \\
& $h_\mathrm{max}$ & Maximum number of HEG retries allowed.\\
& $T_{\text{exp}}$ & The expected duration of a repeater cycle or an iteration.\\
& $T_{\text{attempt}}$ & Time required to attempt to create one entanglement link between QRs. \\
& $T_{\text{corr}}$ & Time needed for Pauli correction on ions after a two-photon BSM \\
& $T_{\text{DBSM}}$ & Time needed for DBSM and Pauli correction on Q-nodes after all two-ion DBSMs \\
& ${D_i^p}(\rho)$ & Depolarizing channel applied on ion $i$ of state $\rho$ with probability $p$  \\
& $p_{\mathrm{MS}}$ & Depolarization probability due to Mølmer–Sørensen (MS) gate\\
& $\theta_i$ & Phase noise in ion $i$ \\
& $\tau$ & Ion-trap qubit coherence time \\
\hline
\multirow{0}{*}{\rotatebox[origin=c]{90}{\parbox{2.5cm}{\centering APE parameters}}} 
&$m$&Number of branches on one side of a RGS\\
&$b_i$& Branching parameters at the $i$-th level of the tree encoding of each RGS core qubit\\
& $R_i$ & Probability of obtaining an outcome from an indirect $Z$ measurement on at least one photon in the i-th level of the tree\\ 
& $T_\mathrm{RGS}$ & Total time to generate the RGS\\
& $MQ_e$& The number of memory qubits at each Q-node\\
\hline
\end{tabular}
\caption{Theoretical analysis notations.}
\label{tab:parameters}
\end{table}

\subsection{Trapped-ion quantum repeaters and networks}

Assume end-to-end entanglement links are generated through continuous iterations between $Q_1$ and $Q_2$ of the QR chain. In each iteration, a two-step parallel HEG scheme is executed to establish short-distance entanglement links along the chain. Afterward, entanglement swapping concatenates these short-distance links into end-to-end entanglement. Each HEG process allows up to $h_\mathrm{max}$ attempts. A HEG process terminates when any single attempt succeeds; otherwise up to $h_\mathrm{max}$ attempts will be made until the process exits with failure. An iteration is considered successful when all HEG and DBSM operations executed along the chain are successful (see Section~\ref{sec:trap_ion_network}).

\subsubsection{Entanglement Rate}

Based on the HEG process discussed in Section~\ref{sec:heg}, the total photon loss probability from a QR, or a Q-node, to a BSM-node can be expressed as
\begin{equation}
    \mu = 1-(1-\mu_{\text{coll}})(1-\mu_{\text{QFC}})(1-\mu_{\text{ch}})(1-\mu_\mathrm{d}) \;,
\end{equation}
where, $\mu_{\text{coll}}$, $\mu_{\text{QFC}}$, $\mu_{\text{ch}}$, $\mu_{d}$ denote photon loss probabilities due to collection, QFC, transmission loss in the fiber channel, and photon detector, respectively. Therefore, the probability that a (single-link) HEG process succeeds in the 1st attempt is $p_\mathrm{BSM}=\frac{(1 - \mu)^2}{2}$, and the probability that it succeeds in the $h$-th attempt is
\begin{equation}
      P(h) = \left( 1 - p_\mathrm{BSM} \right)^{h-1} p_\mathrm{BSM} \;.
   \label{eq:P_h_attempt}
   \end{equation}
For a $n$-repeater chain, there are a total of $n+1$ links to be established, each of which requires a successful HEG process, i.e., the $i$-th link is established if the BSM in the $i$-th BSM node is successful. Define
\[
\mathcal O := \{\, i\in\{1,\dots,n+1\} : i \text{ is odd}\,\},\qquad
\mathcal E := \{\, i\in\{1,\dots,n+1\} : i \text{ is even}\,\}.
\]
The 1st-step HEG corresponds to all links with indices in $\mathcal O$, and the 2nd-step HEG corresponds to all links with indices in $\mathcal E$, as illustrated by Fig.~\ref{fig:HEG_in_parallel}. Let $P_\mathrm{HEG1}$ ($P_\mathrm{HEG2}$) denote the success probability of the 1st (2nd) step of the two-step scheme, which requires successful BSMs in all odd (even) BSM-nodes: 
   \begin{equation}
       P_\mathrm{HEG1} = \left[ 1-(1-p_\mathrm{BSM})^{h_\mathrm{max}} \right]^{|\mathcal O|} \;,\qquad
        P_\mathrm{HEG2} = \left[ 1-(1-p_\mathrm{BSM})^{h_\mathrm{max}} \right]^{|\mathcal E|} \;,
   \end{equation}
where $|\mathcal{O}|=\lceil n/2\rceil$ and $|\mathcal{E}|=\lfloor n/2\rfloor$ are the numbers of odd- and even-indexed links, respectively.
The expected duration 
   of a repeater cycle or an iteration is:
   \begin{flalign}
   \begin{aligned}
       T_{\text{exp}} = \sum_{h_1=1}^{h_\mathrm{max}}\sum_{h_2=1}^{h_\mathrm{max}}{\cdots}\sum_{h_{n+1}=1}^{h_\mathrm{max}}  P(h_1)P(h_2){\cdots} P(h_{n+1}) \left[ T_{\text{attempt}} \left( \max_{i\in\mathcal O}(h_i) {+} \max_{i\in\mathcal E}(h_i) \right) {+} 2T_{\text{corr}} {+} T_{\text{DBSM}}\right] \\
       + \left(1 - P_\mathrm{HEG1}\right) T_{\text{attempt}} h_\mathrm{max} \\
       + \left(1 - P_\mathrm{HEG2}\right)\sum_{h_1=1}^{h_\mathrm{max}}\sum_{h_3=1}^{h_\mathrm{max}}\cdot\cdot\cdot\sum_{h_{|\mathcal O|}=1}^{h_\mathrm{max}} P(h_1)P(h_3)\cdot\cdot\cdot P(h_{|\mathcal O|})  \left[ T_{\text{attempt}} \left(\max_{i\in\mathcal O}(h_i) {+} h_\mathrm{max}\right) {+} T_{\text{corr}}\right] \;,
   \end{aligned}
   && \label{eq:1G-theo_rate}
   \end{flalign}
where $T_\mathrm{attempt}$ is the time it takes to attempt to create entanglement across a single link using entanglement swapping at a BSM-node, $T_{\text{corr}}$ is the time it takes to perform a Pauli correction on an ion after a successful two-photon BSM, and $T_{\text{DBSM}}$ is the time it takes to perform DBSM and a Pauli correction on Q-nodes after all the two-ion DBSMs are completed. In Eq.~\eqref{eq:1G-theo_rate}, the first term corresponds to the cases for which end-to-end entanglement links have been successfully established, while the second (third) term accounts for the cases where entanglement generations fail in the first (second) step of the two-step scheme. Therefore, the end-to-end entanglement generation rate (EGR) of the QR chain is given by:
    \begin{equation}
      \text{EGR} = \frac{P_\mathrm{suc}}{T_{\text{exp}}} \;.
    \end{equation}
where
    \begin{align}
       P_{\text{suc}} =P_\mathrm{HEG1} P_\mathrm{HEG2}
   \end{align}
is the end-to-end entanglement success probability.

\subsubsection{Entanglement Fidelity}

Our analysis begins with a QR chain with $1$ QR, taking the chain in Section~\ref{sec:ent_swapping} as an example. Major sources of fidelity degradation considered in our analysis are ion-photon emission infidelities, imperfect gate operations, and memory decoherence from dephasing. Denote the ion-photon entanglement fidelity by $F_{\rm em-trap}$. To accurately capture ion-photon emission infidelities, we model the state of an ion-photon pair of a successful ion-photon entanglement generation as a Werner state:
    \begin{equation}
    \rho_\mathrm{ion-ph}=w_\mathrm{em}\ket{\phi^+}\bra{\phi^+} + (1-w_\mathrm{em})\frac{I}{4} ,
    \end{equation}
    where
    \begin{align}
    \ket{\phi^+} &= \frac{1}{\sqrt{2}} (\ket{00} + \ket{11}) ,\\
    w_\mathrm{em}&=
    1 - \frac{4}{3} (1 - F_{\rm em-trap}) .
    \end{align}


 Based on the HEG process discussed in Section~\ref{sec:heg} and assuming perfect BSM operations, the resulting state following successful HEG between two ions (e.g., A' and A or B and B' in Fig.~\ref{fig:entanglement_swapping}) is a new Werner state:
\begin{equation}
\rho_{A'-A}=w_{\rm em}^2\ket{\phi^+}\bra{\phi^+} + (1-w_{\rm em}^2)\frac{I}{4} .
\end{equation}


We implement a two-step parallel HEG scheme to establish short-distance entanglement links along the chain (see Section~\ref{sec:parallel_hegs}). During this process, an established entanglement link experiences decoherence. 
We model this decoherence as \textit{collective dephasing}, wherein each qubit undergoes a phase rotation due to Gaussian-distributed temporal noise. Specifically, the phase noise \( \theta_i \) of the $i$-th ion is modeled as a random variable drawn from a normal distribution, $\theta_i \sim \mathcal{N}(0, \sigma^2)$ where \( \sigma^2 \) is the variance and quantifies the extent of dephasing \cite{zwerger2017quantum}. The standard deviation is $\sigma = \frac{2\Delta T}{\tau}$, where $\Delta T$ is the time gap between the successful HEG and DBSM, and \( \tau \) is the ion-trap qubit coherence time. 
Under this dephasing noise, the state becomes 
\begin{align}
\rho_{A'-A} &\rightarrow e^{-i\frac{\theta_A}{2} Z_A} e^{-i\frac{\theta_{A'}}{2} Z_{A'}}\rho_{A'-A} e^{i\frac{\theta_{A'}}{2} Z_{A'}}e^{i\frac{\theta_A}{2} Z_A}\\&=w_{\rm em}^2\ket{\phi^+(\theta_{A'}+\theta_A)}\bra{\phi^+(\theta_{A'}+\theta_A)} + (1-w_{\rm em}^2)\frac{I}{4} \label{eq:ion-ion state HEG} ,
\end{align}
where
\begin{align}
\ket{\phi^+(\alpha)} &\equiv \frac{1}{\sqrt{2}} (\ket{00} + e^{i \alpha}\ket{11}).
\end{align}

After two neighboring entanglement links A'-A and B-B' have been successfully established, DBSM is performed on A and B using the MS gate $U_{AB}=e^{-i\frac{\pi}{4}X_AX_B}e^{i\frac{\pi}{8}Z_A}e^{-i\frac{\pi}{8}Z_B}$, followed by $Z$ measurements on A and B (see Fig. \ref{fig:entanglement_swapping}). Depolarization channels are implemented after $U_{AB}$ to model the MS gate error. Therefore, it has: 
\begin{align}
 \rho_{A'-A}\otimes\rho_{B-B'} &\rightarrow \rho'= D_A^{p_\mathrm{MS}}\otimes D_B^{p_\mathrm{MS}}(U_{AB}\rho_{A'-A}\otimes\rho_{B-B'}U_{AB}^{\dagger}),
\end{align}
and after the $Z$ measurements, assuming perfect Pauli frame adjustment,
\begin{align}
\rho&_{A'-B'}=\hbox{Tr}_{AB}(\Pi_{0,A}\otimes\Pi_{0,B}(\rho'))\\
&\overset{\text{normalized}}{=}\frac{1}{4}\, I_{A'B'}
+ \frac{w_{\rm em}^4}{4}\!\left[
w_{\rm MS} \cos (S-\frac{\pi}{2})\, (X_{A'} X_{B'} - Y_{A'} Y_{B'})\right.\notag\\
&\hspace*{.4\linewidth}\left.+ w_{\rm MS} \sin (S-\frac{\pi}{2})\, (X_{A'} Y_{B'} + Y_{A'} X_{B'})
+ w_{\rm MS}^{2} Z_{A'} Z_{B'}
\right] \label{eq:ion-ion state DBSM}\\
& \;\;\quad=\begin{pmatrix}
\frac{1}{4} + \frac{1}{4}w_{\rm em}^4w_{\rm MS}^2 & 0 & 0 & \frac{1}{2}w_{\rm em}^4w_{\rm MS}\, e^{-i(S-\pi/2)} \\
0 & \frac{1}{4} - \frac{1}{4}w_{\rm em}^4w_{\rm MS}^2 & 0 & 0 \\
0 & 0 & \frac{1}{4} - \frac{1}{4}w_{\rm em}^4w_{\rm MS}^2 & 0 \\
\frac{1}{2}w_{\rm em}^4w_{\rm MS}\, e^{i(S-\pi/2)} & 0 & 0 & \frac{1}{4} + \frac{1}{4}w_{\rm em}^4w_{\rm MS}^2
\end{pmatrix}\; ,\notag
\end{align}
 where 
 \begin{align}
 D_i^p(\rho)&\equiv(1-p)\rho+p\hbox{Tr}_i(\rho)\otimes \frac{I}{2},\\
 \Pi_{0,i}&\equiv\ket{0}_i\bra{0},\\
 S&=\theta_{A'}+\theta_A+\theta_B+\theta_{B'},\\
 w_{\rm MS}&=1-p_{\rm MS}.
\end{align}
For a repeater chain with $n>1$ repeaters, following a similar derivation as above (see Appendix~\ref{Appendix:Derivation}), the entangled state across two Q-nodes Q1 and Q2 is
\begin{align}
\rho=&\frac{1}{4} I_{12}
+ \frac{w_{\rm em}^{2n+2}}{4}\!\left[
w_{\rm MS}^n \cos\!\left(\theta_{\rm tot}-\tfrac{n\pi}{2}\right) (X_1 X_2 - Y_1 Y_2) \right. \notag\\
&\hspace*{.3\linewidth}\left.
+\, w_{\rm MS}^n \sin\!\left(\theta_{\rm tot}-\tfrac{n\pi}{2}\right) (X_1 Y_2 + Y_1 X_2)
+ w_{\rm MS}^{2n} Z_1 Z_2
\right]\label{eq:end-end state}
 \\[6pt]
=&
\begin{pmatrix}
\tfrac{1}{4} + x & 0 & 0 & y\, e^{-i(\theta_{\rm tot} - n\pi/2)} \\
0 & \tfrac{1}{4} - x & 0 & 0 \\
0 & 0 & \tfrac{1}{4} - x & 0 \\
y\, e^{i(\theta_{\rm tot} - n\pi/2)} & 0 & 0 & \tfrac{1}{4} + x
\end{pmatrix}\; ,
\end{align}
where $x = \tfrac{1}{4} w_{\rm em}^{2n+2} w_{\rm MS}^{2n}$, $y = \tfrac{1}{2} w_{\rm em}^{2n+2} w_{\rm MS}^{n}$, $\theta_{\rm tot}=\sum_{i=1}^{2n+2}\theta_i$ and $\theta_i\sim \mathcal{N}(0,\sigma_i^2)$ are independent Gaussian distributions with $\sigma_i=\frac{2\Delta t_i}{\tau}$.

The fidelity w.r.t the noiseless state $(w_{\rm em}=1=w_{\rm MS},\theta_i=0\; \forall i)$ is given by
\begin{align}
 \bar{F} &= \tfrac{1}{4} + x + y \cos(\theta_{\rm tot})
\end{align}

\subsection{APE quantum networks}
\subsubsection{Entanglement generation rate}
Assume end-to-end entanglement links are generated through continuous iterations between $Q_1$ and $Q_2$ of the QR chain. As discussed in Section~\ref{sec:ape_entanglement_gen}, an iteration is considered successful when at least one BSM in each BSM-node is successful and all the logical $X$ and $Z$ measurements of the encoded core qubits in each BSM-node are successful. The success probability is 
\begin{align}
P_{\rm RGS}=\left[(1-(1-P_{\rm BSM})^{m})\right]^{n+1}\left[P^2_{X}P^{2m-2}_{Z}\right]^{n} \;,
\end{align}
where $P_{\rm BSM}$ is the success probability of a photonic BSM on leaf qubits, and $P_{X}$ and $P_{Z}$ are the probabilities of successfully performing the logical $X$ and $Z$ measurements on the tree-encoded core qubits, respectively. These probabilities are given by
\begin{align}
P_{\rm BSM}&=\frac{(1-\mu)^2}{2} \;,\\
P_{X}&=R_0 \label{eq:Psigmax} \;,\\
P_{Z}&=(1-\mu+\mu R_1)^{b_0} \;,
\label{eq:Psigmaz}
\end{align}
where $\mu$ is the single-photon loss probability between an APE QR, or a Q-node, and its neighboring BSM-node:
\begin{align}
\mu&=1-(1-\mu_{\text{coll}})(1-\mu_{\text{QFC}})(1-\mu_{\rm delay})(1-\mu_{\text{ch}})(1-\mu_\mathrm{d}) \;,
    \end{align}
Here, $\mu_{\text{coll}}$, $\mu_{\text{QFC}}$, $\mu_{\rm delay}$, $\mu_{\text{ch}}$, $\mu_{\rm d}$ denote photon loss probabilities due to photon collection, QFC, delay lines for core photons during the RGS generation, transmission loss in the fiber channel, and photon detector loss, respectively. $R_0$ and $R_1$ in Eqs.~\eqref{eq:Psigmax} and \eqref{eq:Psigmaz} are the probabilities of obtaining an outcome from an indirect $Z$ measurement on any given photon in the 0th and 1st level of the tree (which is assumed to have $2$ levels), respectively:
\begin{align}
R_0&=1-[1-(1-\mu)^{b_1+1}]^{b_0} \;,\\
R_1&=1-\mu^{b_1} \;.
\end{align}
The end-to-end entanglement generation rate (EGR) of the QR chain is then 
\begin{equation}
    \text{EGR}= \frac{P_{\rm RGS}}{T_\mathrm{RGS}\times MQ_{e}} \;,
\end{equation}
where $T_\mathrm{RGS}$ is the time it takes to generate an RGS, and 
$MQ_{e}$ is the number of memory qubits at each Q-node.
We divide by $MQ_e$ to normalize the EGR by the total number of matter qubits involved in the Q-nodes. The above equation is based on the assumption that there is a sufficient number of memory qubits in the Q-nodes so that the emitter in each APE QR starts the next RGS generation cycle for the subsequent iteration once the current RGS generation cycle is completed. The required number of memory qubits in each Q-node is 
\begin{equation}
MQ_{e}=m+\left\lceil\frac{L_{\rm seg}}{c\times T_\mathrm{RGS}} \right\rceil m +\left\lceil\frac{L_c-L_{\rm seg}}{c\times T_\mathrm{RGS}} \right\rceil \;,
\label{eq:APE_matter_qubits}
\end{equation}
where $L_{\rm seg}=\frac{L_c}{n+1}$ is the distance between adjacent QRs, with $L_c$ denoting the full length of the network.
The first term in Eq.~\eqref{eq:APE_matter_qubits} corresponds to $m$ memory qubits for the current iteration. It takes time $\frac{L_{\rm seg}}{2c}$ ($c$ is the speed of light) for the transmission of RGS photons from a QR to a BSM-node, and $\frac{L_{\rm seg}}{2c}$ for a Q-node to receive the measurement results from its neighboring BSM-node. During this period, all $m$ memory qubits involved in the current iteration cannot be reused since the Q-node does not yet know the BSM outcomes from its neighboring BSM-node. Hence, the second term on the right-hand side of Eq.~\eqref{eq:APE_matter_qubits} is proportional to the number of RGS generation cycles performed during this period, and each cycle would require $m$ additional memory qubits. Once the Q-node receives the outcomes from its neighboring BSM node, at most one memory qubit cannot be reused if its emitted photon had a successful BSM outcome. This memory qubit is assumed to be reused only after measurement outcomes from all BSM-nodes arrive at the Q-node such that the Pauli correction can be performed. The third term in Eq.~\eqref{eq:APE_matter_qubits} accounts for additional memory qubits needed during this period. 

\subsubsection{Entanglement Fidelity}
 For each successful iteration, a Bell pair is created between Q-nodes at opposite ends of the network up to a Pauli frame. This Pauli frame is affected by the following factors: (a) all the measurement outcomes of the emitters and ancillas, (b) all the measurement outcomes of photons in BSMs and SPDs, and (c) the decoherence noise of the emitters and memory qubits. The outcomes of (a) are readily known in every APE QR node during the generation process. The outcomes of (b) are known in every BSM-node in the case of a successful iteration. The effect of (c) can lead to a Bell pair with a different Pauli frame. Since (a) and (b) are known, they do not affect the fidelity because the corresponding Pauli frame can be calculated, and local Pauli frame corrections can be performed. Here, we study how the error propagated from emitters to photons can affect the final entanglement fidelity in the case of a successful iteration.
The decoherence of the emitter and memory qubits is modeled by the following discrete Pauli $Z$ error: 
\begin{align}
p_Z&=\frac{1-e^{-T_{\rm gate}/T_2}}{2} \;,
\label{eq:zerror_prob}
\end{align}
where $T_{\rm gate}$ is the finite gate time of each quantum gate during the RGS generation and $T_2$ is the dephasing time of the emitter or the memory qubit.

\begin{figure}[h]
\centering
\includegraphics[width=1\textwidth]{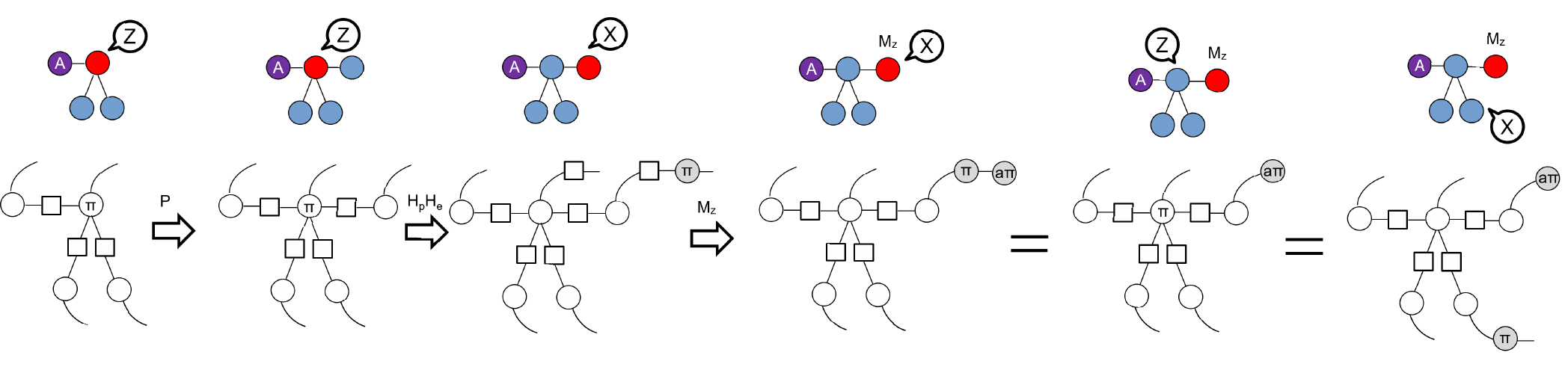}
 \caption{Error propagation during the generation of tree-encoded core qubit and a proof by ZX calculus. We adopt the notation of ZX calculus from~\cite{van2020zx}: a white (grey) dot denotes a $Z$-spider ($X$-spider) and a square denotes a Hadamard box. The steps are as follows: (i) Starting with a $Z$ error at the emitter, a $P$ gate (which is equivalent to a CZ gate between the emitter and a newly emitted photon in $\ket{+}$ state) will not propagate the error because the $Z$ error commutes with CZ. (ii) After Hadamard gates on the emitter ($H_e$) and the newly emitted photon ($H_p$), the $Z$ error before the Hadamard is equivalent to an $X$ error after the Hadamard. (iii) A $Z$-measurement is performed on the emitter qubit. (iv) A $Z$-measurement will unentangle the emitter from the rest of the qubits. To keep track of the error, we can represent the $X$ error on the emitter as a $Z$ error on the neighboring photon (which is a level-1 photon in the tree-encoded core qubit of the RGS). (v) A $Z$ error at the level-1 tree photon is also equivalent to a $X$ error at one of the level-2 tree photons.}
\label{fig:Error propagation1}
\end{figure}

\paragraph{Error propagation during the generation of tree-encoded core qubits}
We first give an explicit example for the emitter error during the CZ gate between the ancilla and the emitter qubits in the green circle in Fig.~\ref{fig:RGS_generation}. This is assumed to happen with probability given by Eq.~\eqref{eq:zerror_prob} with $T_{\rm gate}$ and $T_2$ in Table~\ref{tab:APE-params}. The emitter error during other gates of this cycle can be derived similarly. Since a $Z$ error on the emitter commutes with the CZ gate, it can be represented by the leftmost graph in Fig.~\ref{fig:Error propagation1}. Next, a $P$ gate (which is equivalent to a CZ gate between the emitter and a newly emitted photon in the $\ket{+}$ state) also commutes with the $Z$ error, and hence the state can be represented by the second graph from the left. After Hadamard gates on the emitter ($H_e$) and the newly emitted photon ($H_p$), the $Z$ error before the Hadamard is equivalent to an $X$ error after the Hadamard.  A $Z$-measurement on the emitter will then unentangle the emitter from the rest of the qubits. To investigate how the error could possibly be corrected by the majority vote of RGSs, we notice that the $X$ error on the emitter is also equivalent to a $Z$ error on the neighboring photon (which is a level-1 photon in the tree-encoded core qubit of the RGS), or alternatively, an $X$ error at one of the level-2 tree photons. As a result, this would not affect a logical $Z$ measurement, which involves physical $Z$ measurements of the level-1 tree photon, as defined in Eq.~\eqref{eq:logic_x}. In the case of a loss of the level-1 tree photon, an indirect $Z$ measurement can still be performed by a physical $X$ measurement of one of its level-2 child photons, and hence an $X$ error would not affect this result either. On the other hand, this would flip the outcome of a logical $X$ measurement, which involves a physical $X$ measurement of the level-1 tree photon and physical $Z$ measurements of the second-level tree photons, as defined in Eq.~\eqref{eq:logic_z}. Since there are $b_0$ possible choices of $X_L$, the detection of all photons of each possible $X_L$ will lead to one vote, and the outcome of the logical $X$ measurement depends on a majority vote among at most $b_0$ cast votes. Hence, among all successfully received votes, as long as there are more error-free votes than flipped votes, the error can be corrected.

 \begin{figure}[h]
 \centering
\includegraphics[width=0.7\textwidth]{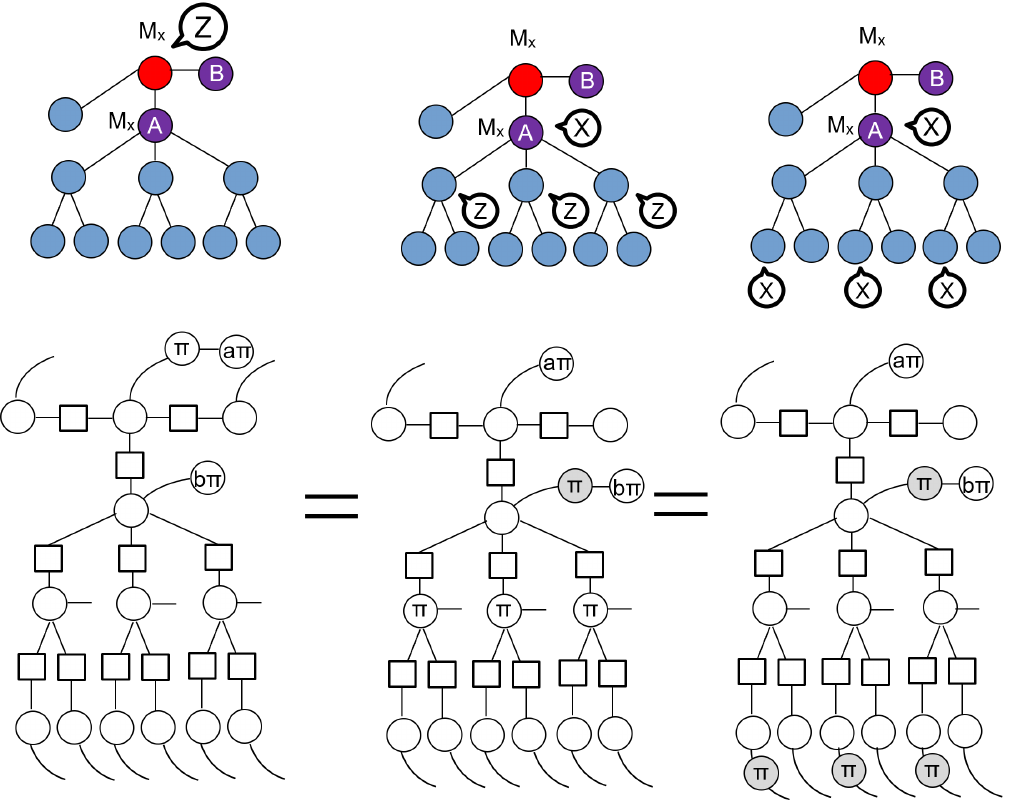}
 \caption{Error propagation during the generation of a leaf qubit and a proof by ZX calculus. The steps are as follows: (i) A $Z$ error on the emitter is equivalent to an $X$ error on ancilla A and $Z$ errors on every level-1 tree photons. (ii) This is also equivalent to an $X$ error on ancilla A and an $X$ error on one of the level-2 photons of every subtree.}
\label{fig:Error propagation2}
\end{figure}

\paragraph{Error propagation during generation of a leaf qubit}
We then give an explicit example for the emitter error during the CZ gate between ancilla A/B and the emitter qubit in the light grey cycle in Fig.~\ref{fig:RGS_generation}. Since the $Z$ error commutes with the CZ gates and the $P$ gate, a $Z$ error appears during the CZ gate between ancilla A/B and the emitter qubit can be represented by the leftmost graph in Fig.~\ref{fig:Error propagation2}, where the $X$ measurements are understood to be performed at the end. A $Z$ error on the emitter is equivalent to an $X$ error on ancilla A and $Z$ errors on every level-1 tree photons, or alternatively, an $X$ error on one of the level-2 photons of every subtree. Since ancilla A is measured in $X$ basis subsequently, the $X$ error will have no effect. In addition, this would not affect a logical $Z$ measurement which involves a physical $Z$ measurement of the first-level tree qubits, as defined in Eq.~\eqref{eq:logic_z}. In the case of a loss of the level-1 tree photon, an indirect $Z$ measurement can still be performed by a physical $X$ measurement of one of its level-2 child photons, and hence an $X$ error would not affect this result either. However, for a logical $X$ measurement, it would flip the parity of every successfully received vote, and hence a majority vote cannot correct for this error.  

 \paragraph{Error propagation of core qubits in RGS to Bell pair}
As shown in Fig. 7c, after at least one BSM succeeds on each BSM node and all logical $Z$ measurements succeed at the core qubits, we are left with a linear cluster state with $2n$ core qubits and 2 end-node qubits, where the successful BSM is equivalent to fusing the cluster state together with the two photons that underwent BSM being removed. Two adjacent $X$ measurements on qubits $i$ and $i+1$, with outcomes $m_i$ and $m_{i+1}$, will correspond to $X_{i-1}$ and $Z_{i-1}$ errors (see derivation in Appendix~\ref{Appendix:Derivation APE fidelity}). We propagate all the error to the memory qubit in the end node Q1, and in the next section, calculate the fidelity of the Bell pair across the end nodes Q1 and Q2. 

\paragraph{Fidelity calculation}
We perform the following entanglement fidelity calculation for a chain of $n$ repeaters. Denote the $j$-th vote of the logical $X$ measurement as $X_{L,j}=X_j\bigotimes_{k\in C_j}Z_k$, where $j$ is any qubit in level 1 of the tree, and hence at most $b_0$ votes can be obtained. The average error probability of each vote is given by Eq. \ref{eq:zerror_prob} with the time $t_c$:
\begin{align}
\bar{e}_{vote}&=\frac{1-e^{-t_c/T_2}}{2} \;,
\label{eq:eI0B}
\end{align}
where $t_c$ is the duration over which the emitter decoheres before the $H_p$ gate, i.e., emitting a total of $b_1+1$ photons followed by a CZ gate between the emitter and ancilla, as seen in Fig.~\ref {fig:RGS_generation}.  
The success probability of obtaining one vote is equal to receiving all photons in $X_{L,j}$:
\begin{align}
S_0&= (1-\mu)^{b_1+1}\;.
\label{eq:s0}
\end{align}
The probability of obtaining a total of $m$ out of $b_0$ votes is: 
\begin{align}
T_0(m)&=\binom{b_0}{m}S_0^m(1-S_0)^{b_0-m} \;.
\label{eq:T0m}
\end{align}
Hence, given that $m$ votes are obtained, the average error probability of the measurement outcome of logical $X$ from the majority vote is: 
\begin{align}
\bar{e}_{X|m}&=\sum^m_{j=\lceil m/2 \rceil}\binom{m}{j}(\bar{e}_{vote})^j(1-\bar{e}_{vote})^{m-j} \;,
\label{eq:eI0m}
\end{align}
where a tie case during a majority vote is considered unsuccessful. 
Then, the average error probability of the logical $X$ measurement due to error during $t_c$ is:
\begin{align}
\bar{e}_{X,c}&=\frac{1}{R}\sum^{b_0}_{m=1}T_0(m)\bar{e}_{X|m} \;,
\label{eq:eXc}
\end{align}
where
\begin{align}
R&=\sum^{b_0}_{m=1}T_0(m) \;.
\end{align}
As the error during the generation of a leaf qubit with duration $t_l$ cannot be corrected by the tree encoding, as illustrated by Fig.~\ref{fig:Error propagation2}, it adds an extra average error probability of 
\begin{align}
\bar{e}_{X,l}&=\frac{1-e^{-t_l/T_2}}{2} \;.
\label{eq:eXl}
\end{align}
Here $t_l$ is the duration over which the emitter decoheres before the $X$
measurement of the emitter qubit, i.e., two CZ gates between the emitter and both ancillas, followed by emission of a photon, as seen in Fig.~\ref{fig:RGS_generation}. Combining Eq.~\ref{eq:eXc} and Eq.~\ref{eq:eXl}, we have the error probability of each $X_L$ of an encoded RGS: 
\begin{align}
\bar{e}_{X}&=\bar{e}_{X,c}+\bar{e}_{X,l} \;.
\label{eq:eX}
\end{align}
Since the error propagation does not lead to error in the logical $Z$ measurement, we have the error probability of each $Z_L$ of an encoded RGS: 
\begin{align}
\bar{e}_{Z}&=0 \;.
\label{eq:eZ}
\end{align}
Propagating all the errors to the memory qubit in the end node Q1, the final entanglement fidelity of the Bell pair across the end nodes Q1 and Q2 with $n$ intermediate repeaters is (see derivation in Appendix~\ref{Appendix:Derivation APE fidelity})
\begin{align}
\bar{E}_{Z}&=\bar{E}_{X}=\frac{1+(1-2\bar{e}_{X})^n}{2}\frac{1-(1-2\bar{e}_{X})^n}{2} \;,\\
\bar{E}_{Y}&=(\frac{1-(1-2\bar{e}_{X})^n}{2})^2 \;,\\
\bar F&=1-\bar{E}_{X}-\bar{E}_{Y}-\bar{E}_{Z} \;.
\end{align}

\section{Design and Modeling using NetSquid}
\label{design}

\begin{table*}[t]
\centering
\caption{Modeling of Trapped-ion and APE Quantum Networks.}
\label{tbl:ape-models}
\small
    \rowcolors{2}{gray!10}{white}
    \begin{tabular}{|m{0.3cm}|m{1.5cm}|m{5.4cm}|m{5.4cm}|}
    \hline 
    \rowcolor{blue!25}
     \thead{\textbf{}} & \thead{\textbf{Model}} & \thead{\textbf{Components}} & \thead{\textbf{Major Functions}} \\
    &
         \begin{minipage}[c][1.6cm]{\linewidth}
            Q-node
        \end{minipage}
    &
    \begin{minipage}[c][1.6cm]{\linewidth}
        \begin{itemize}[leftmargin=*]
            \item 1 \Ca ion
        \end{itemize}
    \end{minipage}
        &
    \begin{minipage}[c][1.6cm]{\linewidth}
        \begin{itemize}[leftmargin=*]
            \item HEG
            \item Pauli frame adjustment based on measurements
        \end{itemize}
    \end{minipage} \\
    \cellcolor{white}
    &
        \begin{minipage}[c][1.6cm]{\linewidth}
            QR
        \end{minipage}
    &
        \begin{minipage}[c][1.6cm]{\linewidth}
        \begin{itemize}[leftmargin=*]
            \item 2 \Ca ions that are trapped in a trap, which is coupled to a cavity
        \end{itemize}
        \end{minipage}
    &
        \begin{minipage}[c][1.6cm]{\linewidth}
        \begin{itemize}[leftmargin=*]
            \item HEG
            \item Deterministic BSM (DBSM)
        \end{itemize}
        \end{minipage} \\
   \cellcolor{white}
    &
        \begin{minipage}[c][1.6cm]{\linewidth}
        BSM-node
        \end{minipage}
    &
        \begin{minipage}[c][1.6cm]{\linewidth}
        \begin{itemize}[leftmargin=*]
            \item A clock that ticks at microsecond resolution
            \item A BSM unit
        \end{itemize}
        \end{minipage}
    &
        \begin{minipage}[c][1.6cm]{\linewidth}
        \begin{itemize}[leftmargin=*]
            \item BSM
            \item Generating clock signals to drive HEG
        \end{itemize}
        \end{minipage} \\
    \cellcolor{white}
   \multirow{-12}{*}{\rotatebox[origin=c]{90}{\textbf{Trapped Ion Network}}}
    &
        \begin{minipage}[c][1.3cm]{\linewidth}
            C-node
        \end{minipage}
    &
        \begin{minipage}[c][1.3cm]{\linewidth}
        \begin{itemize}[leftmargin=*]
            \item Trapped ion quantum network control protocols
        \end{itemize}
        \end{minipage}
    &
        \begin{minipage}[c][1.3cm]{\linewidth}
        \begin{itemize}[leftmargin=*]
            \item Network resource orchestration
        \end{itemize}
        \end{minipage} \\
    \hline
    &
    \begin{minipage}[c][1.6cm]{\linewidth}
       Q-node
    \end{minipage}
    &
    \begin{minipage}[c][1.6cm]{\linewidth}
        \begin{itemize}[leftmargin=*]
            \item \emph{m} matter qubits 
        \end{itemize}
    \end{minipage}
    &
    \begin{minipage}[c][1.6cm]{\linewidth}
        \begin{itemize}[leftmargin=*]
            \item Photon generation
            \item Pauli frame adjustment based on measurements
        \end{itemize}
    \end{minipage} \\
    \cellcolor{white}
    &
    \begin{minipage}[c][1.6cm]{\linewidth}
        QR
    \end{minipage}
    &
    \begin{minipage}[c][1.6cm]{\linewidth}
        \begin{itemize}[leftmargin=*]
            \item 1 emitter and 2 ancillary matter qubits (quantum dot)
            \item A clock that ticks at nanosecond resolution
        \end{itemize}
    \end{minipage}
    &
        \begin{minipage}[c][1.6cm]{\linewidth}
            \begin{itemize}[leftmargin=*]
                \item Repeater graph state generation
                \item Repeater graph state transmission
            \end{itemize}
        \end{minipage} \\
    \cellcolor{white}
    \multirow{-5}{*}{\rotatebox[origin=c]{90}{\textbf{APE Network}}}
    &
        \begin{minipage}[c][2.8cm]{\linewidth}
            BSM-node
        \end{minipage}
    &
        \begin{minipage}[c][2.8cm]{\linewidth}
        \begin{itemize}[leftmargin=*]
            \item A BSM unit
            \item For left or right direction: 1 pair of $X$- and $Z$-basis Single Photon Detector for incoming photons for each level of tree-encoded core qubits
        \end{itemize}
        \end{minipage}
    &
        \begin{minipage}[c][2.8cm]{\linewidth}
        \begin{itemize}[leftmargin=*]
            \item Repeater graph state reception
            \item BSM for leaf photons
            \item Single photon measurement for core photons
        \end{itemize}
        \end{minipage} \\
    \cellcolor{white}
    &
        \begin{minipage}[c][1.3cm]{\linewidth}
            C-node
        \end{minipage}
    &
        \begin{minipage}[c][1.3cm]{\linewidth}
        \begin{itemize}[leftmargin=*]
            \item APE quantum network control protocols
        \end{itemize}
        \end{minipage}
    &
        \begin{minipage}[c][1.3cm]{\linewidth}
        \begin{itemize}[leftmargin=*]
            \item Network resource orchestration
        \end{itemize}
        \end{minipage} \\
    \hline
    \end{tabular}
\end{table*}

In this section, we present our design and modeling of trapped ion and APE quantum networks using NetSquid. We develop building-block node models that correspond to the major entities in a trapped ion or an APE quantum network (i.e., Q-node, QR, and BSM-node), and a control node that supports logically centralized control. The models for each type are listed in Table~\ref{tbl:ape-models}. The details of our design and implementation are described below.

    \subsection{Trapped ion quantum repeaters and networks}
    \label{section:trapped_ion_design}
        
       \subsubsection{BSM-node model} As discussed in Section~\ref{sec:heg}, a trapped ion is driven by a sequence of laser pulses to generate photons. Therefore, we can vary a Q-node's, or a QR's, photon generation time by properly controlling its drive pulses. As such, we implement a clock-assisted scheme to realize heralded entanglement generation. In our design, each BSM-node model includes a local clock and a BSM unit that connects to two neighboring nodes. A neighboring node can be either a Q-node or a QR. In operation, a BSM-node generates and transmits clock signals to each of its neighboring nodes through a classical channel to trigger photon generation. For BSM, because the photon's time-of-flight from the two neighboring nodes to the BSM-node are different, the BSM-node can control its neighboring nodes' photon generation time by adjusting the relative timing between clock signals to each neighboring node accordingly so that the generated photons arrive at the BSM-node simultaneously.
    
        \subsubsection{Q-node and QR models}
\begin{figure*}[t]
        \centering
        \includegraphics[width=1\linewidth]{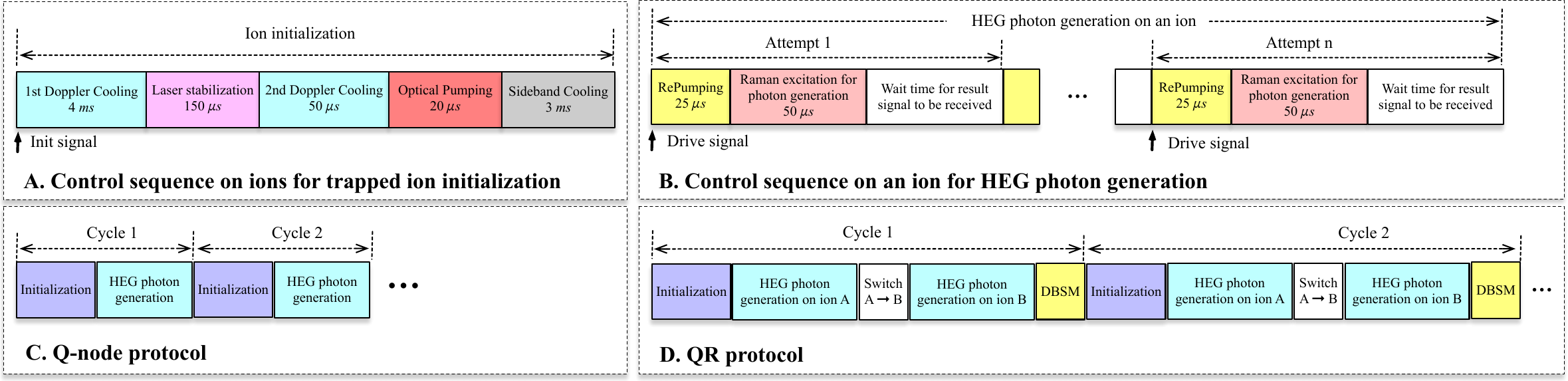}
        \caption{Control sequences and protocols for Q-node and QR models.}
        \label{fig:controlseq}
    \end{figure*}

        Our Q-node and QR models are based on the recent experimental realization of light-matter interfaces~\cite{schupp2021interface} and quantum repeaters that are based on \Ca~\cite{krutyanskiy2023telecom}. Each Q-node model has a single ion in a trap, and each QR model simulates two trapped ions, termed A and B, which are coupled to a cavity. We develop models to simulate the underlying physical processes that govern trapped ion dynamics, which include: (a) Ion initialization that prepares ions for HEG and (b) HEG photon generation that prepares and excites an ion to generate photons for HEG. A photon generation process consists of up to $n$ attempts. Each attempt is triggered by \textit{a clock signal} from a BSM-node. The process terminates when any attempt succeeds; otherwise up to $n$ attempts will be made until the process exits with failure. Based on these building blocks, a Q-node protocol and a QR protocol were developed for entanglement generation (see Fig.~\ref{fig:controlseq}). Both protocols run continuously in cycles. For the Q-node protocol, each cycle consists of two phases: \textit{ion initialization} and \textit{photon generation}. For the QR protocol, each cycle consists of four phases: \textit{ion A and B initialization}, \textit{HEG photon generation on ion A}, \textit{HEG photon generation on ion B}, and \textit{DBSM}.

    \begin{figure}
    \centering
    \vspace{-0mm}
    \includegraphics[width =\linewidth]{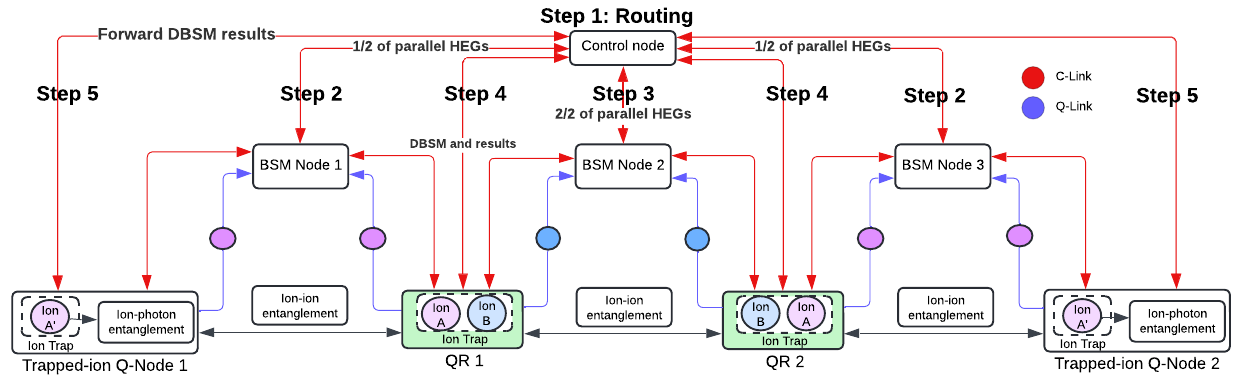}
    \caption{Centralized control for a trapped ion quantum network.}
    \vspace{-4mm}
	\label{fig:trapped_ion_centralized_control}
    \end{figure}

       \subsubsection{Logically centralized control}        
        We design and implement a logically centralized control for a trapped ion quantum network. A control node controls and orchestrates the underlying quantum network devices. Such a design facilitates end-to-end entanglement generation in a trapped ion quantum network. As illustrated in Fig.~\ref{fig:trapped_ion_centralized_control}, to start entanglement generation between two Q-nodes in a network, the control node first performs routing to choose a path between the two Q-nodes, and then activates the two-step approach to conduct HEGs in the path. As soon as HEGs are completed, DBSMs are performed in QRs. DBSM outcomes are transmitted from each BSM-node to the control-node through classical channels. Upon receiving the outcomes of every BSM-nodes, the control node will determine whether the entanglement generation is successful and requires Pauli frame adjustment. This information is then sent to both end Q-nodes through classical channels. In the successful case, each end Q-node performs Pauli frame adjustment if needed to the corresponding matter qubit.

    \subsection{All-photonic quantum repeaters and networks}
    \label{section:APE_design}
        \subsubsection{Graph state generation} In our design, each APE QR model consists of $1$ emitter qubit and $2$ ancillary matter qubits. The emitter qubit is connected to an optical switch, which directs the emitted photons from the emitter into one of the processing units in the QR. There are 6 processing units in total, 3 for left and 3 for right (a leaf processing unit, a 1st-level core processing unit, and a 2nd-level core processing unit), for generation of different types of photon qubits in a RGS: leaf qubit, 1st-level and 2nd-level qubits in the tree-encoded core qubit, respectively. The 1st and 2nd-level core processing units also model the delay lines such that the photons of the encoded core qubit arrive at a BSM-node after the associated leaf qubit. This allows Bell-state measurement of leaf photons to be conducted before any photons of the logical core qubit are measured, since the basis of the latter measurements depends on the BSM outcome on the fly. After passing through each processing unit, the photons are directed to the left or right output quantum channels that connect to neighboring BSM-nodes. Each APE QR model has a local clock that ticks at nanosecond resolution. Driven by the clock signals, the photon qubits of a RGS are generated and transmitted sequentially in a predefined manner. 
         
        \subsubsection{Clock-assisted graph state transmission} 
\begin{figure}[t]
    \centering
    \begin{minipage}[b]{0.66\linewidth}
        \includegraphics[width=\textwidth]{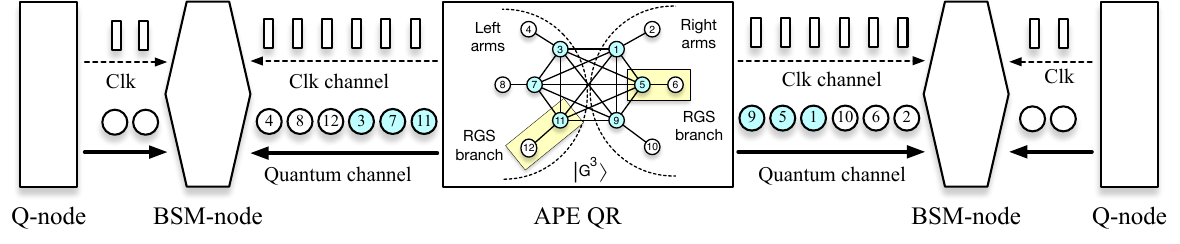}
        \caption{Clock-assisted graph state transmission.}
        \label{fig:transmission}
    \end{minipage}
    \hfill
    \begin{minipage}[b]{0.32\linewidth}
        \includegraphics[width=\textwidth]{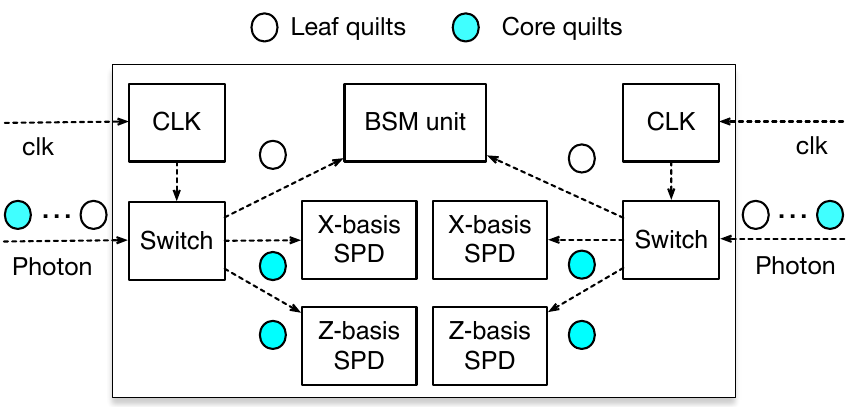}
        \caption{BSM-node model.}
        \label{fig:bsm}
    \end{minipage}
    \end{figure}

Photon loss is an unavoidable phenomenon in a quantum network due to the attenuation of the transmission medium. We implemented a loss-resilient, tree-based encoding scheme to mitigate against photon loss. However, such a scheme will play a role only when photon loss can be detected in time. To this end, we designed and implemented a clock-assisted graph state transmission scheme (see Fig.~\ref{fig:transmission}). As mentioned above, each APE QR model has a local clock that generates clock signals to drive RGS generation. For each photon qubit generation, the QR will also send the driving clock signal through a classic channel to either the left or right BSM-node. This is to herald the arrival of the photon to the BSM-node. It is assumed the clock signal travels at the same speed in the classical channel as the photon qubit in the quantum channel. Given a fixed clock signal interval, the BSM-node will know when to expect the photon to arrive after it receives the clock signal. In the case of a photon loss, the BSM-node will be able to detect the loss event since no photon arrives in the designated time interval, and will take suitable actions to defend against the loss.


        \subsubsection{Graph state reception and processing} Fig.~\ref{fig:bsm} illustrates our BSM-node model. Since each APE QR generates and transmits RGS photon qubits sequentially in a predefined manner, each photon qubit arriving at a BSM-node in the designated time interval can be identified as either a leaf, 1st-level core or 2nd-level core qubit. Next, the leaf qubits are directed to the BSM detector, while the core qubits are directed to the single-photon detectors (SPD). To model a linear optics BSM detector, each pair of leaf qubits of the left and right arms is required to enter the BSM detector simultaneously, and there will be a 50\% probability for a BSM to succeed. Depending on the BSM outcome, the subsequently arriving 1st-level and 2nd-level core qubits are directed to either the $X$-basis SPD or $Z$-basis SPD.
        
         \subsubsection{Logically centralized control} 
         
         We design and implement a logically centralized control for an APE quantum network. A control node controls and orchestrates the underlying quantum network devices. Such a design facilitates end-to-end entanglement generation in an APE quantum network. To start end-to-end entanglement generation between two Q-nodes in a network, the control-node first performs routing to choose a path between the two Q-nodes, and then activates the local clocks in each relevant APE-QR along the path simultaneously. This in turn starts the graph state generation process, which triggers the graph state reception process in the BSM nodes as described in the last section. After all photon qubits are either detected or declared lost in the BSM-nodes, the photonic measurement outcomes are transmitted from each BSM-node to the control node through classical channels. In addition, the measurement outcomes of the emitter qubits of each APE QR-node are also sent to the control node. Upon receiving the outcomes of every BSM-nodes, the control node will determine whether the entanglement generation is successful and requires Pauli frame adjustment. These are then sent to both end Q-nodes through classical channels. In the successful case, each end Q-node performs the suitable Pauli frame adjustment to the corresponding matter qubit. In the failed case, the end Q-nodes allow the involved matter qubits to take part in subsequent entanglement generation attempts.

\section{Simulation and analysis}
\label{analysis}

In this section, we present our simulation of trapped-ion and APE quantum networks using NetSquid. We simulate a QR chain of the form $Q_1 \leftrightarrow BSM_1 \leftrightarrow QR_1 \dots\ QR_n \leftrightarrow BSM_{n+1} \leftrightarrow Q_2$ for each quantum network paradigm. The nodes in the chain are evenly spaced. The chain distance between Q-nodes $Q_1$ and $Q_2$, as well as the number of QRs in the chain, are varied across different simulations. Unless otherwise noted, the parameter choices in our simulations are listed in Tables~\ref{tab:1G-params} and~\ref{tab:APE-params}.

\begin{table}[b]
    \begin{minipage}{.46\linewidth}
      \centering
         \begin{tabular}{|m{5.41cm}|m{0.83cm}|}
            \hline
            \textbf{Simulation parameters} & \textbf{Value} \\
            \hline
            Single qubit gate fidelity & $0.9999$ \\ 
            \hline
            Two qubit gate fidelity & $0.999$ \\ 
            \hline
            Coherence time & $60$ms \\ 
            \hline
            Ion-photon entanglement fidelity & 0.96 \\ 
            \hline
            QFC efficiency & $0.3$ \\ 
            \hline
            Detector efficiency & $0.75$ \\ 
            \hline
            Ion-trap photon collection efficiency & $0.69$ \\ 
            \hline
            Single qubit gate duration & $5$\textmu s \\ 
            \hline
            \text{Mølmer-Sørensen} gate duration & $107$\textmu s \\ 
            \hline
        \end{tabular}  
        \centering
        \caption{Trapped-ion quantum networks}
        \label{tab:1G-params}
        \end{minipage}%
        \begin{minipage}{.54\linewidth}
        \centering

        \begin{tabular}{|m{6.15cm}|m{0.8cm}|}
            \hline
            \textbf{Simulation parameters} & \textbf{Value} \\
            \hline
            Emitter/ancilla CZ gate duration & $100$ns  \\
            \hline
            Emitter/ancilla measurement duration & $20$ns   \\
            \hline
            Emitter coherence time & $3$\textmu s  \\
            \hline
            End node memory coherence time & $20$ms \\
            \hline
            QFC efficiency & $0.95$ \\
            \hline
            Detector efficiency & $1$\\
            \hline
            Single-mode photon emission probability & $0.997$ \\
            \hline
            Photon collection efficiency & $1$ \\
            \hline
        \end{tabular}
        \caption{APE quantum networks}
        \label{tab:APE-params}
    \end{minipage} 
\end{table}

In the simulation of trapped ion QR chains, the matter qubits and light-matter interfaces within each trapped ion QR are simulated using parameters from state-of-the-art experiments~\cite{brown2021materials, krutyanskiy2023telecom, Krutyanskiy2017, schupp2021interface, Poschinger2009}.

In the simulations of APE QR chains, each APE QR generates a ($m,b_0,b_1$) tree-encoded RGS; the quantum emitter and ancilla matter qubits within each APE QR are simulated using parameters chosen to be at the same order of magnitude as state-of-the-art quantum emitters and ancilla matter qubits~\cite{zhan2023performance, schwartz2016deterministic, appel2025many}. Previous research~\cite{hilaire2021resource} has shown that photon losses must be below 15\% in order for the APE QR scheme to outperform both the direct fiber transmission and the memory-based QR scheme. This requirement is mainly due to the fact that the loss tolerance of the tree-encoding only works when the total photon loss is below 50\%. To meet such a requirement, the simulations of APE QR chains were conducted with high values of \textit{quantum frequency conversion efficiency} and \textit{photon collection efficiency}, which are out of reach of today's technology.

Unless otherwise noted, photon loss is modeled at each quantum channel, with a signal attenuation rate of $\alpha=0.2$ dB/km. This gives the overall transmission probability $\sim10^{-\frac{\alpha}L}$, where $L$ is the length of the quantum channel. This is equivalent to a loss probability of $1-$ exp$(-\frac{L}{L_{att}})$, where $L_{att}=22$ km is the attenuation length. 

Through simulation, we study the relative performance and resource requirements of trapped ion and APE quantum networks. The performance metrics include the entanglement generation rate and fidelity between the end nodes of the network. For a fair comparison of resource requirements of different quantum network paradigms, the entanglement generation rate is normalized by the total number of matter qubits involved in the end nodes. The fidelity is calculated as \( F = \langle \Psi^+ | \rho | \Psi^+ \rangle \in [0, 1] \). Here, \( |\Psi^+\rangle \) is a reference Bell state, and \( \rho \) is the density matrix of an entangled pair. It is averaged over multiple trials, and the standard error of the mean is used to quantify the uncertainty in this average fidelity, calculated as the standard deviation of fidelity values divided by the square root of the number of successful trials. The rate is determined by dividing the number of successful events by the total time taken for all trials, giving successful entanglement attempts per second. Please refer to the Appendix for our simulation methodologies.

We carefully verify and validate the simulation results. In particular, the simulation results of trapped ion QR chains are verified and validated using experiment data~\cite{krutyanskiy2023telecom, schupp2021interface} and theoretical analysis in Section~\ref{sec:theoretic_analysis}; the simulation results of APE QR chains are compared and verified with theoretical analysis in Section~\ref{sec:theoretic_analysis} and previous work~\cite{hilaire2021resource}.
        
\subsection{Trapped ion quantum repeater network}

\subsubsection{The chain distance and the number of quantum repeaters}
\begin{figure}[h]
        \centering
        \vspace{-2mm}
        \includegraphics[width =0.8\linewidth]{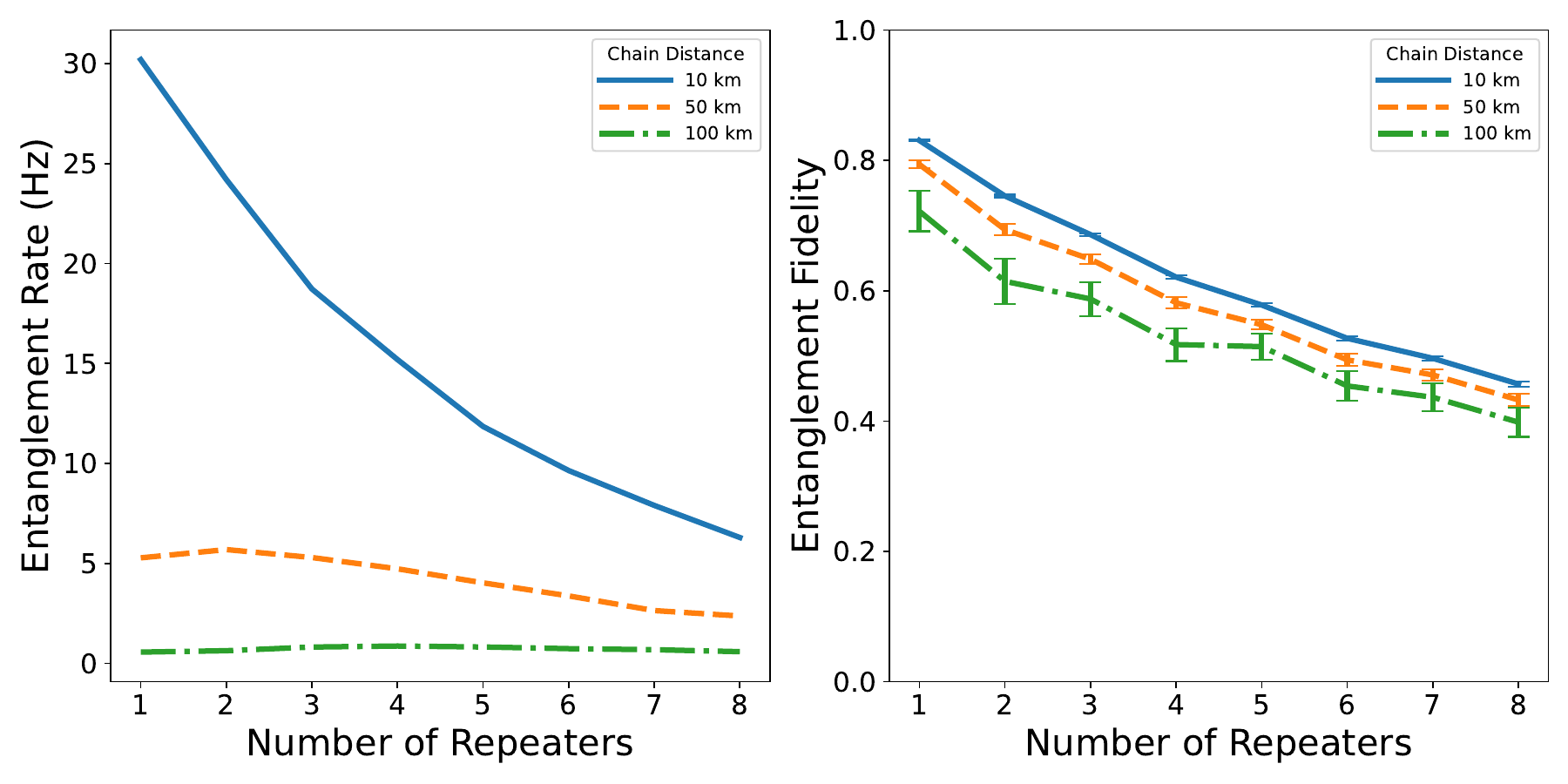}
        \caption{Entanglement generation rate and fidelity with the chain distance and the number of QRs varied. The ion-trap photon collection efficiency is set to $0.69$ (the state of the art).}
        \vspace{-1mm}
	\label{fig:ion_distance_QRs_soa}
        \end{figure}

\begin{figure}[h]
        \centering
        \vspace{-2mm}
        \includegraphics[width =0.8\linewidth]{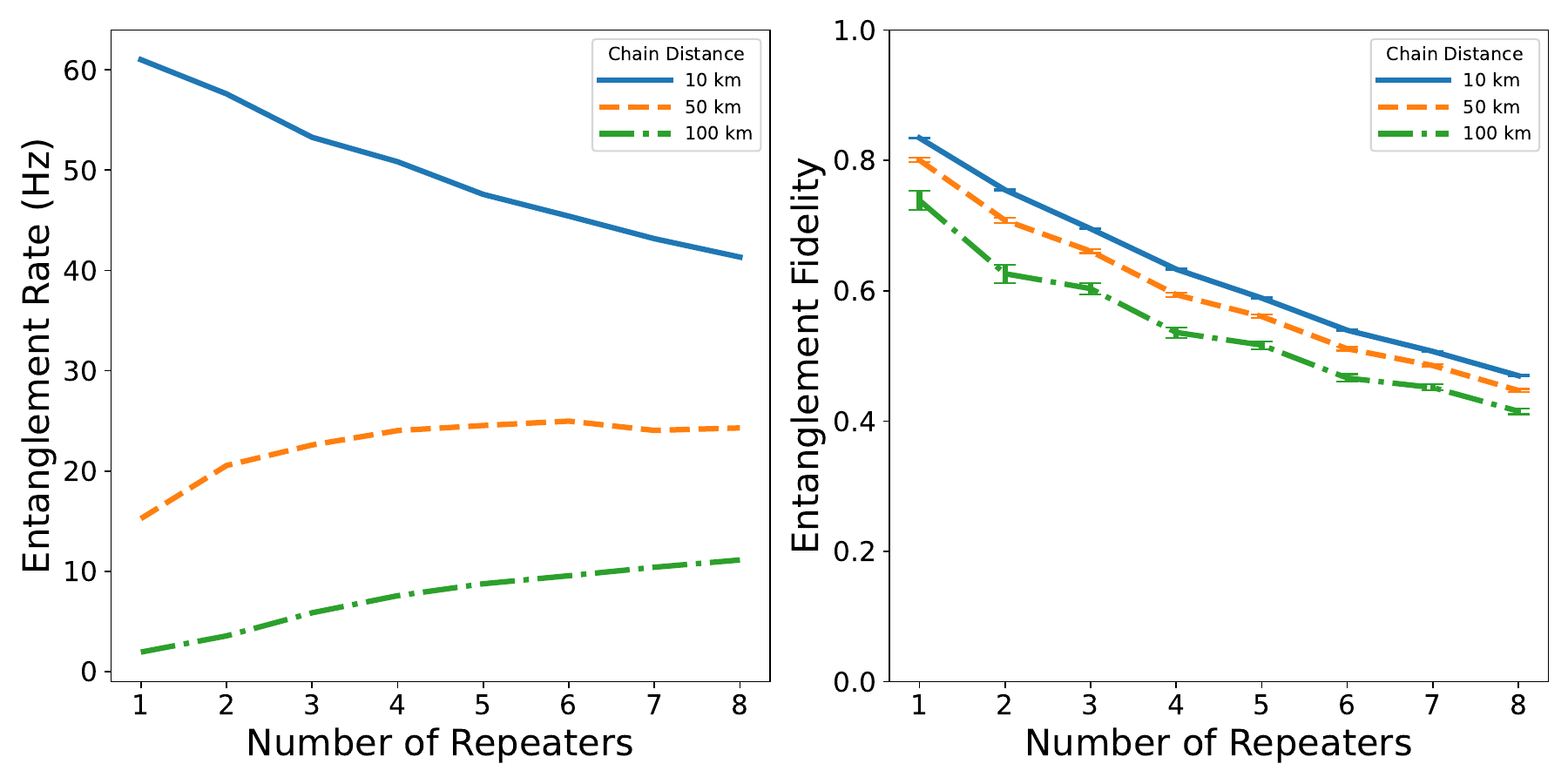}
        \caption{Entanglement generation rate and fidelity with the chain distance and the number of QRs varied. The ion-trap photon collection efficiency is set to 1 (out of reach of today's technology). }
        \vspace{-3mm}
	\label{fig:ion_distance_QRs_ce_1}
        \end{figure}

We study the impact on the entanglement generation rate and fidelity when the chain distance and the number of QRs of a trapped-ion QR chain are varied. We run two scenarios of simulations: (a) using the state-of-the-art parameters (see Table~\ref{tab:1G-params}). In this scenario, the \textit{ion-trap photon collection efficiency} is $0.69$. And (b) with \textit{ion-trap photon collection efficiency} set to $1$. In this scenario, photon loss incurred at each QR is reduced due to the increased \textit{ion-trap photon collection efficiency}. The simulation results are illustrated in Figs.~\ref{fig:ion_distance_QRs_soa} and~\ref{fig:ion_distance_QRs_ce_1}, respectively.

In Fig.~\ref{fig:ion_distance_QRs_soa}, we see that at a chain distance of $10$ km or $50$ km, the entanglement generation rate decreases as the number of QRs in the chain is increased; at a chain distance of $100$ km, the entanglement generation rate is low, close to zero. On the other hand, we can see from Fig.~\ref{fig:ion_distance_QRs_ce_1} that at $10$ km, the entanglement generation rate decreases as the number of QRs in the chain is increased; at $50$ km or $100$ km, the entanglement generation rate increases as the number of QRs is increased; the entanglement generation rates in this scenario are much higher than the corresponding values in the previous scenario, due to the increased ion-trap photon collection efficiency. These simulation results clearly show the impact of QRs on the entanglement generation rate. The major function of QRs is to divide the end-to-end long distance of quantum links into shorter intermediate segments connected by QRs, in which photon loss from fiber attenuation can be corrected. Although adding repeaters in a chain helps to reduce photon loss in quantum links, it also introduces overheads that can be attributed to the collection efficiency of photons, quantum frequency conversion efficiency, and photon detector efficiency, etc. The entanglement rate will decrease when the benefits of adding repeaters in a chain are less than the incurred overheads. 
 
Figs.~\ref{fig:ion_distance_QRs_soa} and~\ref{fig:ion_distance_QRs_ce_1} also show that the fidelity decreases with the number of repeaters, regardless of the chain distance. Similar results were also observed in the simulation of APE QR chains. This is because both QR schemes are mainly designed to mitigate against photon loss in order to improve entanglement generation rates. However, the functioning of QRs necessarily introduces noises, leading to infidelity. The effect of infidelity is cumulative, which increases with the number of repeaters.

        \begin{figure}[h]
        \centering
        \vspace{-2mm}
        \includegraphics[width =0.8\linewidth]{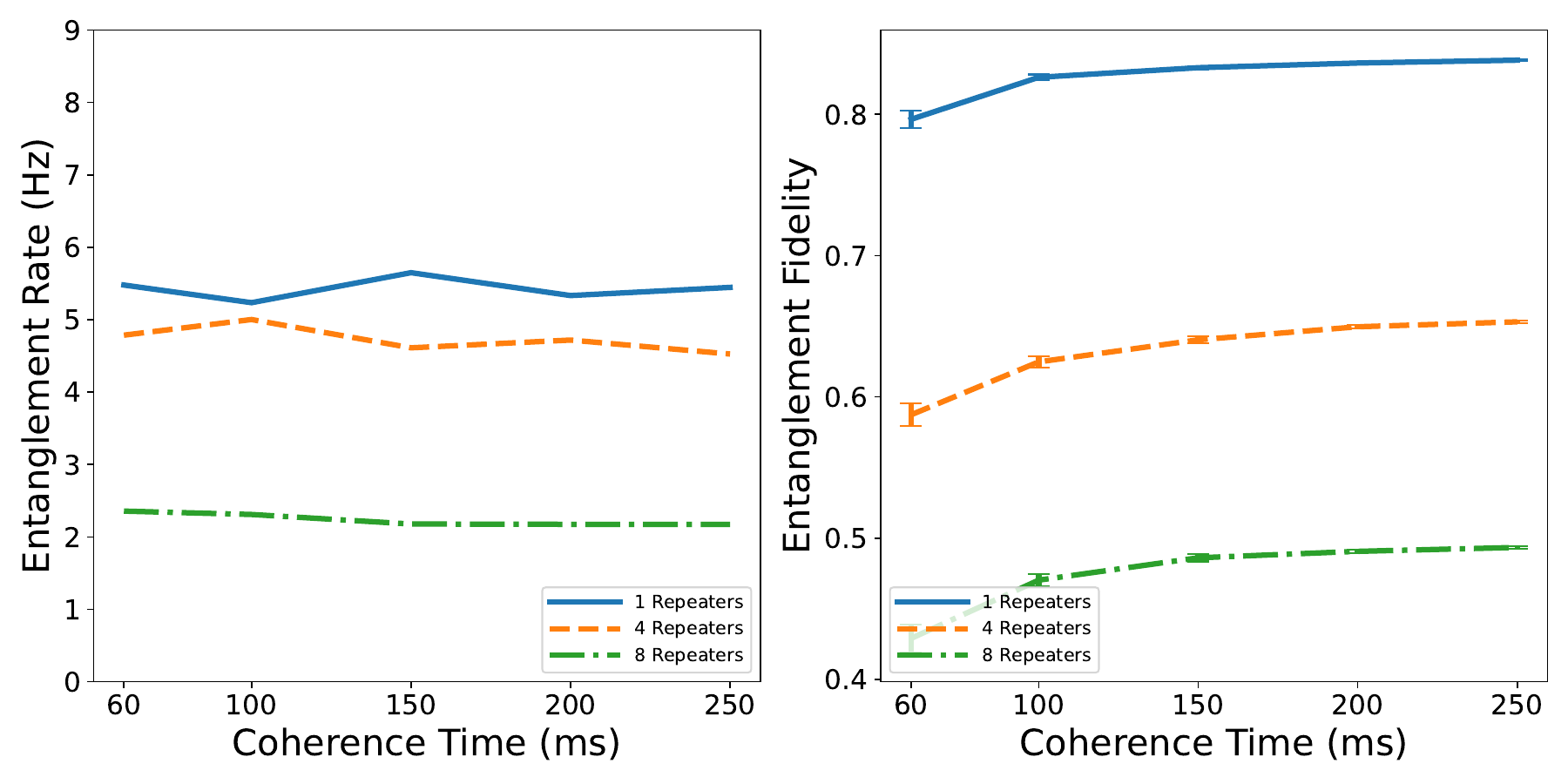}
        \caption{Entanglement generation rate and fidelity with the \Ca coherence time varied. The chain distance is set to $50$ km. Note: The standard error of the mean is small in the figure.}
        \vspace{-2mm}
	\label{fig:ion_coherence_time}
        \end{figure}

 \subsubsection{\Ca Coherence time}
 The state-of-the-art \Ca qubit under working conditions has a coherence time of $60$ ms~\cite{krutyanskiy2023telecom}. In our simulations, we vary the emitter coherence time from $60$ ms to $250$ ms to study the impact of emitter coherence time on the entanglement generation rate and fidelity. As shown in Fig. \ref{fig:ion_coherence_time}, increasing the coherence time improves fidelity for any network configuration, with a larger improvement for longer distances and smaller for shorter distances. This is because longer distances are more sensitive to decoherence, so extending the coherence time has a greater impact on maintaining the entangled state. However, varying coherence time does not affect the entanglement rate.

\subsubsection{Fiber loss}
\label{sec:fiber_loss_trapped_ion}
\begin{figure}[h]
        \centering
        \vspace{-2mm}
        \includegraphics[width =0.8\linewidth]{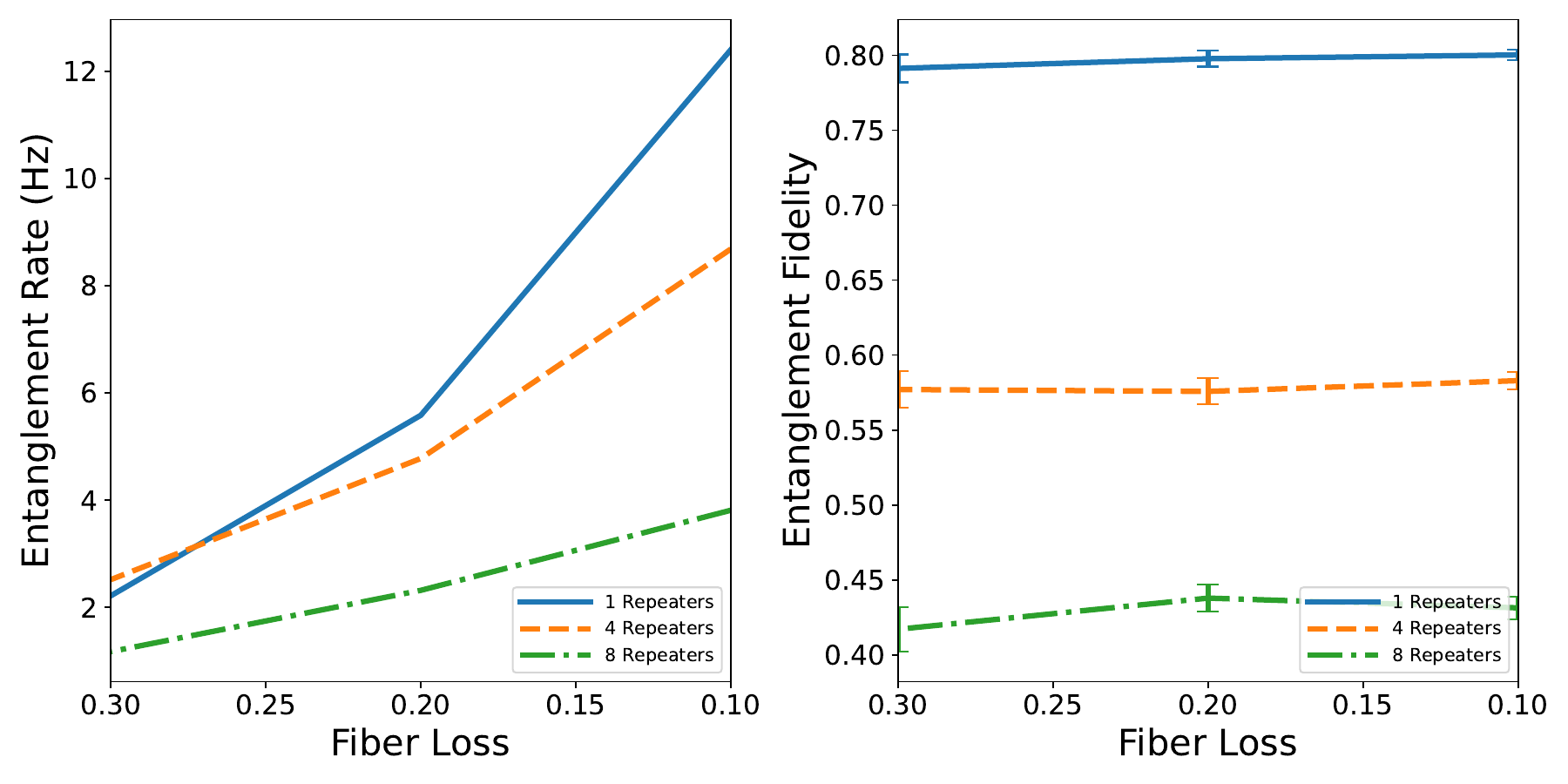}
        \caption{Entanglement generation rate and fidelity with fiber loss varied. The chain distance is set to $50$ km.}
        \vspace{-2mm}
	\label{fig:ion_fiber_loss}
        \end{figure}

The state-of-art fiber technology has a loss below $0.1$ dB/km ~\cite{petrovich2025broadband}. However, most deployed fibers have a fiber loss of ranging from $0.16$ dB/km to $0.3$ dB/km. Specifically, we vary the fiber loss from $0.3$ dB/km to $0.1$ dB/km to study its impact on the entanglement generation rate and fidelity. As shown in Fig.~\ref{fig:ion_fiber_loss}, decreasing fiber loss can significantly increase the entanglement generation rate for any network configuration. However, varying fiber loss has little effect on fidelity. Fiber loss is an important factor that determines quantum network scalability. Ultra-low fiber loss will scale quantum networks to larger distances, reducing the need for QRs.

\subsubsection{Centralized control vs. distributed control}
\begin{figure}[h]
        \centering
        \vspace{-2mm}
        \includegraphics[width =0.8\linewidth]{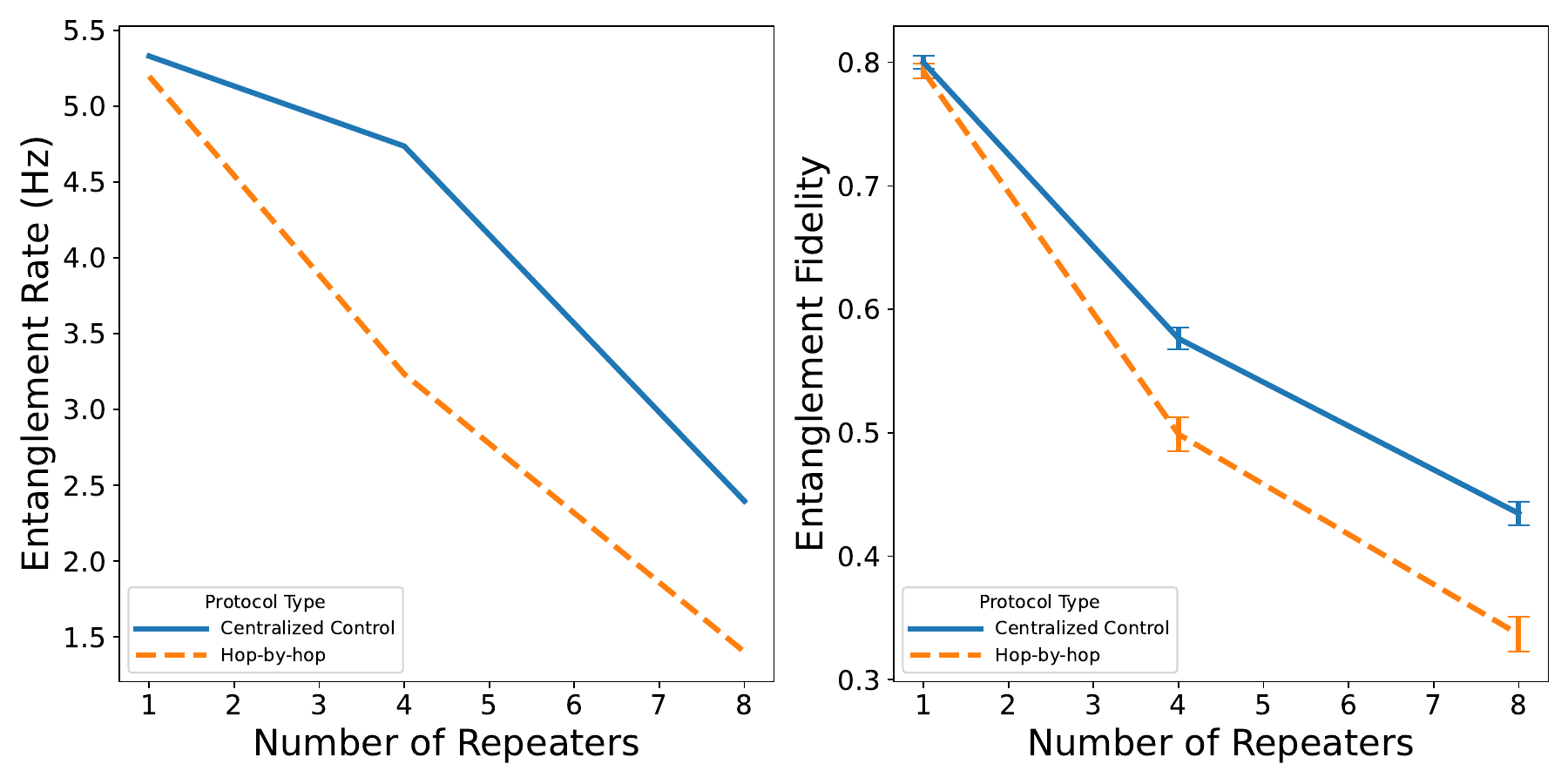}
        \caption{Entanglement generation rate and fidelity with centralized control and distributed control. The chain distance is set to $50km$.}
        \vspace{-3mm}
        \label{fig:ion_centralized_distributed}
        \end{figure}

The centralized control protocol achieves a higher entanglement generation rate than hop-by-hop because it performs HEG in just two steps, reducing delays from sequential operations (Fig. \ref{fig:ion_centralized_distributed}). In the naive hop-by-hop protocol case, entanglement is established sequentially along the repeater chain, leading to longer wait times and lower rates. The rate decreases with more repeaters in the case of the hop-by-hop approach but increases in the centralized control protocol, with the rate gap widening as repeaters increase. The fidelity is lower using hop-by-hop for networks with more than three repeaters due to decoherence from longer wait times. The two protocols have the same overheads for chains of length 2 or less, but these results highlight the importance of optimizing the control protocol when scaling repeater networks.

     \begin{figure}[h]
        \centering
        \vspace{0mm}
        \includegraphics[width =0.8\linewidth]{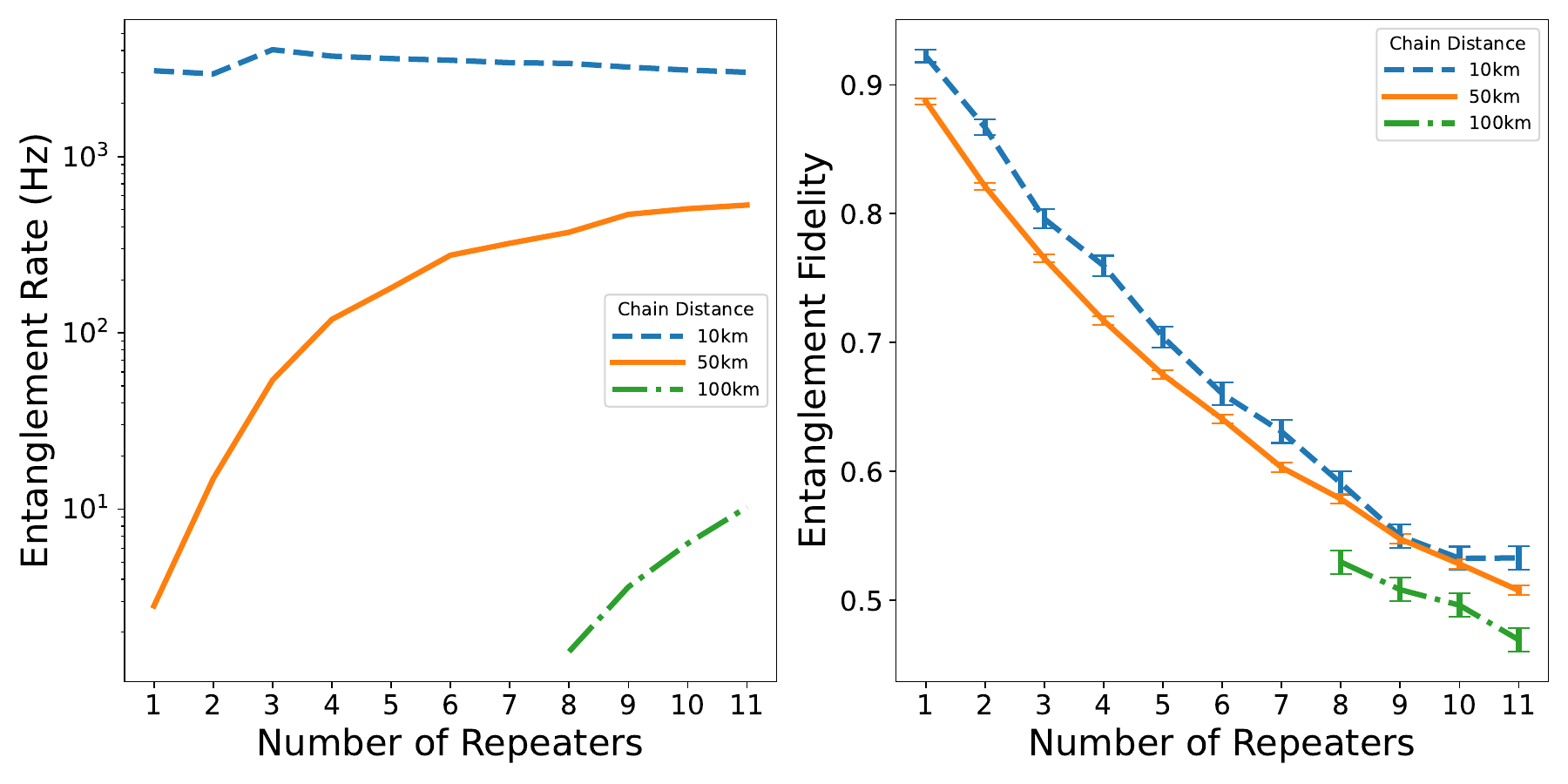}
        \caption{Entanglement generation rate and fidelity with the chain distance and the number of QRs varied. The RGS is set to $(m,b_0,b_1)=(6,6,3)$.}
        \vspace{-2mm}
	\label{fig:Perf vs num of repeaters}
    \end{figure}

\subsection{APE quantum repeaters and networks}
  
\subsubsection{The chain distance and the number of quantum repeaters} 
We study the impact on the entanglement generation rate and fidelity when the chain distance and the number of QRs of an APE QR chain are varied. As illustrated in Fig.~\ref{fig:Perf vs num of repeaters}, at a chain distance of $10$km, the entanglement generation rate decreases as the number of QRs in the chain is increased; at a chain distance of $50$km or $100$km, the entanglement generation rate increases as the number of QRs is increased. Fig.~\ref{fig:Perf vs num of repeaters} also shows that the fidelity decreases with the number of repeaters, regardless of the chain distance. Similar results have been observed in the simulations of trapped ion QR chains.

\subsubsection{Varying RGS size and tree encodings}
RGSs are important resources in an APE quantum network. Given a two-level tree-based encoding, larger RGSs are typically more resilient to photon loss but require longer generation time. As a result, for a smaller RGS, each end-to-end entanglement generation attempt may be less likely to succeed, but attempts can be made more frequently; for a larger RGS, each end-to-end entanglement generation attempt has a higher success probability, at the expense of fewer attempts. To achieve an optimum entanglement generation rate in an APE quantum network, RGS selection is thus not straightforward. To shed light on the meaningful RGSs to investigate in our simulation, five RGSs $(1,25,1)$, $(5,4,2)$, $(6,6,3)$, $(7,8,4)$ and $(8,11,4)$ are studied in our simulations, which correspond to $102$, $130$, $300$, $574$ and $896$ photons, respectively. Specifically, $(6,6,3)$ and $(8,11,4)$ are chosen to showcase the optimized RGS parameters for a QR chain with a distance of $50$ km and 8 repeaters (see Appendix B). $(5,4,2)$ is chosen to illustrate the effect of half the size of $(6,6,3)$; and $(7,8,4)$ is selected to illustrate an intermediate size between $(6,6,3)$ and $(8,11,4)$. Finally, $(1,25,1)$, the optimized RGS size with 1 repeater, is also included. This serves as an illustration of the exotic RGS shape with the fewest number of branches and a large number of subtrees. 

\begin{figure}[h]
        \centering
        \vspace{-2mm}
        \includegraphics[width =0.8\linewidth]{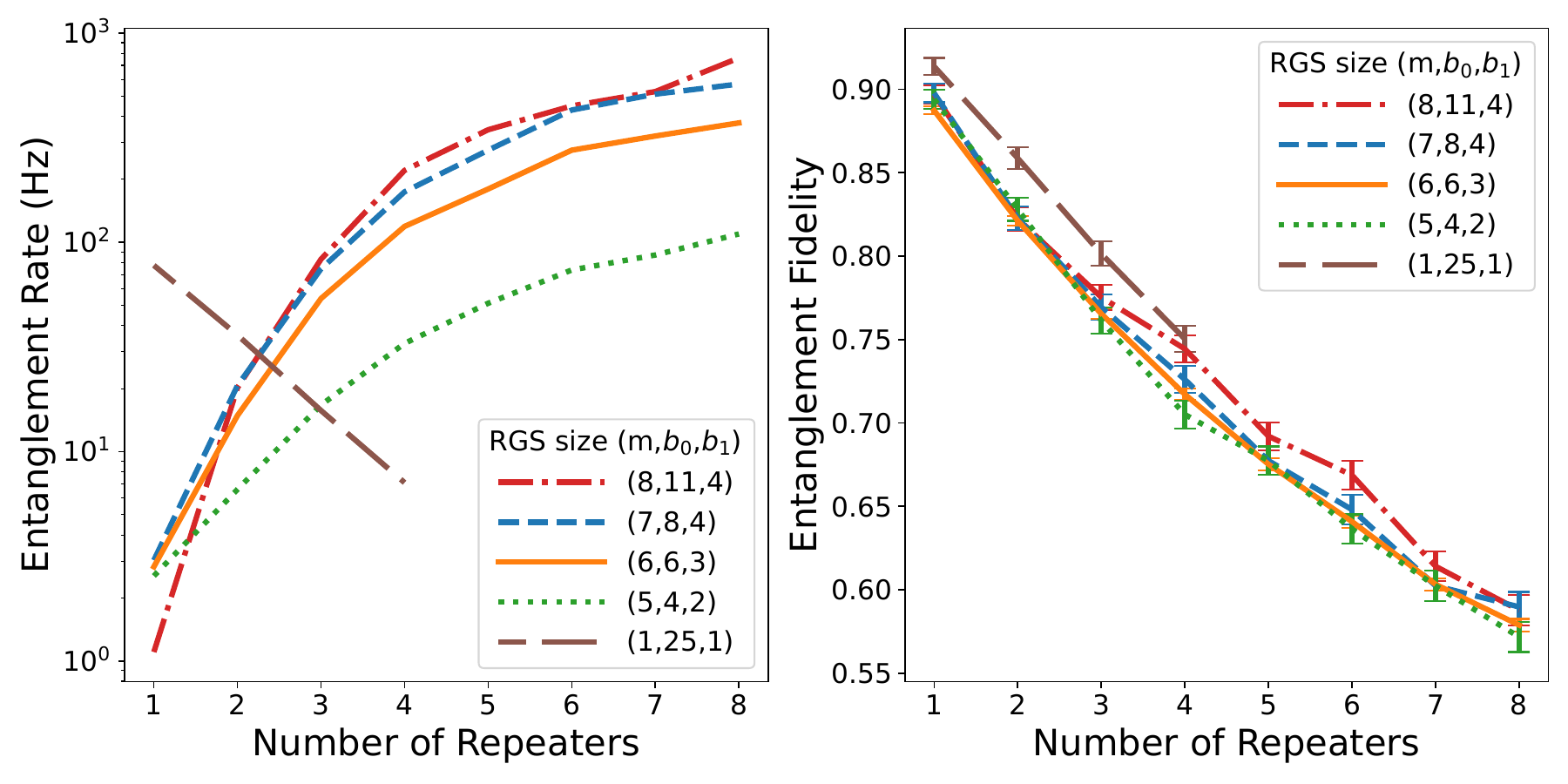}
        \caption{Entanglement generation rate and fidelity with the RGS size and shape and the number of QRs varied. The chain distance is set to $50$ km.}
        \vspace{-2mm}
	\label{fig:RGSsize}
        \end{figure}

The following observations can be drawn from the simulation results shown in Fig.~\ref{fig:RGSsize}: (a) In the case of 1 QR, entanglement generation rates decrease with the RGS sizes; $(1,25,1)$ has the highest rate, which confirms the theoretical analysis in Appendix B. At a chain distance of $50$ km, APE networks with one single repeater have a low success probability for each entanglement generation attempt due to the relatively high fiber loss across node-to-node distance, regardless of the RGS size. However, larger RGSs require longer generation time, which plays a major role in determining the final rates (see Eq. \ref{equ:EGR}). (b) For more than 1 QR, the reduction in node-to-node distance and the associated fiber loss give larger RGSs a huge benefit over smaller ones in mitigating against photon loss. Therefore, entanglement generation rates increase with the RGS size, with the benefit gradually diminishing with larger RGS size, due to longer RGS generation time. (c) For the exotic RGS $(1,25,1)$, $m=1$ means every BSM must be successful in the BSM node in order for the end-to-end entanglement generation to succeed. As a result, the success probability quickly drops with the number of repeaters, and so does the entanglement rate. Due to the low success probability and the long simulation time required, the simulation of (1,25,1) is only run up to 4 repeaters.

Fig.~\ref{fig:RGSsize} also shows that the impact of the RGS size on entanglement fidelity is not obvious among the three ordinary RGSs $(5, 4, 2)$, $(6, 6, 3)$ and $(7, 8, 4)$. This could be explained by the trade-off between the emitter decoherence effect and the error correction effect of RGSs. For a RGS of $(m, b_0, b_1)$, a larger $m$ required the RGS to remain entangled with the ancilla qubits for a longer time; a larger $b_0$ leads to a larger number of CZ gates between the emitter and ancillas. This results in an overall longer time the RGS being entangled with the emitter and ancilla, which in turn increases the emitter and ancilla decoherence effect on the RGS. On the other hand, increasing $b_0$ and $b_1$ also means a larger encoding of each logical core qubit of the RGS. The tree encoding can correct certain Pauli errors by taking majority votes among subtrees when performing the logical $X$ or $Z$ measurements of the encoded core qubits. With a two-level tree with more subtrees ($b_0$) or larger subtrees ($b_1$), the chance of successfully correcting a Pauli error on the photons increases. The RGS (8,11,4) has higher entanglement fidelity when more than 3 QRs are used and photon loss is small enough that more arriving photons mean it is more likely to be able to perform majority votes. The RGS (1,25,1) requires fewer CZ gates during generation and hence exhibits less noise and higher fidelity.

    \begin{figure}[h]
        \centering
        \vspace{-2mm}
        \includegraphics[width =0.8\linewidth]{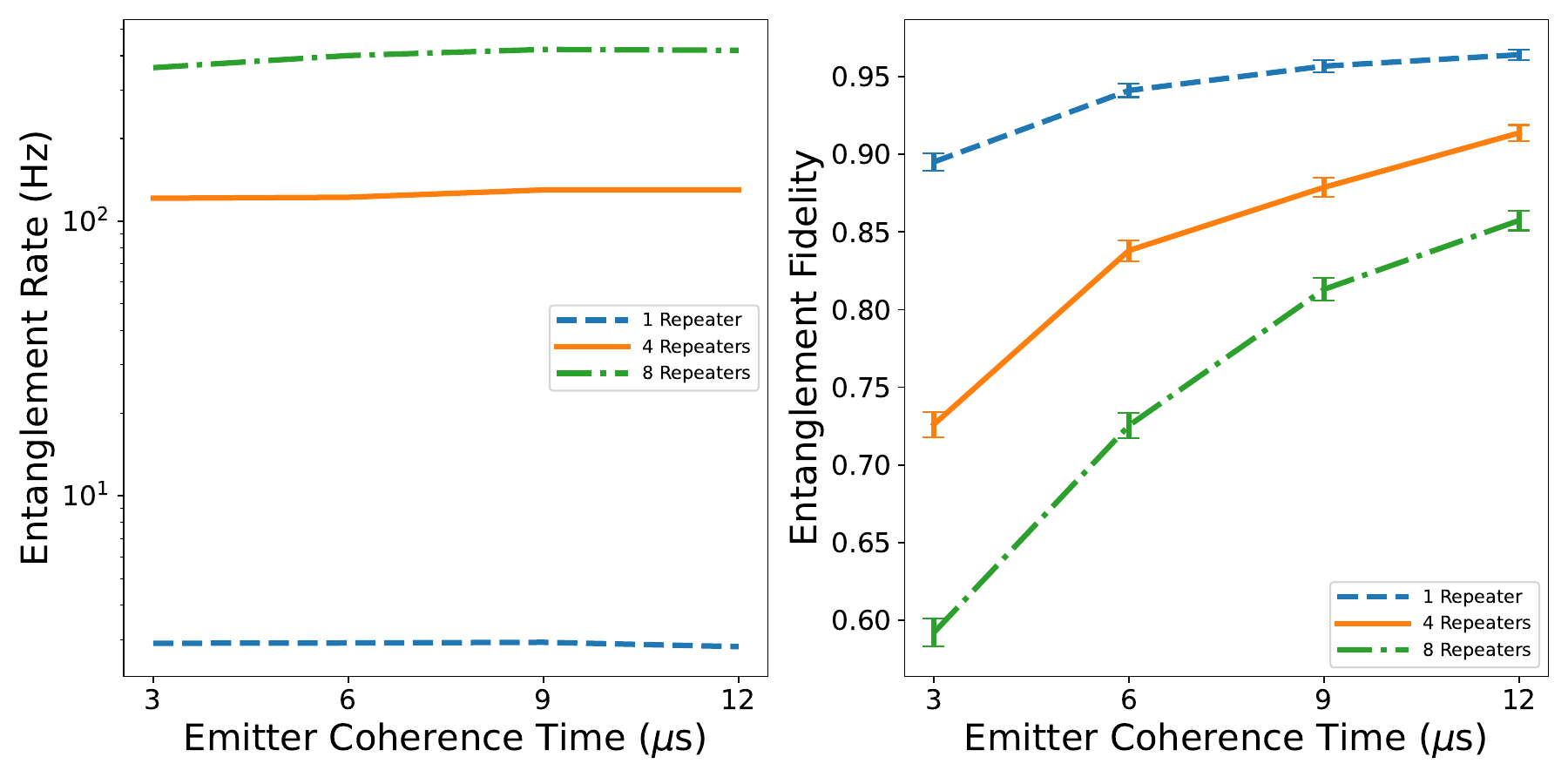}
        \caption{Entanglement generation rate and fidelity with the emitter coherence time varied. The RGS is set to $(m,b_0,b_1)=(6,6,3)$ and the chain distance is set to $50$ km.}
        \vspace{-2mm}
	\label{fig:Fidelity vs emitterT2}
        \end{figure}

\subsubsection{Emitter coherence time} State-of-the-art quantum dot emitters exhibit coherence times on the order of \textmu s \cite{huthmacher2018coherence,nguyen2023enhanced,gillard2022harnessing,hogg2025fast} . In our simulations, we varied the emitter coherence time from $3$ \textmu s to $12$ \textmu s to study the impact of emitter coherence time on the entanglement generation rate and fidelity. The simulation results show that varying emitter coherence time does not have much impact on the entanglement generation rate while it does have a big impact on the entanglement fidelity (see Fig.~\ref{fig:Fidelity vs emitterT2}). Therefore, improving quantum emitter coherence time is critical to increase entanglement fidelity of APE quantum networks.

\subsubsection{Fiber loss}

\begin{figure}[h]
        \centering
        \vspace{-2mm}
        \includegraphics[width =0.8\linewidth]{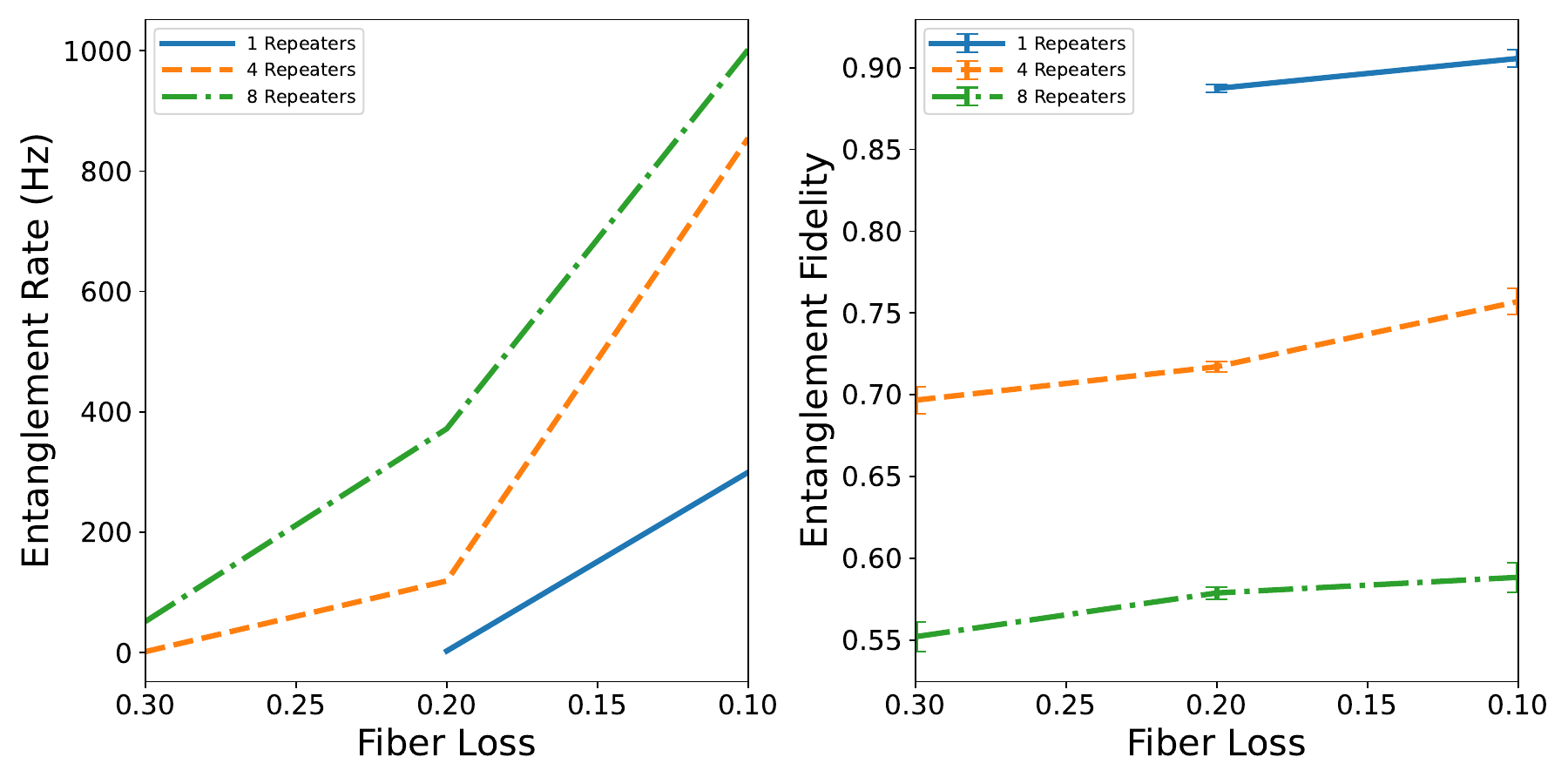}
        \caption{Entanglement generation rate and fidelity with the fiber loss varied. The RGS is set to $(m,b_0,b_1)=(6,6,3)$ and the chain distance is set to $50$ km.}
        \vspace{-2mm}
	\label{fig:Fidelity vs fiber loss}
        \end{figure}

In our simulations, we vary fiber loss from $0.3$ dB/km to $0.1$ dB/km to study its impact on the entanglement generation rate and fidelity of APE quantum networks. As shown in Fig.~\ref{fig:Fidelity vs fiber loss}, decreasing fiber loss can significantly increase the entanglement generation rate for any network configuration. In addition, it is interesting to note that decreasing fiber loss helps to improve fidelity. This is because the tree encoding of RGSs in the APE scheme can correct errors to some extent by majority vote. Smaller fiber loss increases the chance of photons arriving at BSM-nodes, ensuring improved ability to perform majority voting, and finally leading to increased fidelity.

\section{Comparison and discussion}
\begin{figure}[h]
        \centering
        \vspace{-2mm}
        \includegraphics[width =0.8\linewidth]{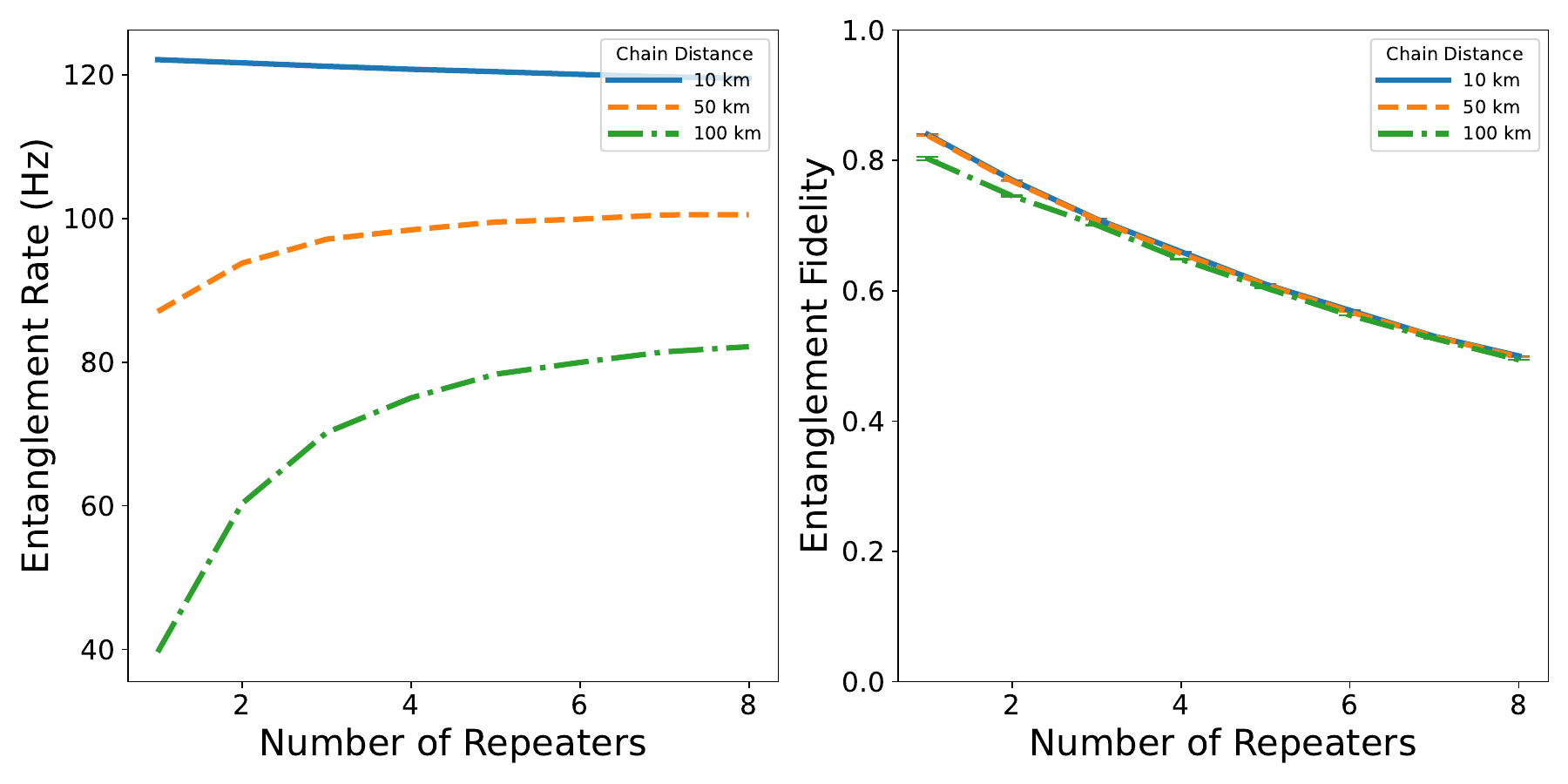}
        \caption{Entanglement generation rate and fidelity for 1G trapped-ion repeater networks with photon collection, QFC and detector efficiencies matched to the APE parameters in Table ~\ref{tab:APE-params}.}
        \label{fig:ion_matching_APE}
        \end{figure}
\begin{figure}[h]
        \centering
        \vspace{-2mm}
        \includegraphics[width =0.8\linewidth]{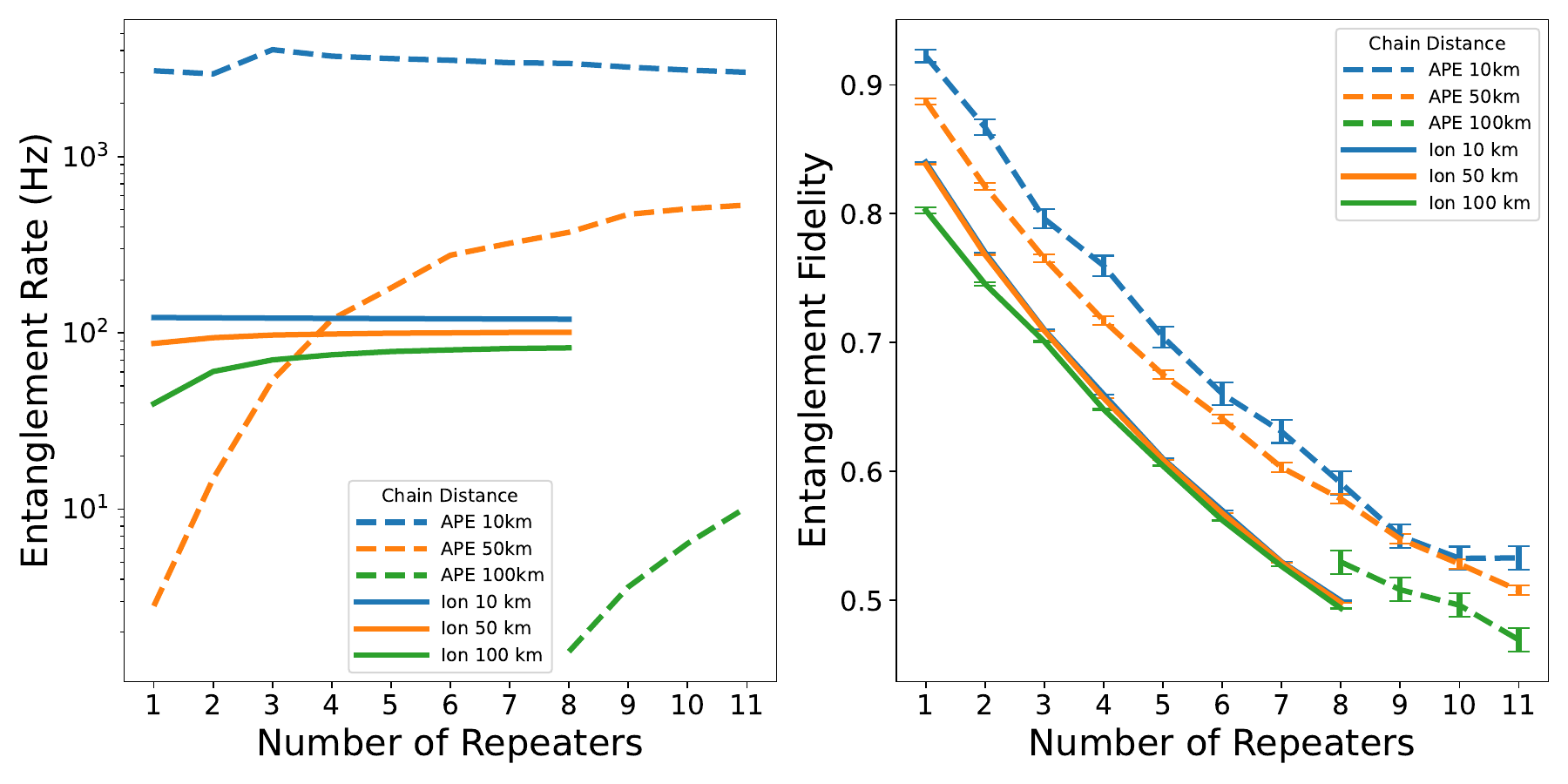}
        \caption{Entanglement generation rate and fidelity for 1G trapped-ion and APE repeater networks at different chain distances, under the same photon collection, QFC and detector efficiencies. The RGS is set to $(m,b_0,b_1)=(6,6,3)$. We simulated 1G networks up to 8 repeaters due to scalability constraints with existing simulation tools.}
        \vspace{-4mm}
	\label{fig:comparison}
        \end{figure}
\label{comparison} To compare trapped-ion and APE repeater networks, we simulate the former one with apparatus parameters matching with those of APE (i.e, the same photon collection efficiency, QFC efficiency and detector efficiency), as shown in Fig.~\ref{fig:ion_matching_APE}. This sheds light on the benefits of the two QR schemes in the regime when apparatus parameters fulfill the requirements imposed by the error correction mechanism of the RGS. In Fig.~\ref{fig:comparison}, we see that: (a) At $10$ km, APE outperforms trapped-ion repeater networks at all numbers of repeaters in terms of entanglement rate. (b) At $50$ km, APE with one or two repeaters has a lower rate mainly due to the relatively high fiber loss of node-to-node distance. Beyond three repeaters, the reduction in node-to-node distance and the associated fiber loss provides APE networks a huge benefit over the trapped-ion repeater network in entanglement rate. (c) At $100$ km, the APE rate is highly hindered by the fiber loss and a huge computation time is required for sufficient successful iterations to be collected in the NetSquid simulation. As illustrated in Fig.~\ref{fig:comparison}, the trapped-ion repeater network exhibits similar entanglement generation rates for $10$ km and $50$ km with a small number of repeaters. It is less sensitive to the number of repeaters and outperforms the APE repeater network in the regime of below nine repeaters at $100$ km. 

These rate comparison results show that the APE QR scheme performs well in networks with short node-to-node distances. Under such conditions, the RGS mechanism can effectively mitigate against photon loss, at the cost of complex RGS generation. Larger RGSs are typically more resilient to photon loss. Increasing the sizes of RGSs may help APE at larger distances, but this benefit is eventually canceled by the longer time required for the RGS generation cycle, and by the added cost and complexities required for APE QR design and implementation. On the other hand, the trapped ion QR scheme performs better than APE in networks with large node-to-node distances. This is because the error correction mechanism of RGSs can only tolerate individual photon loss up to a lower threshold. Therefore, the APE QR scheme is a preferred choice when sufficient QRs are available. Otherwise, the 1G trapped ion QR scheme is a preferred choice, which appears to have better scaling properties at distance in terms of rate.

Fig.~\ref{fig:comparison} also illustrates APE outperforms trapped-ion repeater networks in terms of entanglement fidelity under various conditions, which may be attributed to the distinct inherent sources of noise in the two approaches and the error correction capability in APE under our simulation environment. 

\section{Conclusions}
\label{conclusion}
In this paper, we present our research findings on theoretic analysis and simulation of memory-based trapped ion QRs and networks, and all-photonic APE QRs and networks. Major research findings of this study include: \textit{(a) The entanglement rate may not increase with the number of QRs.} Although adding repeaters in a chain helps to reduce photon loss in quantum links, it also introduces overheads that can be attributed to the collection efficiency of photon, quantum frequency conversion efficiency, photon detector efficiency, etc. The entanglement rate will decrease when the benefits of adding repeaters in a chain are less than the incurred overheads. \textit{(b) The entanglement fidelity decreases with the number of QRs, regardless of the chain distance.} Both QR schemes in our study are mainly designed to mitigate against photon loss in order to improve the entanglement generation rate. However, the functioning of QRs necessarily introduces noise, leading to infidelity. The effect of infidelity is cumulative, which increases with the number of repeaters. Using tree encodings in RGSs in the APE scheme can correct errors to some extent by majority voting. However, for longer distances, the high fiber loss reduces the number of successfully transmitted photons, and hence the ability to perform majority voting. To improve entanglement fidelity, advanced mechanisms such as code concatenation or quantum distillation are required. 

Our study and analysis identify important quantum network performance metrics including quantum network capacity, resource consumption requirements, and scalability. Such results can guide the research and development of QR technologies and the design and construction of large-scale quantum networks.

Our comparison of memory-based trapped-ion and APE QRs reveals their relative merits, which vary across different parameter regimes and length scales. This raises the intriguing possibility of a hybrid quantum network, with APE QRs at city scales and memory-based QRs for inter-city connections. It would be interesting for future research to study such QRs and networks in a quantitative way.

\ack{This document was prepared as an account of work sponsored by the United States Government. While this document is believed to contain correct information, neither the United States Government nor any agency thereof, nor the Regents of the University of California, nor any of their employees, makes any warranty, express or implied, or assumes any legal responsibility for the accuracy, completeness, or usefulness of any information, apparatus, product, or process disclosed, or represents that its use would not infringe privately owned rights. Reference herein to any specific commercial product, process, or service by its trade name, trademark, manufacturer, or otherwise, does not necessarily constitute or imply its endorsement, recommendation, or favoring by the United States Government or any agency thereof, or the Regents of the University of California. The views and opinions of authors expressed herein do not necessarily state or reflect those of the United States Government or any agency thereof or the Regents of the University of California.}

\funding{This research is supported by the Quantum Internet to Accelerate Scientific Discovery ASCR Research Program funded through Berkeley Lab FWP FP00013429, by Berkeley Lab LDRD 25-153, and by the Virginia Commonwealth Cyber Initiative (CCI). EB acknowledges support from the National Science Foundation, grant no. 2137953. SEE acknowledges support from the National Science Foundation, grant no. 2137645.}

\bibliography{biblio}
\bibliographystyle{unsrt} 

\clearpage
\centerline{\textbf{Appendix}}

\appendices
\numberwithin{equation}{section}

\section{Trapped-ion quantum networks}

\renewcommand{\thefigure}{S.\arabic{figure}} 
\setcounter{figure}{0}

\renewcommand{\theequation}{A.\arabic{equation}} 
\setcounter{equation}{0} 

\subsection{Simulation methodology}

In our simulations, end-to-end entanglements are generated through iterations between the Q-nodes. A successful iteration is defined as one in which an end-to-end entanglement generation process succeeds. An entanglement cycle is the time taken until either a process fails or succeeds. The average entanglement generation rate (EGR) for trapped ion quantum networks is calculated using the following formula:

\[
\text{EGR} = \frac{\text{successful iterations}}{T_{\text{tot}}}
\tag{A.1}
\]

where \( T_{\text{tot}} \) is the total entanglement cycles across iterations.


Each simulation task runs for approximately 1500 iterations. The average fidelity is then calculated over all successful iterations. 

\subsection{HEG retry limit in trapped-ion quantum networks}

\begin{figure}[h]
        \centering
        \includegraphics[width =0.8\linewidth]{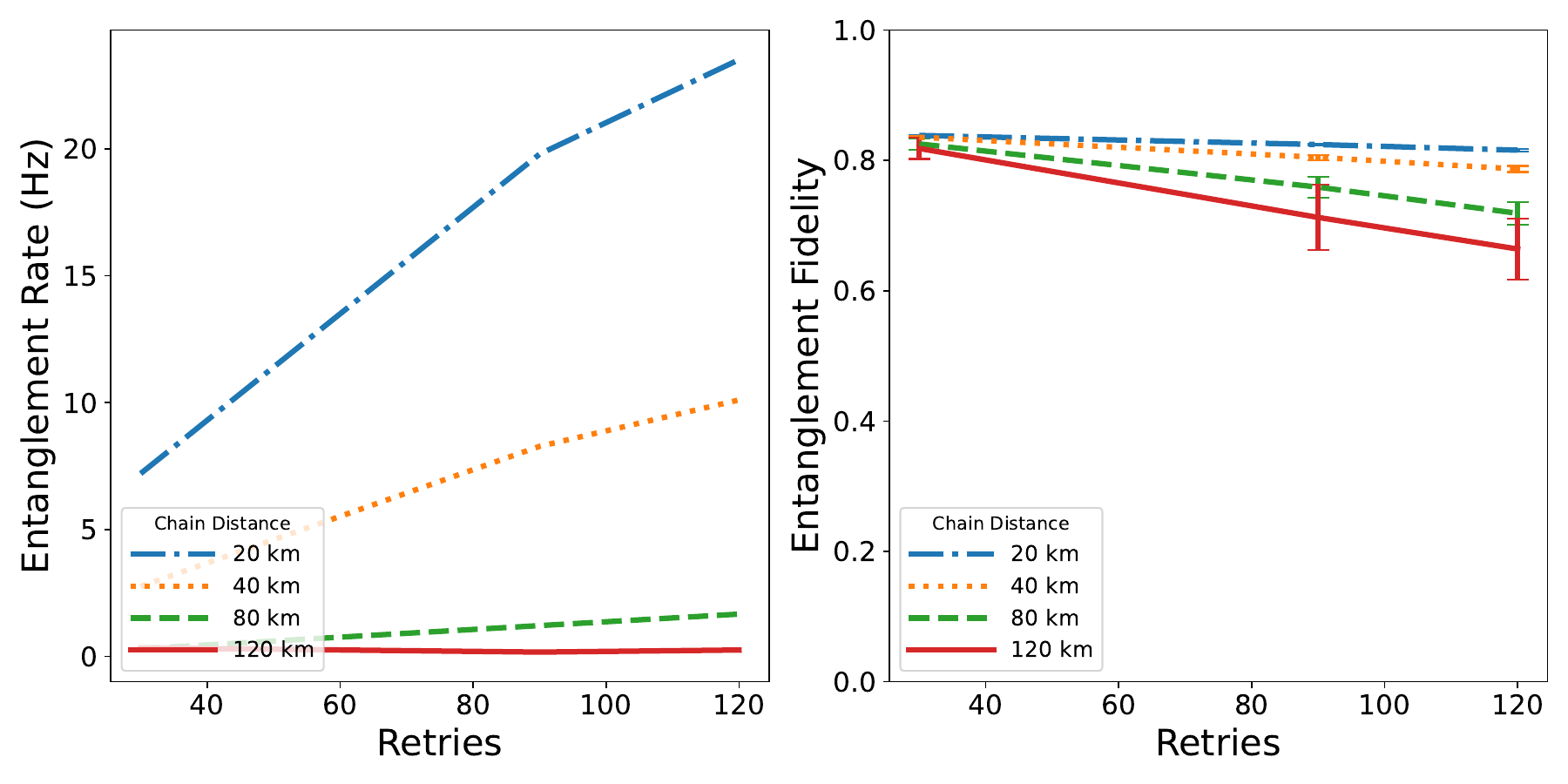}
        \caption{Entanglement generation rate and fidelity with the HEG retry limit and the chain distance varied. The number of QR is set to 1.}
    	\label{fig:ion_HEG retries}
        \end{figure}

In our trapped-ion quantum network design, HEG is implemented with retries, up to a retry limit. As illustrated in Fig.~\ref{fig:ion_HEG retries}, at a chain distance of $20$ km or $40$ km, entanglement rate significantly increases as the HEG retry limit is increased; at $80$ km or $120$ km, HEG retry limit has little impact on the entanglement rate. This is because in a larger quantum network, transmitting photons and passing messages takes more time, and increasing the HEG retry limit will significantly increase the overall entanglement generation process overheads. As a result, it won't improve the overall generation rate. Fig.~\ref{fig:ion_HEG retries} also shows that the fidelity decreases with HEG retry limit, regardless of the chain distance. Therefore, HEG retry limit is a difficult parameter to select. A trade-off between rate and fidelity should be considered. In our simulations, the HEG retry limit is set to 90. 

\clearpage

\subsection{Simulation vs. theoretical analysis cross-validation}
\label{Appendix:trapped_ion_validation}

\begin{figure}[h]
        \begin{subfigure}[b]{1\linewidth}
        \centering
        \includegraphics[width =0.72\linewidth]{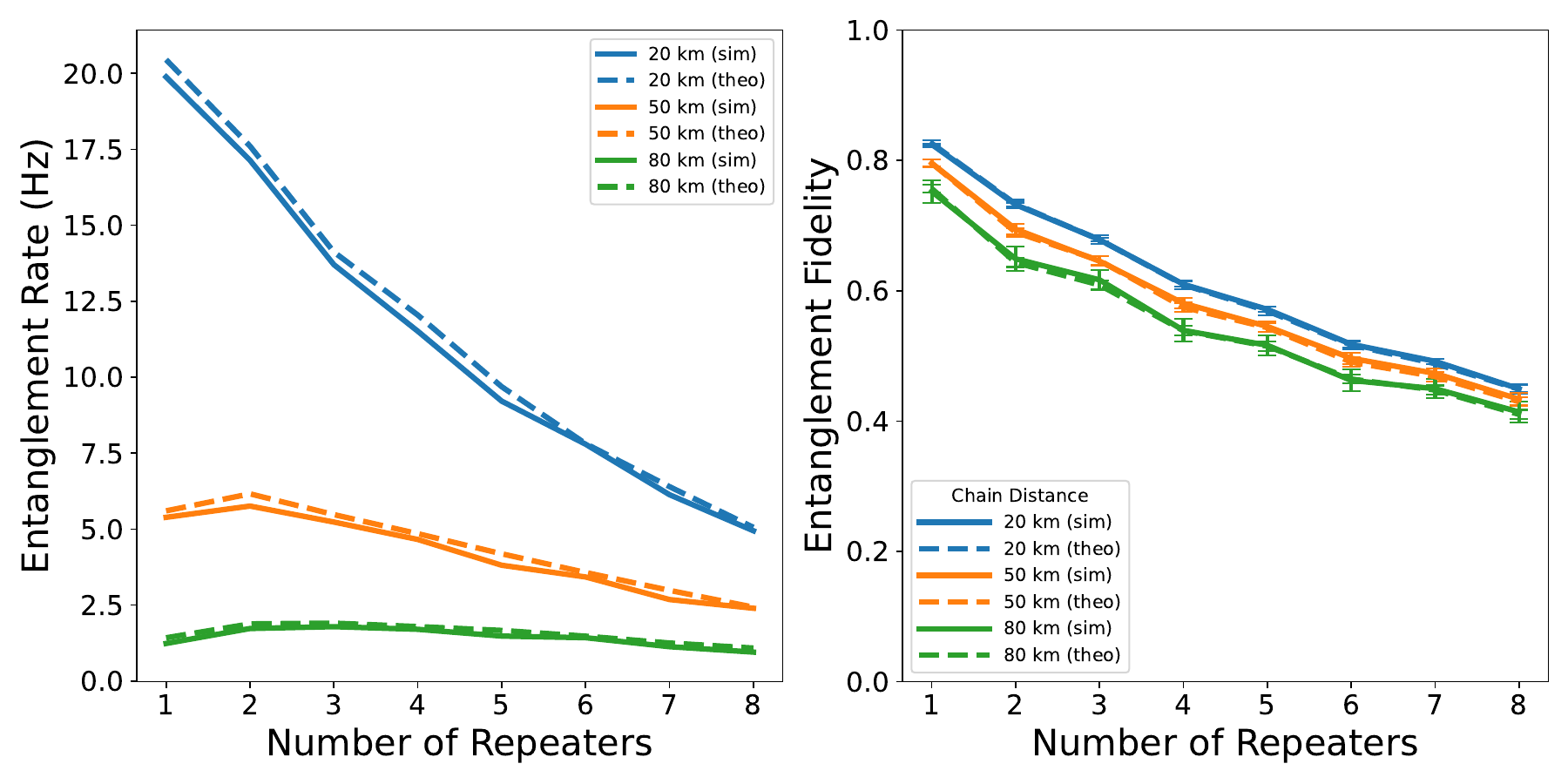}
        \caption{Entanglement generation rate and fidelity with the chain distance and the number of QRs varied.}
        \label{fig:trapped_ion_validation_rate_fid}
        \end{subfigure}

        \begin{subfigure}[b]{1\linewidth}
        \centering
        \includegraphics[width =0.72\linewidth]{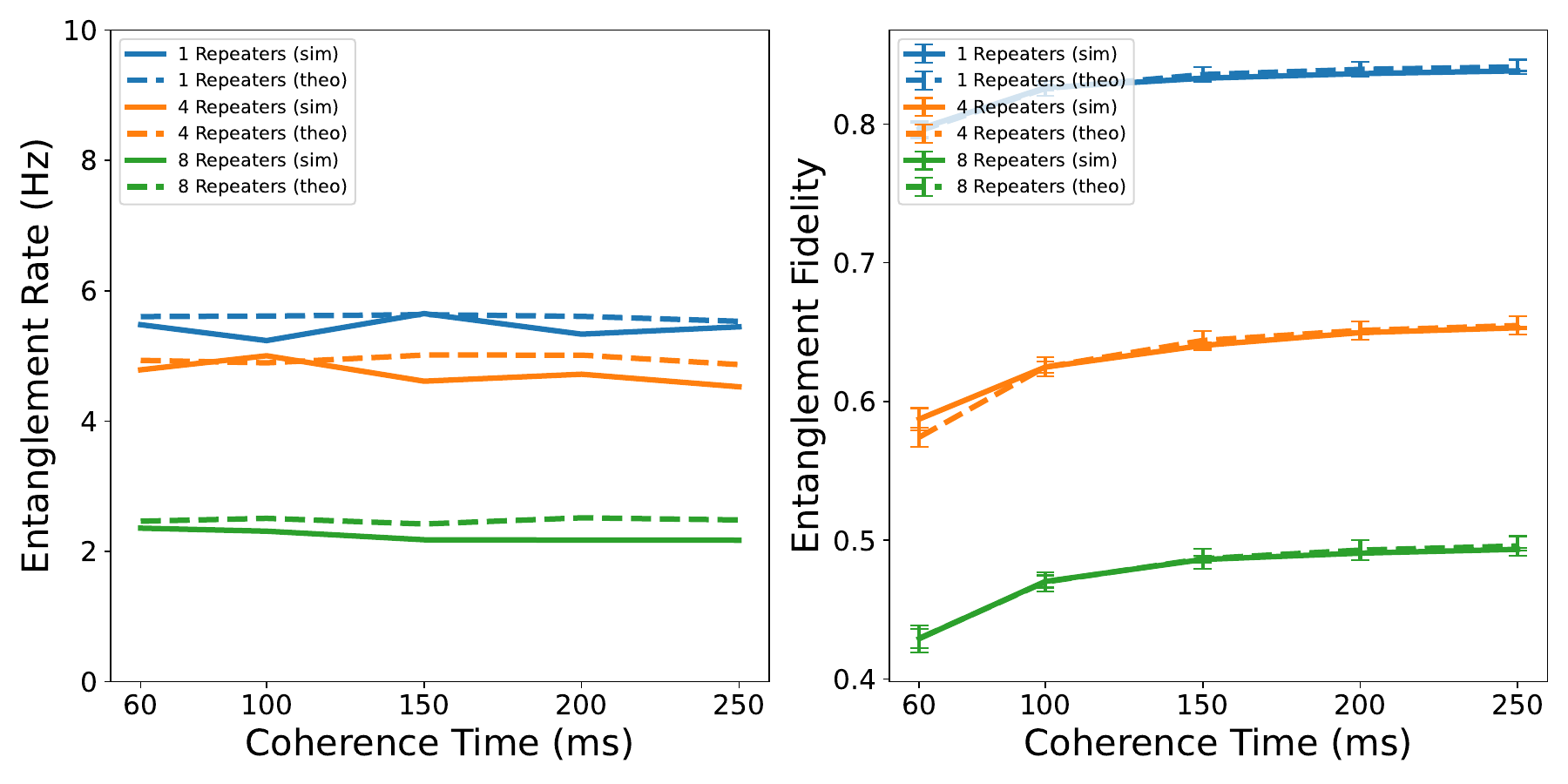}
        \caption{Entanglement generation rate and fidelity with the \Ca coherence time varied. The chain distance is set to $50$ km.}
        \label{fig:trapped_ion_validation_ct}
        \end{subfigure}

        \begin{subfigure}[b]{1\linewidth}
        \centering
        \includegraphics[width =0.72\linewidth]{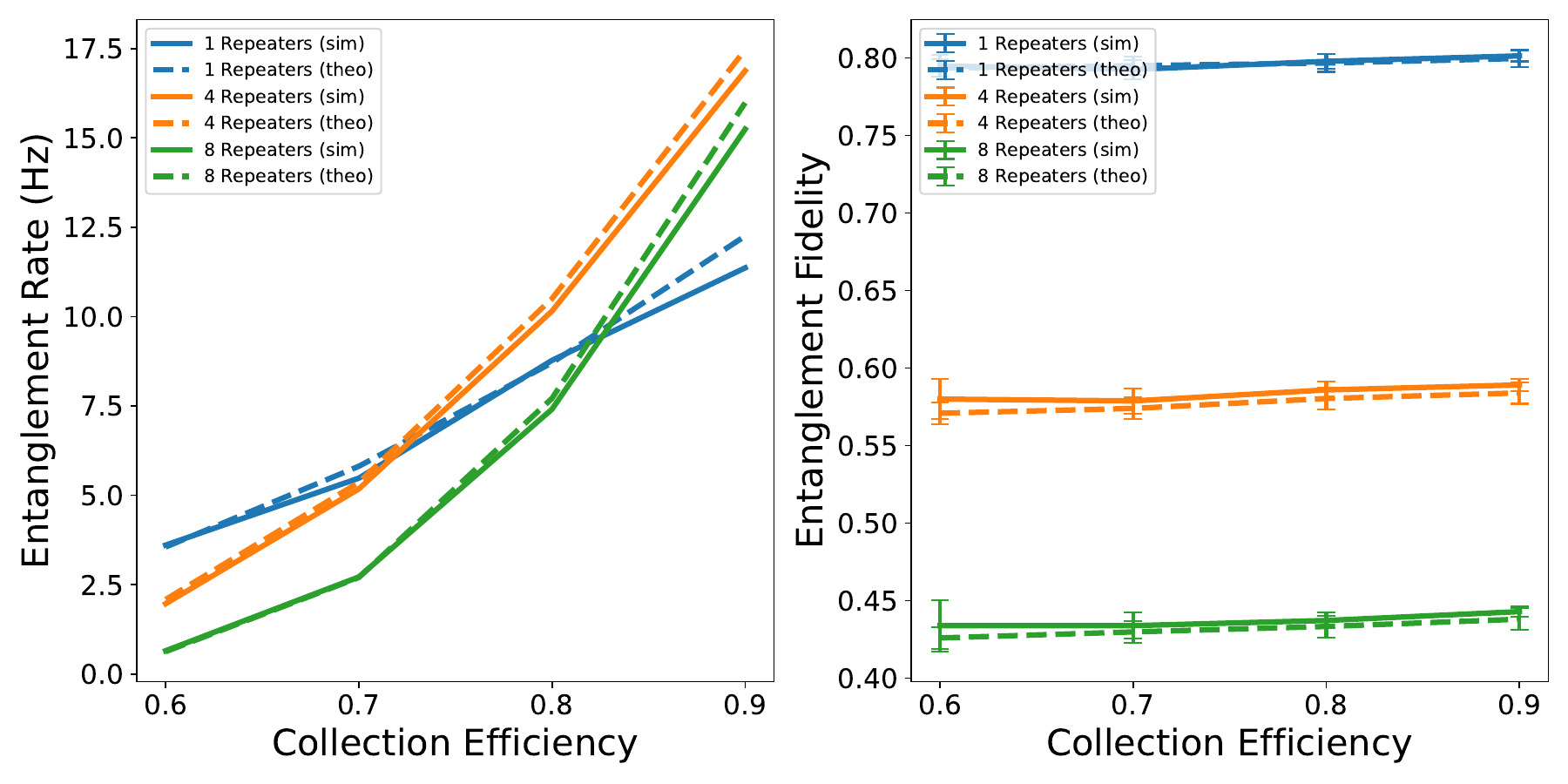}
        \caption{Entanglement generation rate and fidelity with the ion-trap photon collection efficiency varied. The chain distance is set to $50$ km.}
        \end{subfigure}
        \caption{Simulation vs. theoretical analysis cross-validation for trapped-ion quantum networks}
        \label{fig:trapped_ion_validation}
        \end{figure}

We carefully compare and cross-validate simulation results with theoretical analysis to ensure correctness and accuracy.  Fig.~\ref{fig:trapped_ion_validation} shows that simulation study of trapped-ion quantum network is in align with theoretical analysis.

\clearpage

\subsection{Derivation of ion-ion entangled state across two end nodes with n repeaters}
\label{Appendix:Derivation}
\begingroup
\setlength{\parindent}{0pt}
 Define DBSM between ions j and k as the following operations:
\begin{enumerate}
    \item \(e^{i\frac{\pi}{8}Z_j} e^{-i\frac{\pi}{8}Z_k}\)
    \item \(e^{-i\frac{\pi}{4} X_j X_k}\)
    \item Depolarization channels \(D_j^{p_{\rm MS}} \otimes D_k^{p_{\rm MS}}\) to model the MS gate error.
    \item \(Z\) measurements on \(j\) and \(k\).
\end{enumerate}

Define $\rho_{ij}(w,\lambda,S)\equiv\frac{1}{4}\, I_{ij}
+ \frac{w}{4}\!\left[
\lambda \cos S\, (X_i X_j - Y_i Y_j)
+ \lambda \sin S\, (X_i Y_j + Y_i X_j)
+ \lambda^{2} Z_i Z_j
\right]$.\\\medskip

 Lemma 1: Given the states $\rho_{ij}(w_1,\lambda_1,\phi_1)$ and $\rho_{kl}(w_2,\lambda_2,\phi_2)$. After DBSM between j and k, the resultant state is $\rho_{il}(w_1w_2,\eta\lambda_1\lambda_2,\phi_1+\phi_2-\frac{\pi}{2})$, where $\eta=1-p_{\rm MS}$.\\
 Proof:
  After operation 1, the state becomes 
\begin{equation*}
\rho'_{ijkl}
= \bigl[ \rho_{ij}(w_1,\lambda_1,\phi_1+\tfrac{\pi}{4}) \bigr]
\otimes
\bigl[ \rho_{kl}(w_2,\lambda_2,\phi_2-\tfrac{\pi}{4}) \bigr]  
\end{equation*}
Conjugating the state by $U=e^{-i\frac{\pi}{4}X_jX_k}$ and using the relations 
\begin{align*}
U X_j U^{\dagger} &= X_j, \qquad & U Y_j U^{\dagger} &= Z_j X_k, \qquad & U Z_j U^{\dagger} &= -\,Y_j X_k,\\
U X_k U^{\dagger} &= X_k, \qquad & U Y_k U^{\dagger} &= X_j Z_k, \qquad & U Z_k U^{\dagger} &= -\,X_j Y_k.
\end{align*}\\the state becomes

\begin{align*}
\rho''_{ijkl}=&\frac{1}{16}\, I_{ijkl}+
\frac{w_1}{16}\!\left[
  \alpha_1\!\left( X_i X_j - Y_i Z_j X_k \right)
  + \beta_1\!\left( X_i Z_j X_k + Y_i X_j \right)
  + \gamma_1\!\left( - Z_i Y_j X_k \right)
\right]\\
&+\frac{w_2}{16}\!\left[
  \alpha_2\!\left( X_k X_l - X_j Z_k Y_l \right)
  + \beta_2\!\left( X_k Y_l + X_j Z_k X_l \right)
  + \gamma_2\!\left( - X_j Y_k Z_l \right)
\right]\\
&+\frac{w_1 w_2}{16}\Big[
\alpha_1\alpha_2\big(
 X_i X_j X_k X_l
 - X_i Z_k Y_l
 - Y_i Z_j X_l
 + Y_i Y_j Y_k Y_l
\big)\\
&\quad+ \alpha_1\beta_2\big(
 X_i X_j X_k Y_l
 + X_i Z_k X_l
 - Y_i Z_j Y_l
 - Y_i Y_j X_k X_l
\big)\\
&\quad+ \alpha_1\gamma_2\big(
 - X_i Y_k Z_l
 - Y_i Y_j Z_k Z_l
\big)\\
&\quad+ \beta_1\alpha_2\big(
 X_i Z_j X_l
 - X_i Y_j Y_k Y_l
 + Y_i X_j X_k X_l
 - Y_i Z_k Y_l
\big)\\
&\quad+ \beta_1\beta_2\big(
 X_i Z_j Y_l
 + X_i Y_j Y_k X_l
 + Y_i X_j X_k Y_l
 + Y_i Z_k X_l
\big)\\
&\quad+ \beta_1\gamma_2\big(
 X_i Y_j Z_k Z_l
 - Y_i Y_k Z_l
\big)
+ \gamma_1\alpha_2\big(
 - Z_i Y_j X_l
 - Z_i Z_j Y_k Y_l
\big)\\
&\quad+ \gamma_1\beta_2\big(
 - Z_i Y_j Y_l
 + Z_i Z_j Y_k X_l
\big)
+ \gamma_1\gamma_2\big(
 Z_i Z_j Z_k Z_l
\big)
\Big]
\end{align*}
with the shorthand\\
\begin{align*}
\alpha_1 &= \lambda_1 \cos\!\left(\phi_1+\tfrac{\pi}{4}\right), &
\beta_1  &= \lambda_1 \sin\!\left(\phi_1+\tfrac{\pi}{4}\right), &
\gamma_1 &= \lambda_1^{2},\\[2pt]
\alpha_2 &= \lambda_2 \cos\!\left(\phi_2-\tfrac{\pi}{4}\right), &
\beta_2  &= \lambda_2 \sin\!\left(\phi_2-\tfrac{\pi}{4}\right), &
\gamma_2 &= \lambda_2^{2}.
\end{align*}\\
Applying an independent single-qubit depolarizing channel of rate $p_{\rm MS}$ on ions j and k simply shrinks any non-identity Pauli on the acted-on qubit(s). If a term has a non-identity on j or k, its coefficient is multiplied by 
$\eta=1-p_{\rm MS}$:
\begin{align*}
\rho'''_{ijkl}=&\frac{1}{16}\, I_{ijkl}+
\frac{w_1}{16}\Big[
\eta\,\alpha_1 X_i X_j
-\eta^{2}\alpha_1 Y_i Z_j X_k
+\eta^{2}\beta_1 X_i Z_j X_k
+\eta\,\beta_1 Y_i X_j
-\eta^{2}\gamma_1 Z_i Y_j X_k
\Big]\\
&+\frac{w_2}{16}\Big[
\eta\,\alpha_2 X_k X_l
-\eta^{2}\alpha_2 X_j Z_k Y_l
+\eta\,\beta_2 X_k Y_l
+\eta^{2}\beta_2 X_j Z_k X_l
-\eta^{2}\gamma_2 X_j Y_k Z_l
\Big]\\
&+\frac{w_1 w_2}{16}\Big[
\alpha_1\alpha_2\big(
\eta^{2} X_i X_j X_k X_l
+\eta(-X_i Z_k Y_l)
+\eta(-Y_i Z_j X_l)
+\eta^{2} Y_i Y_j Y_k Y_l
\big)\\
&+ \alpha_1\beta_2\big(
\eta^{2} X_i X_j X_k Y_l
+\eta X_i Z_k X_l
+\eta(-Y_i Z_j Y_l)
+\eta^{2}(-Y_i Y_j X_k X_l)
\big)\\
&+ \alpha_1\gamma_2\big(
\eta(-X_i Y_k Z_l)
+\eta^{2}(-Y_i Y_j Z_k Z_l)
\big)\\
&+ \beta_1\alpha_2\big(
\eta X_i Z_j X_l
+\eta^{2}(-X_i Y_j Y_k Y_l)
+\eta^{2} Y_i X_j X_k X_l
+\eta(-Y_i Z_k Y_l)
\big)\\
&+ \beta_1\beta_2\big(
\eta X_i Z_j Y_l
+\eta^{2} X_i Y_j Y_k X_l
+\eta^{2} Y_i X_j X_k Y_l
+\eta Y_i Z_k X_l
\big)\\
&+ \beta_1\gamma_2\big(
\eta^{2} X_i Y_j Z_k Z_l
+\eta(-Y_i Y_k Z_l)
\big)
+ \gamma_1\alpha_2\big(
\eta(-Z_i Y_j X_l)
+\eta^{2}(-Z_i Z_j Y_k Y_l)
\big)\\
&+ \gamma_1\beta_2\big(
\eta(-Z_i Y_j Y_l)
+\eta^{2} Z_i Z_j Y_k X_l
\big)
+ \gamma_1\gamma_2\big(
\eta^{2} Z_i Z_j Z_k Z_l
\big)
\Big]
\end{align*}
\\Measuring Z on ions j and k with outcomes (0,0)  kills any term containing $X_j$,$Y_j$,$X_k$,$Y_k$. The normalized post-measurement state on 
ions i and l is
\begin{align*}
\rho_{il}^{(00)}=& \tfrac{1}{4} I_{il}
+ \tfrac{w_1 w_2}{4}\!\left[
  \eta\,\lambda_1\lambda_2\!\left(
    \sin(\phi_1+\phi_2)\,(X_i X_l - Y_i Y_l)\right.\right.\\
    &\hspace*{.2\linewidth}\left.\left.- \cos(\phi_1+\phi_2)\,(X_i Y_l + Y_i X_l)\right) + \eta^{2}\lambda_1^{2}\lambda_2^{2}\, Z_i Z_l\right]\\
=&\; \rho_{il}\!\left(w_1 w_2,\; \eta\,\lambda_1\lambda_2,\; \phi_1+\phi_2-\tfrac{\pi}{2}\right)
\end{align*}\\
where we have used the trigonometric identities
\[
\alpha_1 \beta_2 + \beta_1 \alpha_2 = \lambda_1 \lambda_2 \sin(\phi_1+\phi_2),
\qquad
-\alpha_1 \alpha_2 + \beta_1 \beta_2 = -\,\lambda_1 \lambda_2 \cos(\phi_1+\phi_2).
\]\\
Here, there are a total of 4 different measurement outcomes, each of equal probability. We assume perfect Pauli frame adjustments such that we only need to consider the (0,0) post-measurement state. \\
Theorem 1: For $n$ repeaters, the final ion-ion entangled state across two end nodes is given by \eqref{eq:end-end state}.\\
Proof: We first consider a sequential entanglement swapping from the left end node to the right end node. After each successful entanglement swapping between two ions at different nodes, the resultant state is given by the Werner state  \eqref{eq:ion-ion state HEG}, which can be rewritten in the Pauli basis as 
\begin{align*}
\rho_{12} &=w_{\rm em}^2\ket{\psi^+(\theta_1+\theta_2)}\bra{\psi^+(\theta_1+\theta_2)} + (1-w_{\rm em}^2)\frac{I}{4}\\
&=\frac{1}{4}\, I_{12}
+ \frac{w}{4}\!\left[
 \cos (\theta_1+\theta_2)\, (X_1 X_2 - Y_1 Y_2)
+  \sin (\theta_1+\theta_2) (X_1 Y_2 + Y_1 X_2)
+  Z_1 Z_2
\right]\\
&=\rho_{12}(w=w^2_{\rm em},\lambda=1,S=\theta_1+\theta_2)
\end{align*}
where
\begin{align*}
\ket{\phi^+(\alpha)} &\equiv \frac{1}{\sqrt{2}} (\ket{00} + e^{i \alpha}\ket{11})
\end{align*}
After DBSM between ions $2$ and $3$ at the same repeater node with state $\rho_{12}(w=w^2_{\rm em},\lambda=1,S=\theta_1+\theta_2)$ and $\rho_{34}(w=w^2_{\rm em},\lambda=1,S=\theta_3+\theta_4)$ respectively, by Lemma 1, we obtain 
\begin{align*}
\rho_{14}\!\left(w^4_{\rm em},\; w_{\rm MS},\; \theta_1+\theta_2+\theta_3+\theta_4-\tfrac{\pi}{2}\right)
\end{align*}\\
which is \eqref{eq:ion-ion state DBSM}. Repeat the above steps n-1 times, each with an extra ion-ion state after successful HEG, i.e. $\rho_{56}(w=w^2_{\rm em},\lambda=1,S=\theta_5+\theta_6)$, $\rho_{78}(w=w^2_{\rm em},\lambda=1,S=\theta_7+\theta_8)$, et cetera,  we obtain the final state across two end nodes
\begin{align*}
\rho_{Q1-Q2}\!\left(w^{2n+2}_{\rm em},\; w^n_{\rm MS},\; \theta_{\rm tot}-\tfrac{n\pi}{2}\right)
\end{align*}
where $\theta_{\rm tot}=\sum_{i=1}^{2n+2}\theta_i$ and Q1(Q2) corresponds to the 1st(2n+2-th) ion. This is the state in \eqref{eq:end-end state}. Since Lemma 1 applies to two arbitrary states in the form $\rho(w_1,\lambda_1,\phi_1)$ and $\rho(w_2,\lambda_2,\phi_2)$, the above result for sequential entanglement swapping from the left end node to the right end node holds for an arbitrary sequence of entanglement swapping.

\clearpage

\section{APE quantum networks}

\renewcommand{\theequation}{B.\arabic{equation}} 
\setcounter{equation}{0} 

\subsection{Simulation methodology}
In our simulation, end-to-end entanglements are generated through iterations between the Q-nodes. An iteration is considered successful when at least one BSM in each BSM node is successful and all the logical $X$ and $Z$ measurements of the encoded core qubits in each BSM node are successful. The successful probability is calculated from the successful iterations divided by the total iterations. The entanglement generation rate (EGR) is then calculated as follows:

\begin{align}
\label{equ:EGR}
\text{EGR} = \frac{P_{\mathrm{succ}}}{T_{\mathrm{rgs}} \times MQ_{e}}
\end{align}

where  
$T_{\mathrm{rgs}}$ is the total time to generate the RGS, and  
$MQ_{e}$ (Eq \eqref{eq:APE_matter_qubits}) is the number of memory qubits at each endpoint.

The division by $MQ_e$ is to normalize the EGR to match the ion trap repeater with one memory qubit at each endpoint. For each successful iteration, a Bell pair is created across the two ends up to a Pauli frame. As mentioned in the main text, this Pauli frame would be affected by the following: (a) all the measurement outcomes of the emitters and ancilla, (b) all the measurement outcomes of photons in BSM and SPD, and (c) the decoherence noise of the emitters and memory qubits. To determine the fidelity, we used (a) and (b) to calculate the outcomes of every successful logical $X$ or $Z$ measurement of each logical core qubit. A calculation with RGS without tree encoding in each APE node was then performed to obtain the expected Pauli frame of the Bell pair, which was used to calculate its fidelity w.r.t. the one under the noise model. The average fidelity was then calculated after 3000 successful iterations. For (6,6,3)-RGS across a chain distance of 50km, simulations were repeated until 18000 successful iterations had been reached.

\clearpage

\subsection{Simulation vs. theoretical analysis cross-validation}
\label{Appendix:ape_validation}

We carefully compare and cross-validate simulation results with theoretical analysis to ensure correctness and accuracy. Fig.~\ref{fig:ape_validation} shows that simulation study of APE quantum networks is in align with theoretical analysis.

\begin{figure}[h]
    
    \begin{subfigure}[b]{1\linewidth}
    \centering
    \includegraphics[width =0.72\linewidth]{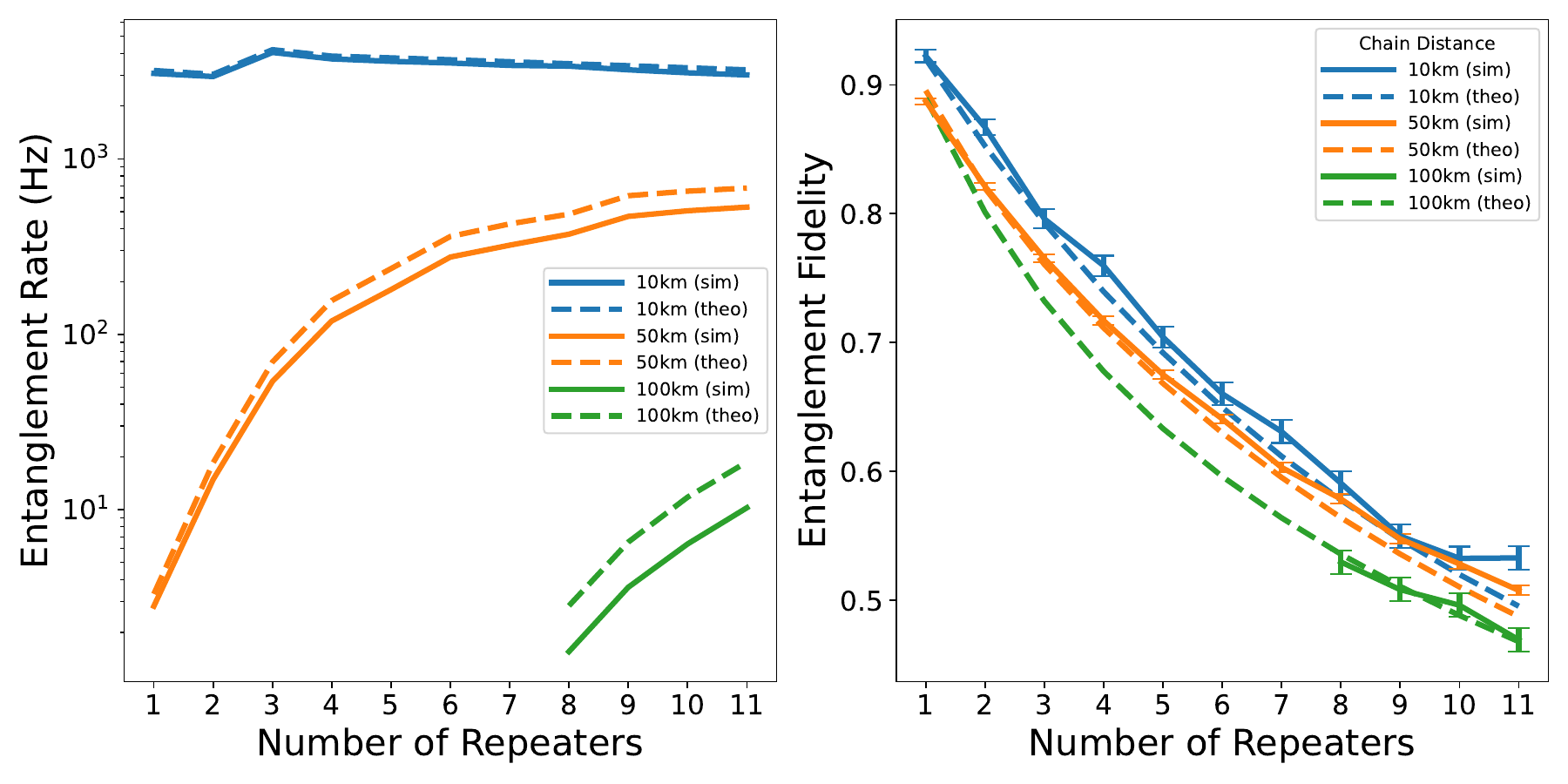}
    \caption{APE Entanglement generation rate and fidelity with the chain distance and the number of QRs varied.}
    \end{subfigure}
        
    \begin{subfigure}[b]{1\linewidth}
    \centering
    \includegraphics[width =0.72\linewidth]{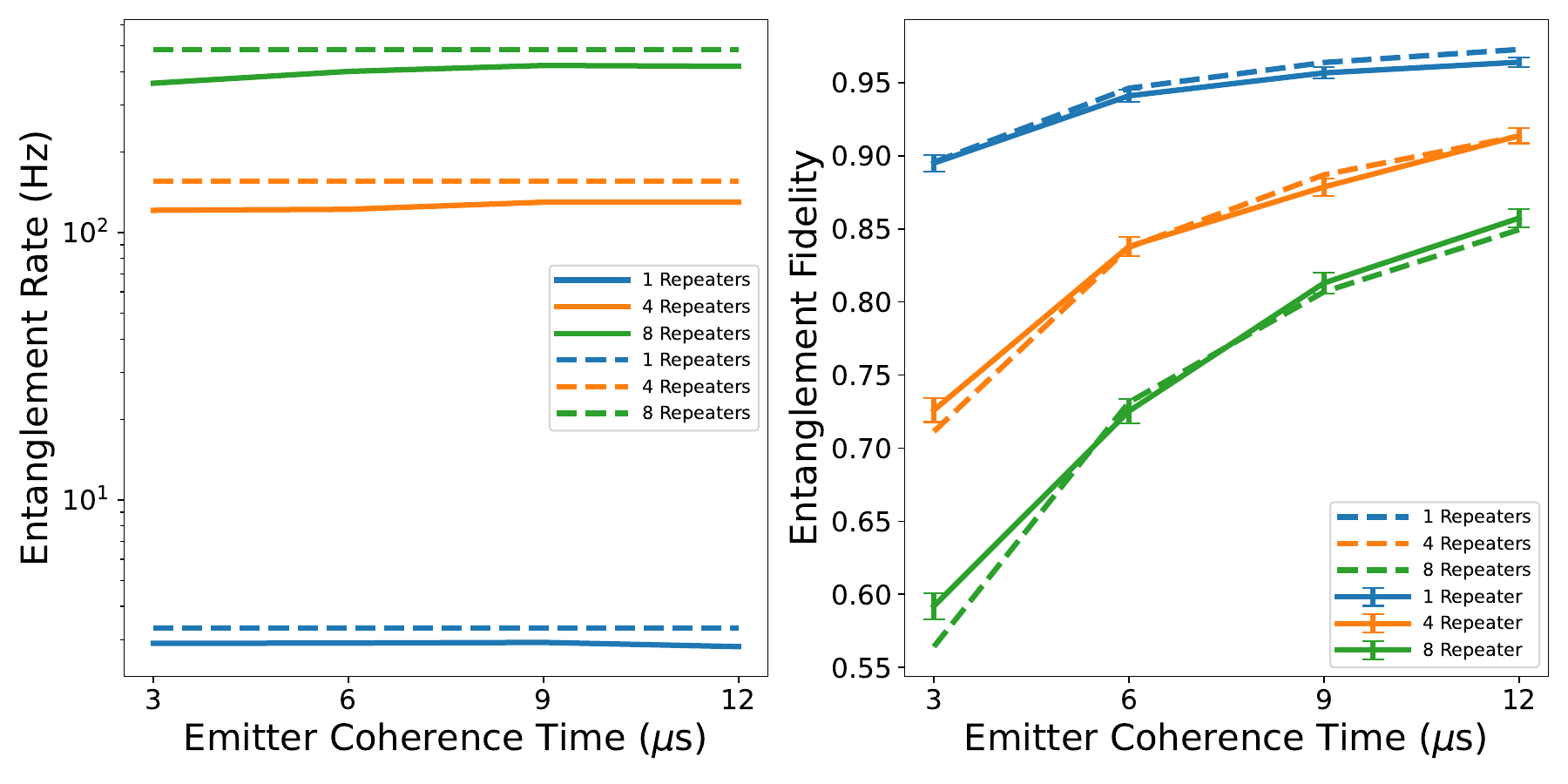}
    \caption{APE Entanglement generation rate and fidelity with the emitter coherence time and the number of QRs varied.}
    \end{subfigure}

    \caption{Simulation vs. theoretical analysis cross-validation for APE quantum networks}
    \label{fig:ape_validation}
\end{figure}

\begin{figure*}[h]
    \begin{subfigure}{\linewidth}
        \centering
        \vspace{-2mm}
        \includegraphics[width =1.0 \linewidth]{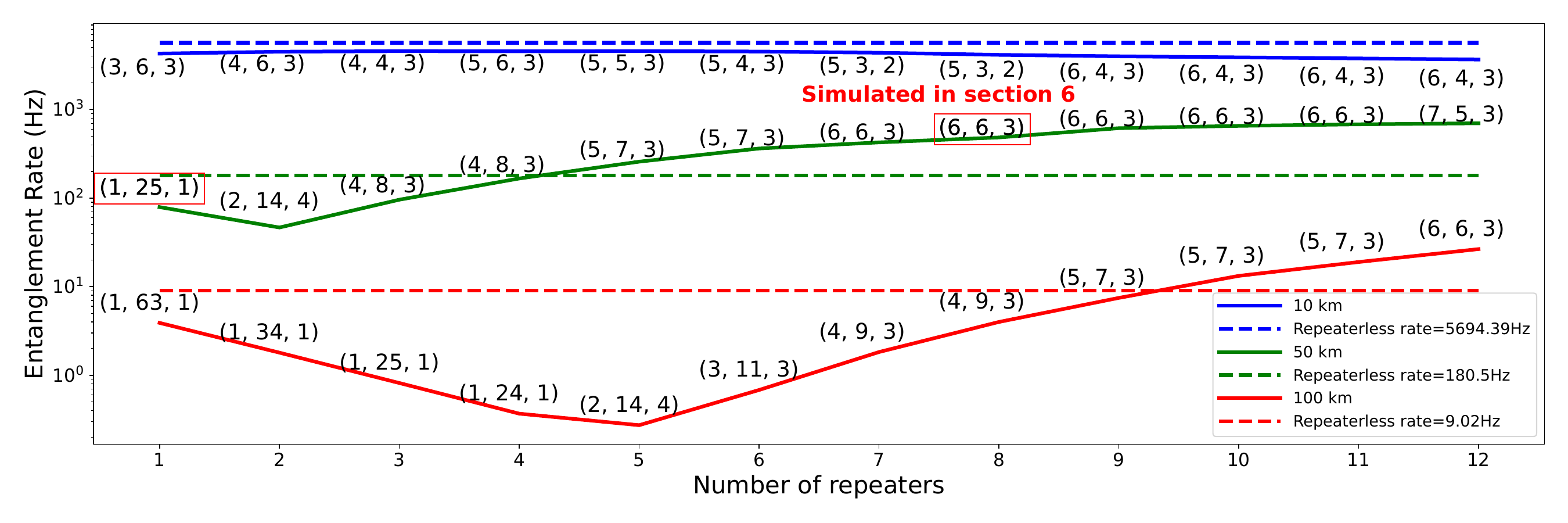}
        \vspace{-2mm}
        \caption{Optimized entanglement rate and its corresponding RGS parameters (m,$b_0$,$b_1$), optimizing over all RGS sizes with $\leq$300 photons and photon loss parameters in Table ~\ref{tab:APE-params}.}
        \vspace{2mm}
	    \label{fig:optRGS300}
        \end{subfigure}
    \vspace{\baselineskip}   
    \begin{subfigure}{\linewidth}
        \centering
        \vspace{2mm}
        \includegraphics[width =1.0 \linewidth]{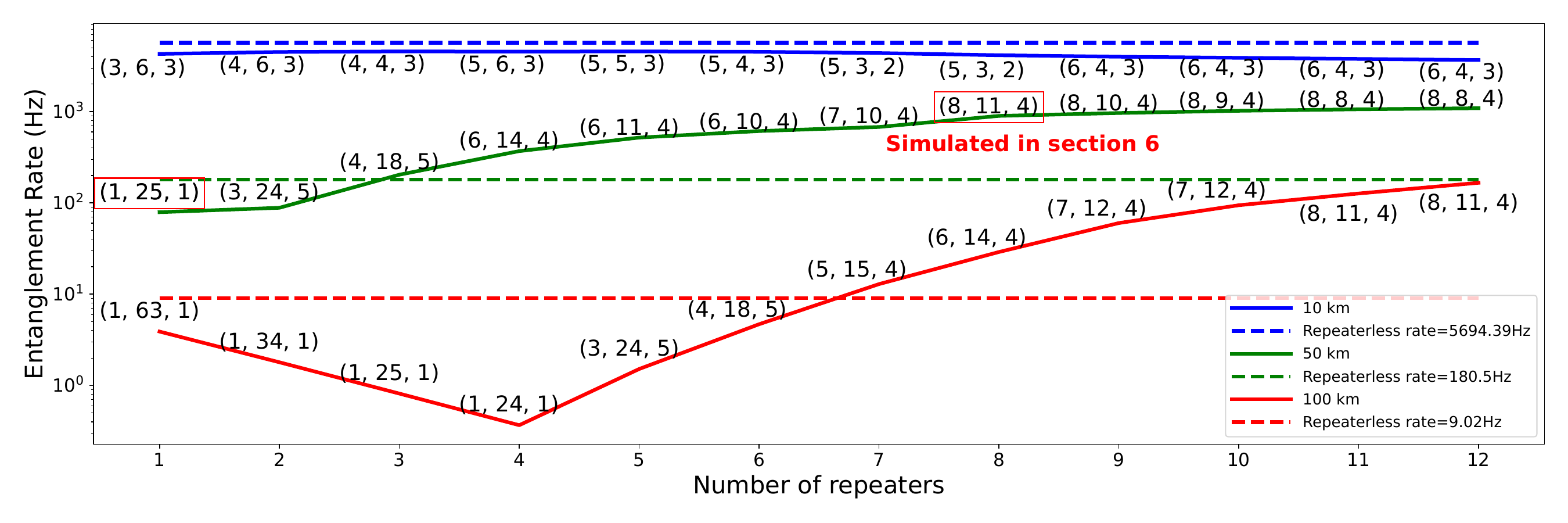}
        \vspace{-2mm}
        \caption{Optimized entanglement rate and its corresponding RGS parameters (m,$b_0$,$b_1$), optimizing over all RGS sizes with $\leq$900 photons and photon loss parameters in Table ~\ref{tab:APE-params}.}
        \vspace{-2mm}
	    \label{fig:optRGS900}
        \end{subfigure}
\end{figure*}

\subsection{Optimized RGS size}
To shed light on the meaningful RGSs to investigate in our NetSquid simulation, a preliminary analysis is conducted to determine the optimized theoretical EGR and the corresponding RGS, under maximum photon number 300(900) as shown in Fig.~\ref{fig:optRGS300}(~\ref{fig:optRGS900}). EGR is calculated with the theoretical successful probability as listed in \cite{azuma2015all} rather than the simulated successful probability; $T_\mathrm{RGS}$ is based on the parameter choices in Table \ref{tab:APE-params}. The repeaterless entanglement rates are also plotted as a baseline. We study the case with the photon loss parameters listed in Table ~\ref{tab:APE-params}. At 10 km, the optimized RGSs are the same in both $<300$ photons and $<900$ photons cases. At 50 km, the APE repeaters outperform the repeaterless entanglement rate starting at 5 repeaters for RGS $size<300$ photons, and at 3 repeaters for RGS $size<900$ photons. At 100 km, the APE repeaters outperform the repeaterless entanglement rate starting at 10 repeaters for RGS $size<300$ photons, and at 7 repeaters for RGS $size<900$ photons. 

\subsection{Derivation of APE fidelity from logical measurement error rate}
\label{Appendix:Derivation APE fidelity}
Given the logical $X$ and logical $Z$ error rate in Eq.\eqref{eq:eX} and \eqref{eq:eZ} due to emitter decoherence, we want to know their effect on the fidelity of the final entangled Bell pair upon success of the repeater scheme. As shown in Fig. 7c, after at least one BSM succeeds on each BSM node and all logical $Z$ measurements succeed at the core qubits, we are left with a linear cluster state with $2n$ core qubits and 2 end-node qubits, where the successful BSM is equivalent to fusing the cluster state together with the two photons that underwent BSM being removed. Since $\bar{e}_Z=0$, there is no error introduced by the logical $Z$ measurement so far. 

We next study how the logical $X$ measurement error propagates. For a linear cluster state with the stabilizers $<...,Z_{i-2}X_{i-1}Z_i,Z_{i-1}X_{i}Z_{i+1},Z_{i}X_{i+1}Z_{i+2},Z_{i+1}X_{i+2}Z_{i+3},...>$, two adjacent $X$ measurements on qubits $i$ and $i+1$, with outcomes $m_i$ and $m_{i+1}$, will convert it to the state $<...,(-1)^{m_i}X_i,(-1)^{m_{i+1}}X_{i+1},(-1)^{m_{i+1}}Z_{i-2}X_{i-1}Z_{i+2},(-1)^{m_{i}}Z_{i-1}X_{i+2}Z_{i+3},...>$. Hence, a flip in the outcomes $m_i$ and $m_{i+1}$ correspond to $X_{i-1}$ and $Z_{i-1}$ error.

Denote an error channel with Pauli error $\sigma$ as
\begin{align}
\mathcal{E}^{\sigma}_{1-2e}(\rho) \coloneq & (1-e)\rho +e \sigma \rho \sigma \\
=& \frac{1+(1-2e)}{2}\rho+\frac{1-(1-2e)}{2}\sigma \rho \sigma
\label{equ:error channel}
\end{align}
Since an even number of the same Pauli will cancel, we notice $\mathcal{E}^{\sigma}_{\lambda}\circ \mathcal{E}^{\sigma}_{\mu}=\mathcal{E}^{\sigma}_{\lambda \mu}$ and hence $(\mathcal{E}^{\sigma}_{1-2e})^n=\mathcal{E}^{\sigma}_{(1-2e)^n}$. Denote qubit $1$ and $2n+2$ as the memory qubit at the end nodes Q1 and Q2. With a total of $2n$ logical $X$ measurements of the linear cluster state, the final Bell pair across the two end nodes is $<(-1)^{m_{odd}}X_{1}Z_{2n+2},(-1)^{m_{even}}Z_{1}X_{2n+2}>$, where $m_{odd}=\sum_{i\,odd}m_i$ and $m_{even}=\sum_{i\,even}m_i$. This gives the following overall error channel:
\begin{align}
\mathcal{E}_{overall}(\rho)=&(\mathcal{E}^{X}_{1-2\bar{e}_X})^n(\mathcal{E}^{Z}_{1-2\bar{e}_X})^n(\rho) \\
=& (1-\bar{E}_X-\bar{E}_Y-\bar{E}_Z)\rho+\bar{E}_X X\rho X +\bar{E}_Y Y\rho Y
+ \bar{E}_Z Z\rho Z \\
\label{equ:error channel overall}
\end{align}
where 
\begin{align}
\bar{E}_X=&\frac{1+(1-2\bar{e}_X)^n}{2}\frac{1-(1-2\bar{e}_X)^n}{2}=\bar{E}_Z\\
\bar{E}_Y=&(\frac{1-(1-2\bar{e}_X)^n}{2})^2
\label{equ:error channel overall}
\end{align}
The average fidelity is given by 
\begin{align}
\bar{F}=1-\bar{E}_X-\bar{E}_Y-\bar{E}_Z
\end{align}
\end{document}